\documentclass{article}

\usepackage[english]{babel}
\usepackage[a4paper,top=2cm,bottom=2cm,left=3cm,right=3cm,marginparwidth=1.75cm]{geometry}
\usepackage{algorithm}
\usepackage{algpseudocode}
\usepackage{amsmath}
\usepackage{amsfonts}
\usepackage{graphicx}
\usepackage{epstopdf}
\usepackage[dvipsnames,table,xcdraw]{xcolor}
\usepackage[colorlinks=true, allcolors=blue]{hyperref}
\usepackage{longtable}
\usepackage{booktabs}
\usepackage[small,bf,singlelinecheck=off]{caption}%
\usepackage[square,numbers,compress]{natbib}
\usepackage{wrapfig}
\usepackage{tikz}
\usetikzlibrary{shapes, arrows, positioning, matrix, fit, graphs, graphs.standard}
\usepackage{authblk}
\usepackage{placeins}
\usepackage{xspace}
\usepackage{makecell}
\usepackage{tabularx}
\usepackage{caption}

\usepackage{mathtools}
\usepackage{hyperref}

\mathtoolsset{showonlyrefs=true}

\usepackage[most]{tcolorbox}
\newtcolorbox{newtextbox}[1][]
{colback=gray!10, arc=3pt, boxrule=0.8pt, 
 colframe=black!70,  left=8pt,
 right=8pt, boxsep=1pt, toptitle=1mm, title={\textcolor{blue!100}{#1}},
 colbacktitle=gray!20,enhanced,
 bottomtitle=1mm, coltext=blue, fonttitle=\bfseries
}

\newcommand{\ZZ}{\mathbb{Z}}
\newcommand{\NN}{\mathbb{N}}
\newcommand{\NP}{$\textbf{NP}$}
\newcommand{\RN}{\cite{gitlab_repo}}
\newcommand{\gurobi}{Gurobi\xspace}
\newcommand{\abstwo}{ABS2\xspace}

\newcommand{\customcaptiontext}{}
\newcommand{\customcaption}[2][]{\caption[#1]{\customcaptiontext#2}}

\graphicspath{{figures/}}

\title{
Quantum Optimization Benchmarking Library \\
The Intractable Decathlon}

\author[1, 2]{Thorsten Koch \thanks{koch@zib.de}}
\author[3]{David E. Bernal Neira}
\author[4]{Ying Chen}
\author[5]{Giorgio Cortiana}
\author[6]{Daniel J. Egger}
\author[7]{Raoul Heese}
\author[8]{Narendra N. Hegade}
\author[8,9]{Alejandro Gomez Cadavid}
\author[10]{Rhea Huang}
\author[11]{Toshinari Itoko}
\author[12]{Thomas Kleinert}
\author[3, 13]{Pedro Maciel Xavier}
\author[5]{Naeimeh Mohseni}
\author[14]{Jhon A. Montanez-Barrera}
\author[15]{Koji Nakano}
\author[10]{Giacomo Nannicini}
\author[5]{Corey O'Meara}
\author[16]{Justin Pauckert}
\author[6]{Manuel Proissl}
\author[3]{Anurag Ramesh}
\author[1]{Maximilian Schicker}
\author[11]{Noriaki Shimada}
\author[11]{Mitsuharu Takeori}
\author[17]{V{\'i}ctor Valls}
\author[18]{David Van Bulck}
\author[6]{Stefan Woerner \thanks{wor@zurich.ibm.com}}
\author[6]{Christa Zoufal \thanks{ouf@zurich.ibm.com}}

\affil[1]{Zuse Institute Berlin}
\affil[2]{Technische Universit{\"a}t Berlin}
\affil[3]{Davidson School of Chemical Engineering, Purdue University}
\affil[4]{National University of Singapore}
\affil[5]{E.ON Digital Technology GmbH}
\affil[6]{IBM Quantum, IBM Research Europe -- Zurich}
\affil[7]{NTT Data}
\affil[8]{Kipu Quantum GmbH}
\affil[9]{University of the Basque Country UPV/EHU}
\affil[10]{University of Southern California}
\affil[11]{IBM Quantum, IBM Research Tokyo}
\affil[12]{Quantagonia}
\affil[13]{Federal University of Rio de Janeiro}
\affil[14]{Forschungszentrum J{\"u}lich}
\affil[15]{Hiroshima University}
\affil[16]{T-Systems International GmbH}
\affil[17]{IBM Quantum, IBM Research Europe -- Dublin}
\affil[18]{Ghent University}

\begin{document}

\maketitle
\begin{abstract}
Through recent progress in hardware development, quantum computers have advanced to the point where \textit{benchmarking} of (heuristic) quantum algorithms at scale is within reach.
Particularly in combinatorial optimization--where most algorithms are heuristics--it is key to empirically analyze their performance on hardware and track progress towards quantum advantage.
To this extent, we present \textit{ten optimization problem classes} that are difficult for existing classical algorithms and can (mostly) be linked to practically relevant applications, with the goal to enable systematic, fair, and comparable benchmarks for quantum optimization methods.
Further, we introduce the \textit{Quantum Optimization Benchmarking Library} \RN{} 
where the problem instances and solution track records can be found.
The individual properties of the problem classes vary in terms of objective and variable type, coefficient ranges, and density. Crucially, they all become challenging for established classical methods already at system sizes ranging from less than $100$ to, at most, an order of $100\,000$ decision variables, allowing to approach them with today's quantum computers.
We reference the results from state-of-the-art solvers for instances from all problem classes and demonstrate exemplary baseline results obtained with quantum solvers for selected problems. 
The baseline results illustrate a standardized form to present benchmarking solutions, which has been designed to ensure \textit{comparability} of the used methods, \textit{reproducibility} of the respective results, and \textit{trackability} of algorithmic and hardware improvements over time. 
We encourage the optimization community to explore the performance of available classical or quantum algorithms and hardware platforms with the benchmarking problem instances presented in this work toward demonstrating quantum advantage in optimization.
\end{abstract}

\pagebreak

\addtocontents{toc}{\protect\setcounter{tocdepth}{2}}

\tableofcontents

\pagebreak
\section{Introduction}
\label{sec:int}

Optimization is subject to extensive research due to its importance in science and industry.
Applications include experiment design, protein folding, logistics, and finance, where better algorithms can improve performance, accuracy, and cost efficiency.
Optimization algorithms fall into three categories: provably exact algorithms, provably approximate algorithms with a priori performance guarantees, and heuristic algorithms without such guarantees, although posterior performance bounds may sometimes be available. 
Complexity theory classifies optimization problems as ``easy'' or ``difficult'', determining what types of algorithms may exist.
Many relevant optimization problems are \NP-hard, i.e., the computational resources required to find provably optimal solutions grow exponentially with the problem size. We do not expect the existence of provably exact efficient algorithms for such problems in general \cite{Abbas_2024}.
For some of them, the possible performance of provably approximate algorithms is limited by known inapproximability bounds cf.~\cite{goemans1995maxcut, khot2002uniquegames, khot2007inapproximability}.
Here, inapproximability bounds imply that solving a problem better than the given bound is \NP-hard, i.e., as hard as solving the problem to optimality.
In practice, it is often sufficient to find a good solution instead of a provably optimal one.
Thus, classic heuristics and ML-based solvers are viable approaches for many problems of interest.
Since heuristics lack a priori guarantees, rigorous benchmarking is crucial to assess the corresponding performance and resource requirements. 
To enable a fair comparison of algorithms and hardware platforms and to guarantee reproducibility of reported results, a suitable set of problem instances and metrics needs to be defined.
This also allows tracking progress in algorithm and hardware improvements over time.

Quantum computing offers a novel toolbox to approach optimization problems--see \cite{Abbas_2024} for an in-depth discussion of quantum optimization algorithms, their challenges, and potential. 
With recent advances in quantum hardware \cite{Kim2023}, systematic benchmarking of quantum optimization algorithms and comparison with classical solvers have become feasible. 
This paper proposes ten classes of combinatorial optimization problems, see Table~\ref{tab:intro:problem_classes}, and corresponding problem instances, which can be found in an open-source repository \textit{Quantum Optimization Benchmark Library} \RN{}.
These problem classes were selected because they include instances that are difficult for state-of-the-art classical solvers with only a hundred to a hundred thousand variables, have diverse characteristics, and many of them have connections to industrial applications.
This makes them suitable candidates for benchmarking and exploring the potential of quantum advantage in optimization.
Further, for each class, we provide instances of varying complexity, ranging from those already feasible for today's quantum hardware to those challenging for state-of-the-art classical solvers. 

\begin{table}[hb]
    \centering
    \begin{tabular}{ll}
     Problem & Description \\
    \toprule
    \hyperref[sec:problems:marketshare]{Market Split} & Multi-dimensional subset-sum\\
    \hyperref[sec:problems:labs]{LABS} & Low autocorrelation binary sequences\\
    \hyperref[sec:problems:birkhoff]{Birkhoff}    & Minimum Birkhoff decomposition\\
    \hyperref[sec:problems:steiner]{Steiner}   & Steiner tree packing in graphs (VLSI Design/Wire Routing) \\
    \hyperref[sec:problems:sports]{Sports}  & Sports tournament scheduling (STS)\\
    \hyperref[sec:problems:portfolio]{Portfolio}  & Multi-period portfolio optimization with transaction costs\\ 
    \hyperref[sec:problems:stable-set]{Independent Set} & Unweighted maximum independent set (MIS)\\
    \hyperref[sec:problems:network]{Network}   & Communications network design problem\\
    \hyperref[sec:problems:routing]{Routing}    & Capacitated vehicle routing problem (CVRP)\\
    \hyperref[sec:problems:topology]{Topology}  & Graph topology design (Node-Degree-Diameter problem)\\
    \bottomrule
    \end{tabular}
    \customcaption{
    The intractable decathlon: 
    Ten combinatorial optimization and feasibility problem classes that are difficult to solve with state-of-the-art solvers and can already be studied with today's quantum computers. They are introduced in detail in Section~\ref{sec:problem_classes}.}
    \label{tab:intro:problem_classes}
\end{table}

Notably, we promote model-independent benchmarking, which allows for maximal flexibility in choosing models, algorithms, and hardware platforms.
A prerequisite for our goal is the establishment of a fair and systematic benchmarking effort to advance toward demonstrating quantum advantage in optimization.
Given the large number of possibilities, exhaustively comparing algorithms is an enormous challenge that requires the scientific community's collective effort.

In addition to identifying the best approach to solving a given problem, benchmarking is critical for verifying algorithms' correctness, identifying bottlenecks, improving existing approaches, and tracking progress over time.
This includes the progress in algorithms, software libraries, and hardware platforms \cite{nation2024benchmarkingperformancequantumcomputing}.
The goal, scope, and limitations of a benchmarking effort must be clearly defined, as this determines the type of conclusions it may support.
An overview of the main goals of benchmarking is presented below in the box ``Goals in Benchmarking'', cf.~\cite{Finzgar22QUARK} for a related discussion.

The remainder of this work is structured as follows. 
First, we discuss related work focusing on quantum application benchmarking in Section~\ref{sec:rel_work}.
Then, in Section~\ref{sec:methods}, we present our benchmarking methodology, covering different modeling approaches and key metrics to ensure reproducibility, comparability, and trackability of results.
Afterward, Section~\ref{sec:problem_classes} introduces the ten benchmarking problem classes. In addition to defining the problem, this section also provides additional background, such as explaining the relevance of these problems for quantum optimization benchmarking and presenting classical baseline results.
To kick off this collaborative benchmarking effort, Section~\ref{sec:illustrative_quantum_baselines} discusses how to contribute solutions to the open-source repository and provides illustrative quantum baseline results for selected problem instances.
We conclude in Section~\ref{sec:discussion_conclusion}.

\begin{newtextbox}[Goals in Benchmarking]
\textit{Applications Benchmarking:} The goal is to find the best possible algorithms--classical or quantum--to solve a given problem instance. Thus, benchmarks must be model-independent to allow all possible approaches to solve a problem. This is the only level that ultimately allows demonstrating quantum advantage.
\\
\\
\textit{Algorithm Benchmarking:}
The goal is to identify suitable strategies for setting hyperparameters, identifying bottlenecks, improving algorithms, and tracking progress over time. Since algorithm benchmarking does not entail comparing against all possible algorithms, it will not allow demonstrating quantum advantage. Nevertheless, it can be used to estimate an algorithm's scaling, which may facilitate the identification of potential asymptotic scaling advantages and help track progress toward quantum advantage.
\\
\\
\textit{System Benchmarking:}
The goal is to identify the best way to run a fixed algorithm for a fixed problem on a given platform. This includes tuning algorithmic hyperparameters or parameters of the execution environment, e.g., for error suppression and mitigation, and to confirm that the algorithm is working as expected. It can also be used for application-centric hardware benchmarking.
\end{newtextbox}

\section{Related Work}
\label{sec:rel_work}

A general introduction to benchmarking can be found in~\cite{dunning2018works, JainComputerPerformance91, KochBertholdPedersenetal2022,bartzbeielstein2020benchmarking, Muller2011Spec,dongarra1979linpack}.
Many important classical optimization benchmarks have been established via competitions and open-source libraries. Notable examples are the Discrete Mathematics and Theoretical Computer Science (DIMACS) implementation challenges \cite{dimacs23}, the Satisfiability (SAT) competitions \cite{froleyks2021sat,SATcompetition}, and the Mixed Integer Programming Library (MIPLIB) \cite{GleixnerHendelGamrathetal2021}. 
Another set of optimization benchmarks is regularly conducted by Hans Mittelmann \cite{mittelmann2023bench}. Here, different optimization pipelines are executed using the same computational setup.

Most benchmarks in quantum computing focus on the performance of individual components of the quantum computing stack and their integration~\cite{proctor2024benchmarkingquantumcomputers, acuaviva2024benchmarkingqc, lorenz2025systematicbenchmarkingquantumcomputers}:
Randomized benchmarking focuses on gate quality~\cite{magesan2012interleaved}.
The layer fidelity represents a quality metric of layers of two-qubit gates~\cite{mckay2023benchmarking}.
Circuit Layer Operations Per Second (CLOPS) measures the speed of a quantum computer~\cite{nation2024benchmarkingperformancequantumcomputing, Wack2021}.
Speed and quality of compiling quantum circuits from high-level abstract representations to machine-level instructions have been investigated and reported in \emph{Benchpress}~\cite{nation2024benchmarkingperformancequantumcomputing}.
In addition, application-driven benchmarks have been proposed as full-stack system benchmarks. They usually focus on problems that can be solved to optimality classically and track the progress of quantum computers to achieve a known optimal solution \cite{granet2025appqsimapplicationorientedbenchmarkshamiltonian, Martiel_2021, lubinski2024optimizationapplicationsquantumperformance, tomesh2022supermarq, Santra2024, Finzgar22QUARK}.
While this allows tracking the progress of hardware and algorithms, it is not sufficient to demonstrate quantum advantage.

This work presents a selected set of optimization problem instances that are difficult for state-of-the-art algorithms and corresponding benchmarking metrics that enable a fair and reliable comparison.
See \cite{mcgeoch2024foolmassesgivingperformance, bernal2024benchmarking, lall2025reviewcollectionmetricsbenchmarks} for further discussion on the importance of appropriately chosen benchmarking metrics. The presented set of problem instances enables investigating potential advantages of novel algorithms and computational platforms.
Furthermore, it facilitates tracking the progress towards potential quantum advantage in optimization. 
This is comparable to the curated set of many-body quantum systems presented in \cite{Wu_2024_VariationalBenchmarksQuantumManyBody}--but for optimization.

\section{Methods}\label{sec:methods}

In this section, we introduce our methodology for quantum optimization benchmarking. 
More specifically, we discuss what general problem types are considered, different ways to model them, and metrics to measure the performance of a specific optimization workflow.

\subsection{Problem Types}

We consider ten classes of combinatorial optimization problems. Schrijver \cite{Schrijver2003} describes combinatorial optimization as follows: 
\emph{``Combinatorial Optimization searches for an optimum object in a finite collection of objects. Typically, the collection has a concise representation, while the number of objects is huge--more precisely, it grows exponentially in the size of the representation. So scanning all objects one by one and selecting the best one is not an option''.} 
There are other classes of optimization problems for which quantum solvers have been proposed, including several classes of polynomial-time solvable convex optimization problems; see~\cite{Abbas_2024} for an overview. We focus on combinatorial optimization for the following two reasons: 
First, quantum algorithms for convex optimization usually require fault-tolerant quantum computers. Hence, they cannot be tested on existing quantum devices.
For combinatorial optimization, on the other hand, multiple (heuristic) approaches are known that can be implemented and tested already today.
Second, classical computers excel at solving convex optimization problems even at massive scales, while combinatorial optimization problems can be intractable for classical algorithms already at a relatively small number of variables.
This increases the possibility that near-term quantum computers outperform classical ones in certain cases.

More specifically, we consider combinatorial optimization problems with $n \in \NN$ decision variables of the form
\begin{equation}
x^*=\operatorname*{argmin}_{x\in X} f(x),
\label{eq:comboptprob}    
\end{equation}
with feasible space $X = \{x \in \ZZ^n \mid g(x) \leq b,\, l\leq x \leq u\}$, where $g: \ZZ^n \rightarrow \ZZ$, $b \in \ZZ, l, u \in \ZZ^n$, and objective function $f: X \rightarrow \ZZ$.
It should be noted that the coefficients are (if necessary) re-scaled and rounded to integers to mitigate potential numerical accuracy issues and to simplify the solution.
We distinguish between \emph{feasibility problems} and \emph{optimization problems}. For feasibility problems, we can assume $f$ is constant, and we are interested in finding any solution $x^\ast\in X$ that satisfies all given constraints, or in showing that $X=\emptyset$. For optimization problems, we have a non-constant objective function to be minimized or maximized, in addition to potential constraints to be satisfied. If an algorithm returns a solution $\hat{x}\in X$, we call the solution \emph{feasible}. If in addition, $f(\hat{x})=\min_{x\in X} f(x)$, we call the solution \emph{optimal}. Multiple optimal solutions may exist.

From the theoretical side, both problem types are closely related, as we can solve optimization problems via a sequence of feasibility problems by adding a constraint for a certain objective value $k$, i.e., $X_k=\{x \in X \mid f(x)\le k \}$, and asking whether $X_k = \emptyset$. The value of $k$ can then be iteratively adjusted, for example, using binary search, until the optimal value is found.
We can also make the problem unconstrained, for instance, by removing the constraint and instead minimizing
$\bar{f}(x)=f(x) + M\cdot \max\{0, g(x) - b\}$, $X = \{x \in\ZZ^n \mid l\leq x \leq u\}$ for a suitably large $M$, see Section~\ref{ssec:modeling} for a more detailed discussion.

In practice, a good solution without proof of optimality or provable bound on the approximation ratio is often sufficient.
If a formal guarantee on the solution quality is not required, finding a feasible solution with a good objective value is often achievable with classical heuristics, although challenging problem instances exist.
We restrict our benchmarking study to computing good solutions without requiring a proof of optimality. First, because we want to assess practical performance, and second, because most known quantum optimization algorithms are heuristics and do not provide any a priori or a posteriori performance guarantees.

\subsection{Modeling}\label{ssec:modeling}

Mathematical programming became a fundamental tool in many disciplines thanks to the existence of high-performance solvers that can find good--often even optimal--solutions for large classes of problems. Mathematical programming provides a way to classify models according to their structure and match model families with their respective specialized algorithms.
The invention of more advanced solvers led to a shift in how problems are modeled and solved.
As an example, the impactful success of solvers for \emph{Mixed Integer Programming} (MIP) resulted in many applications that were reformulated into linear formulations to benefit from highly performant implementations. Modern MIP solvers can solve--often to proven optimality--Mixed Integer Linear and Quadratic Programs (MILP/MIQP), as well as their pure Integer and Binary variants, ILP, IQP, BLP, and BQP, respectively. 

The choice of problem formulation, even in the same model class, can have a tremendous impact on the solvability of a problem \cite{AchterbergKochTuchscherer2008}. 
While the global optimum is typically the same for different problem formulations\footnote{Variations may occur, for example, in relaxed, e.g., linear or semi-definite programming formulations.}, one might be better suited for a particular problem instance or algorithm than another.
The development of novel hardware and algorithmic devices to solve \emph{Quadratic Unconstrained Binary Optimization} (QUBO) problems led to a renewed interest in \emph{formulations} of problems as QUBO~\cite{Lucas2014,glover2018tutorial}.
While all \emph{Integer Linear Programs} (ILP), or \emph{Integer Quadratic Programs} (IQP)
can be arbitrarily approximated by some QUBO, naive reformulations will often contain a large number of densely connected variables and will require outstanding precision for the coefficients to accurately represent the original structure~\cite{alessandroni2024alleviatingquantumbigmproblem}. 
For example, ILP can explicitly represent different constraints or variables that can be difficult to encode into a QUBO, since many constraints can only be incorporated implicitly into QUBO via penalty terms that vanish for feasible solutions.
In addition, to realize a QUBO representation, integer variables must be decomposed into binary ones--a useful library to handle this conversion can be found at \cite{QUBO++}. 
Direct translation from a general ILP to a QUBO requires making decisions on hyperparameters that define the encoding and penalty functions~\cite{glover2018tutorial} and can impact the quality of a solution depending on the considered algorithm.
Furthermore, comparing the results received from multiple formulations of a problem requires careful interpretation.
The objective values of two QUBOs representing the same problem using different penalty factors cannot be directly compared. 
Therefore, solutions must be mapped back to the original problem to understand and compare their performance.
Higher-level modeling languages exist that provide additional structure to accommodate a problem's original semantics more naturally, and multiple tools can automate reformulating from a high-level representation into, for instance, ILP or QUBO; see, e.g.,~\cite{Koch2004,xavier:2023.qubo,volpe:2024}.
Thus, the fact that a problem can be cast into a certain form does not necessarily mean that, in practice, it is a good idea to solve it in that form.
Given the inherent formulation differences, we focus on \textit{model-independent benchmarking} such that one may find the best formulation and corresponding algorithm for a given problem instance. 
Another type of reformulation allows mapping \emph{Higher-Order Unconstrained Binary Optimization} (HUBOs), which may either be solved directly \cite{Pelofske:2024swm} or converted to QUBOs \cite{Anthony2017QuadraticReformulation, ROSENBERGIG1975}. To this extent, higher-order binary terms are broken into multiple quadratic terms by introducing additional binary variables. Depending on the number and degree of the higher-order terms, this can significantly increase the number of binary variables, potentially leading to impractical problem formulations. Some quantum optimization algorithms can handle higher-order terms natively, giving them a potential advantage over those algorithms (quantum and classical) that cannot.
Notably, many quantum optimization algorithms are designed for QUBOs. A common approach is, thus, to start with a higher-level problem formulation and reformulate it as QUBO.  
Tutorials on QUBO reformulations can be found, e.g., in \cite{glover2018tutorial,xavier:2024.disjunctive}.

\subsection{Metrics}\label{ssec:metrics}

The different possible goals of running a benchmark, as described in Section~\ref{sec:int}, require different metrics to be considered. 
To ensure that benchmarking results allow for a fair comparison, the reported metrics must be clearly defined.
In the following, we will focus on \textit{solution quality} and required \textit{computational resources} and discuss the corresponding metrics.
Examples of how to report benchmarking results for selected problem instances and quantum algorithms are presented in Section~\ref{sec:illustrative_quantum_baselines}.

\subsubsection{Solution Quality}

The metrics for evaluating solution quality differ between feasibility and optimization problems. 
A feasibility problem is completely solved as soon as a feasible solution is found.
However, proving that no such solution exists is usually significantly more challenging and impossible for most heuristics.
For optimization problems, the key metric to be reported is the \emph{best objective value} corresponding to a \emph{feasible solution} found by an algorithm. If the algorithm also provides a lower bound (for minimization) or an upper bound (for maximization) for the optimal objective value, this bound can be provided as well.
To solve an optimization problem to \emph{proven optimality}, an algorithm must first find an optimal $x^*\in X$ and then prove that no better $x\in X$ exists with $f(x) < f(x^*)$, e.g., by providing a tight bound on the objective value.
Usually, the second step is the more difficult one.

The above respective algorithms may run deterministically or, as many classical and most quantum algorithms do, stochastically. 
In the latter case, additional details should be reported.
Stochastic algorithms should be executed multiple times to ensure the reported results describe the typical behavior of the algorithm. 
Then, for both feasibility and optimization problems, the number of repetitions yielding feasible solutions should be reported.
For optimization problems, additionally, the best objective value achieved by a feasible solution found across all repetitions should be stated, as well as the number of \emph{successful} repetitions. Given a number of repetitions of a stochastic algorithm and a given tolerance $\epsilon \geq 0$, we consider a single run successful if it finds a feasible solution with an objective value that is $\epsilon$-close to the best-found feasible solution across all repetitions. 
More precisely, suppose the best found solution over all repetitions, denoted by $x'$ with corresponding objective value $f(x')$, which may differ from the global optimum $f(x^*)$. Then, a solution $x$ corresponding to a successful run must satisfy $f(x) \leq (1 + \epsilon) f(x')$ (minimization) or $f(x) \geq (1 - \epsilon)f(x')$ (maximization)--assuming non-negative objective values. 

Notably, one may also want to track metrics that present combined information about solution quality and the time it takes to arrive at a particular quality. One example thereof is the \textit{primal integral} \cite{BERTHOLD2013PrimalIntegral}, which tracks the primal bound towards the optimal (or best known) solution throughout the iterations executed by a solver. This approach has been found to be suitable for comparing the performance of solvers that do not necessarily return optimal solutions. 
This includes machine learning-based optimization solvers \cite{kool2019attentionlearnsolverouting, bi2024learninghandlecomplexconstraints, berto2025rl4coextensivereinforcementlearning, luo2024neuralcombinatorialoptimizationheavy}--trained on a large number of known instances--which can, in certain cases, predict good solutions quickly.  

\subsubsection{Computational Resources}
\label{sec:resources}

To compare different methods and algorithms, it is crucial to transparently report the type of computational resources used and the duration for which they were utilized.
Thus, software versions and hardware specifications, such as the type of processing units used and the amount of memory available, must be reported, as discussed in more detail below.

First, the \textit{total runtime (s)}, i.e., the total wall-clock time for running the considered algorithm using the reported hardware, excluding idling times, e.g., queuing for execution on a (quantum) computer while the remaining processing units are waiting. 
For stochastic algorithms that are repeated multiple times, the average total runtime over all repetitions should be reported.
If the algorithm is parallelized and uses multiple processing units simultaneously, this should be noted in the hardware description but not translated into a single-core runtime. The definition of the total runtime, excluding idling times, can be easily measured and represents the runtime most relevant for practical applications.
In certain cases, it may also be useful to report the time-to-solution (TTS), i.e., the time from the algorithm's start until the returned solution was found. 

Second, to better understand how different hardware is utilized, reported runtimes should be split into the execution time on \textit{CPUs}, \textit{GPUs}, \textit{quantum computers}, and \textit{alternative hardware}, such as, for instance, Field Programmable Gate Arrays (FPGAs) or Ising machines.
How to accurately define, measure, and report runtime on a quantum computer is discussed in Section~\ref{sssec:qc_runTime} below.

There are many additional details of interest, such as the intermediate problem sizes after pre-solving or the time and resources spent in different stages of the workflow: pre-solving, pre-processing (circuit preparation, etc.), training of machine learning-based solvers, post-processing, etc. While it is difficult to capture all of these details in a submission table (see Table~\ref{tbl:metrics}), we strongly encourage benchmark submitters to provide such details in an accompanying reference, such as a paper or repository.

\subsubsection{Measuring Quantum Computing Runtimes}
\label{sssec:qc_runTime}

\begin{figure}[ht!] 
\centering
    \includegraphics[width=0.6\textwidth]{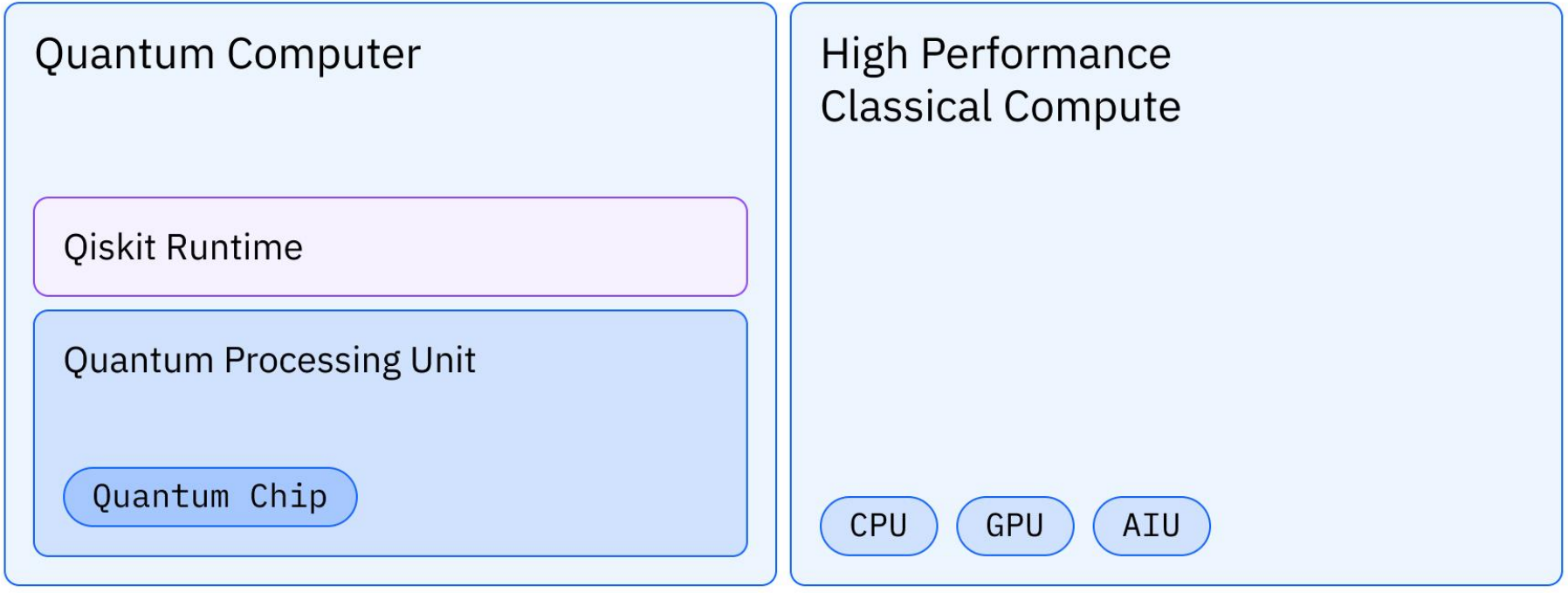}
\customcaption{Illustration of a computational environment that provides access to quantum and classical resources. On the left side, it shows the components of an IBM Quantum computer: the Qiskit Runtime environment \cite{qiskit}, the Quantum Processing Unit (QPU), and the actual quantum chip. The QPU consists of the quantum chip and control electronics such as interconnects, amplifiers, and signal filtering. The right side illustrates typical classical high-performance computational resources such as central processing units (CPU), graphical processing units (GPU), and artificial intelligence units (AIU), e.g., tensor processing units. These classical resources may be employed by an algorithm in synergy with the quantum resources to achieve the best possible result. Credits: IBM Quantum~\cite{gambetta24_blog}.}
\label{fig:quantum_computer}
\end{figure}

In the following, we define quantum computing runtime and how to measure it on the example of the IBM Quantum Platform \cite{ibm_quantum_platform, mandelbaum24_blog}; see Figure~\ref{fig:quantum_computer}.
A quantum computer consists of multiple stages, including a runtime environment, such as the \textit{Qiskit Runtime}, the control electronics, and the actual quantum chip. 
The payload sent to a quantum computer needs to pass these stages, and even though some steps are performed using classical computing, such as, e.g., compiling a quantum circuit to microwave pulses, we attribute them to the runtime on the quantum computer.
Thus, we define the runtime of a \textit{single run} of a quantum (sub-) routine as the total time it takes to receive the payload, prepare it, run the circuit on the QPU, and extract and return samples from the system by measuring the qubits.
We would like to highlight that the number of samples drawn from a quantum circuit corresponds to a hyperparameter that has to be chosen as part of the considered quantum algorithm.
Furthermore, we assume that circuits are provided using a topology and gate set that is natively supported by the quantum computer. Thus, while we include the time it takes to prepare such a circuit for execution, e.g., to microwave pulses, we explicitly exclude the time it takes to map a given circuit to the instruction set supported by a particular quantum computer. This additional transpilation time should be considered during classical pre-processing. 
Similarly, we attribute time spent on error mitigation to the quantum computer, if it happens as part of the runtime environment, and to the classical processing units, if it is handled via post-processing.

Last, we outline how to measure the quantum computing runtime in the most reliable way. When using an IBM Quantum computer, one should run quantum circuits using the Qiskit Runtime \emph{Session}-mode~\cite{qiskit} instead of \emph{Batch}- or \emph{Job}-mode. Once a session has been started, it blocks the whole quantum computer for a user, in contrast to batch and single-job mode, which may share resources--such as circuit preparation pipelines--across different users and, thus, may not provide accurate runtime estimates.
After a job has been executed, the results returned from the IBM Quantum Platform \cite{ibm_quantum_platform} contain the session start and end times, as well as the times individual jobs were submitted, started, and completed.
In addition, the results report actual usage, which refers to the time that was actually spent on the QPU, excluding classical pre- and post-processing inside the quantum computer.
Due to the potential parallelization of circuit preparation pipelines before executing them on the QPU, we cannot just add the runtimes of the individual jobs.
Instead, we take the total time the session is active, i.e., the difference between start and end time excluding initial queuing time, and further subtract all times where the session was active but idling, i.e., while no jobs were submitted and we were not waiting for previously submitted jobs to be completed. 
This is what we define as the quantum computer runtime, while the idling time in between jobs should be attributed to, e.g., the classical runtime.

\section{Problem Classes \& Instances}
\label{sec:problem_classes}

This section presents ten problem classes and corresponding instances that are non-trivial to solve with state-of-the-art methods--for most of them already at relatively small scales. Furthermore, the respective problem classes are mostly related to practically relevant problems.
Such problems are hard to find for two reasons: First, unsolved problems of industrial relevance are rarely published.
Second, problems of practical relevance that cannot be solved are usually discarded and replaced by artificial simplifications of the original problem. 
This leads to a \textit{selection bias} and may be one reason why a benchmarking framework that includes truly difficult (for reasonable system sizes) and relevant (ideally in practical settings) optimization problems that enable a meaningful and fair comparison of quantum and classical algorithms has not been established until now.

Problem instances can be loosely classified into three categories, which are not always mutually exclusive: 
\textit{Real-world instances} arise from real-world problems, and are therefore crucial for demonstrating a practical quantum advantage. Since there is already a demand for a solution, such instances have often been modeled (and possibly simplified) such that they can be tackled by existing solvers. The availability of real-world instances to the scientific community is often limited, even more so for unsolved problems as discussed above. 
\textit{Random instances} are based on randomly generated data, which may be motivated by real-world problems or completely artificial. Depending on the generation method, such instances often do not capture the structural peculiarities of a real-world setting, and may be easy or hard to solve.
\textit{Crafted instances} are most likely the best way to construct artificial examples that are hard to solve. However, there is a risk that the instances are of no practical relevance, as they often represent borderline pathological cases; see, e.g., \cite{HougardyZhong2021,VercesiEtal2023}.
A quantum advantage for random or crafted instances is interesting, as it may help to identify real-world problems with similar structure.
The problem instances presented here were mostly crafted to fit a range of criteria we identified as crucial for a useful optimization benchmarking framework, while they are still mostly related to real-world problems.

Problem instances can be further classified as \emph{primal} difficult, \emph{dual} difficult, or both.
Primal difficulty refers to instances where it is difficult to \textit{find} an optimal solution, dual difficulty refers to instances where it is difficult to \textit{prove} a solution is optimal. 
An astonishing number of instances are primal easy but dual hard.
Before solving an instance, it is a priori unclear which is the case. 
One may, however, gain intuition about the difficulty of a problem instance from solutions of other instances of the same problem class, since instances within the same class often behave similarly.
Since most currently known quantum optimization algorithms are heuristics, we consider the chances for demonstrating a near-term quantum advantage higher for primal difficult problems and do not yet expect a benefit for proving optimality, although that might be possible later. 
Hence, we mainly present problems that are considered primal difficult. 
Further, if it is easy to find a very good or even optimal solution, the problem is not difficult from a practical point of view, and thus, there is less potential for demonstrating practical quantum advantage.

We applied the following criteria for the selection of the problems presented in the remainder of this section. First, we consider problems that can be modeled using only integer variables, preferably binary ones, and all coefficients can be represented as integers.
In an all-integer setting, checking the feasibility of solutions is easy and does not suffer from issues with rounding or numerical accuracy, and integer variables are more amenable for known quantum optimization algorithms without requiring too many qubits. 
Second, problem classes should contain instances that are hard to solve for state-of-the-art classical approaches, preferably even at a relatively small size, and preferably because of primal difficulty. This enables testing known quantum optimization algorithms already today.
Third, all selected instances have a feasible solution. For those problem classes that are not genuinely feasible, we generated the instances in a way that guarantees feasibility, as discussed in the following sections.  

Table~\ref{tab:problem_classes} provides an overview of the problem classes and a summary of their key properties.
In particular, it highlights challenges that may be introduced when translating problem instances from Integer Programming to QUBO. This includes increasing the number of variables, conversion of sparse to dense problems, or significantly larger ranges of coefficients. 
This shows the importance of selecting the right modeling approach for a problem.

\begin{table}[hb!]
\centering
\resizebox{\textwidth}{!}{%
    \begin{tabular}{l|c|c|r|c|c|c|c|c|r|c}
    \toprule
    & \multicolumn{7}{c}{\textbf{MIP}} & \multicolumn{3}{|c}{\textbf{QUBO}}\\
    \hline
    \textbf{Problem} & \textbf{Type}  & \textbf{Dense} & \textbf{\#Vars} & \textbf{Coeffs} & \textbf{Constr} &\textbf{Feas} & \textbf{Bin} 
    & \textbf{Dense} & \textbf{\#Vars} & \textbf{Coeffs} \\
    \hline
    \hyperref[sec:problems:marketshare]{Market Split} & F / L  & d     & 78 & $48$  & $1$ & no  & yes & d & 70 & $\sim 5\cdot10^{6}$\\
    \hyperref[sec:problems:labs]{LABS}            &  Q / Q     & s     & 81 & 4
    & $1$ & yes & yes & s & 820 & $\sim 4\cdot10^4$    \\
    \hyperref[sec:problems:birkhoff]{Birkhoff}    & L / L    & s  & 240 & $10^4$ & $3$ & yes & no & d & 3,480 & $\sim 3\cdot 10^{10}$  \\
    \hyperref[sec:problems:steiner]{Steiner}      &  L / L    & s & 423,360 & $3$
    & $9$ & no  & yes & - & - & -  \\
    \hyperref[sec:problems:sports]{Sports} 	      &  F / L  & s & 8,608 & $2$ & $>10$ & no  & no & s & $11,791$ & $\sim 4.5\cdot10^{3}$ \\
    \hyperref[sec:problems:portfolio]{Portfolio}  &  Q / L    & \phantom{$^*$}s$^*$  & 690 & $\sim 3 \cdot 10^{4}$ & $2$ & yes & yes & d & 690 & $\sim 2 \cdot 10^9$  \\
    \hyperref[sec:problems:stable-set]{Independent Set} &  L / L & s & 500 & $1$ & $1$  & yes & yes & d & 500 & 2    \\
    \hyperref[sec:problems:network]{Network}      &  L / L & s & 1,211 & $10^6$ & $5$ & yes & no & s  & 46,330 & $\sim 2.5 \cdot 10^{19}$ \\
    \hyperref[sec:problems:routing]{Routing}      &  L / L & s  & - & - & $<10$ & yes & no & s & - & - \\
    \hyperref[sec:problems:topology]{Topology} 	  &  L / Q  & s  & 2,176 & $2$ & $4-7$ & yes & no & s & - & - \\
    \bottomrule
    \end{tabular}
}  
\customcaption{
    Problem classes overview - details on illustrative MIP and QUBO formulations, as applicable. 
    \textbf{Type} denotes the objective/constraint type as linear (L) or quadratic (Q) for either if the problem corresponds to an optimization problem. For feasibility problems, the objective type is indicated as feasibility (F). 
    \textbf{Dense} denotes whether the model is dense (d), defined as $|A|\geq\frac{1}{4}nm$, where $A \in \mathbb{R}^{m \times n}$ is the constraint matrix of a MIP and $n, m$ denote the number of variables and constraints, respectively, or sparse (s), otherwise. For QUBO models, density is defined as $|Q| > \frac{1}{8}n^2$, where we assume $Q$ is an upper-triangular cost matrix (due to symmetry, and thus, the reduced pre-factor of $\frac{1}{8}$ instead of $\frac{1}{4}$ used for MIP). Again, $n$ denotes the number of variables. Portfolio has a sparse constraints matrix but a dense quadratic objective, which we indicate by $s^*$.
    \textbf{\#Vars} denotes the approximate number of decision variables of the models corresponding to instances where the problems can become difficult for classical methods to solve to proven optimality in MIP formulation. Hence, the number of decision variables in the QUBO column denotes the size of the QUBO formulation, where solvers struggle to solve the corresponding MIP. This is to highlight how certain problems become significantly larger when represented as QUBO instead of an MIP. 
    \textbf{Coeffs} is the approximate coefficient range, i.e., the maximum range of all coefficients in the models. Again, we can have a significant discrepancy between MIP and QUBO representations for certain classes.    
    \textbf{Constr} Number of different constraint types in the MIP formulation ($\neq$ total number of constraints). This indicates the complexity of modeling the problems.
    \textbf{Feas} denotes whether it is trivial to find or construct a (initial) feasible solution to the problem or not. For QUBO, all solutions are feasible, and thus, we drop the column.
    \textbf{Bin} indicates whether all variables in the MIP problem are naturally binary or not. This is always true for QUBO and, thus, not reported. If this is not the case, we see a discrepancy between the number of variables in the MIP and the QUBO models. For LABS, we have natural binary variables; however, we receive quartic terms in the objective from adding squared quadratic equality constraints as penalties that need to be decomposed into quadratic terms by adding binary variables.
    QUBO instances requiring more than $1.6\cdot 10^5$ variables have not been generated due to computational restrictions. We indicate this in the respective columns (-). Further, no values are given for Routing problems since we were able to find the optimal solution for all tested instances.
}
\label{tab:problem_classes}
\end{table}

The set of problems was selected to be diverse and can be grouped into three categories:
\begin{itemize}
    \item Market Split, Maximum Independent Set, and Network Design are classic binary optimization problems. They have been studied intensively before the year 2000, cf.~corresponding sections. Since then, only limited efforts have been put into improving the state-of-the-art in targeted exact solvers or heuristics. We attribute this to the fact that general MIP solvers and known heuristics are good enough for many practically relevant settings. 
    However, these problem classes are well-established, can be effectively solved at small sizes, and become difficult to solve exactly for growing dimensions. The latter is particularly true for Market Split and Network Design.
    It follows that these problems are great candidates for testing, tracking, and pushing the capabilities of quantum algorithms.
    
    \item  LABS, Minimum Birkhoff Decomposition, Steiner Tree Packing, Sports Scheduling, and Topology Design are problem classes that received limited attention compared to, e.g., Traveling Sales Person, Independent Set, or Knapsack problems. Thus, it may be that the known state-of-the-art classical algorithms could still be significantly improved. At the same time, quantum algorithms may offer an interesting approach to solving these types of practically relevant problems, especially because LABS and Minimum Birkhoff Decomposition are already difficult at very small sizes. Steiner Tree Packing and Sports Scheduling become interesting at larger system sizes, but they have many practically relevant applications and, especially for the latter, even heuristics are struggling. 
    
    \item Portfolio Optimization and Vehicle Routing problems have obvious practical applications. Companies face diverse variations of these problems depending on industry and operational execution. Hence, there are indefinitely many variations of these problems depending on individual constraints and/or objectives, resulting in problem instances of varying difficulty. For this reason, it is difficult to define a small set of problem instances to benchmark and compare established approaches, which are representative of these problem classes. In addition, many established solvers from commercial software vendors are not accessible for open-source benchmarks. Despite the practical relevance of these problem classes, no well-established benchmark set exists for testing the progress of quantum algorithms. We aim to fill this gap by providing a selection of relevant problem instances.
    However, it should be noted that all given Vehicle Routing problem instances can be solved to optimality with existing methods. This is a shortcoming that we aim to address in the future.
\end{itemize}

\begin{figure}[htb!]
\centering
\includegraphics[width=0.85\textwidth]{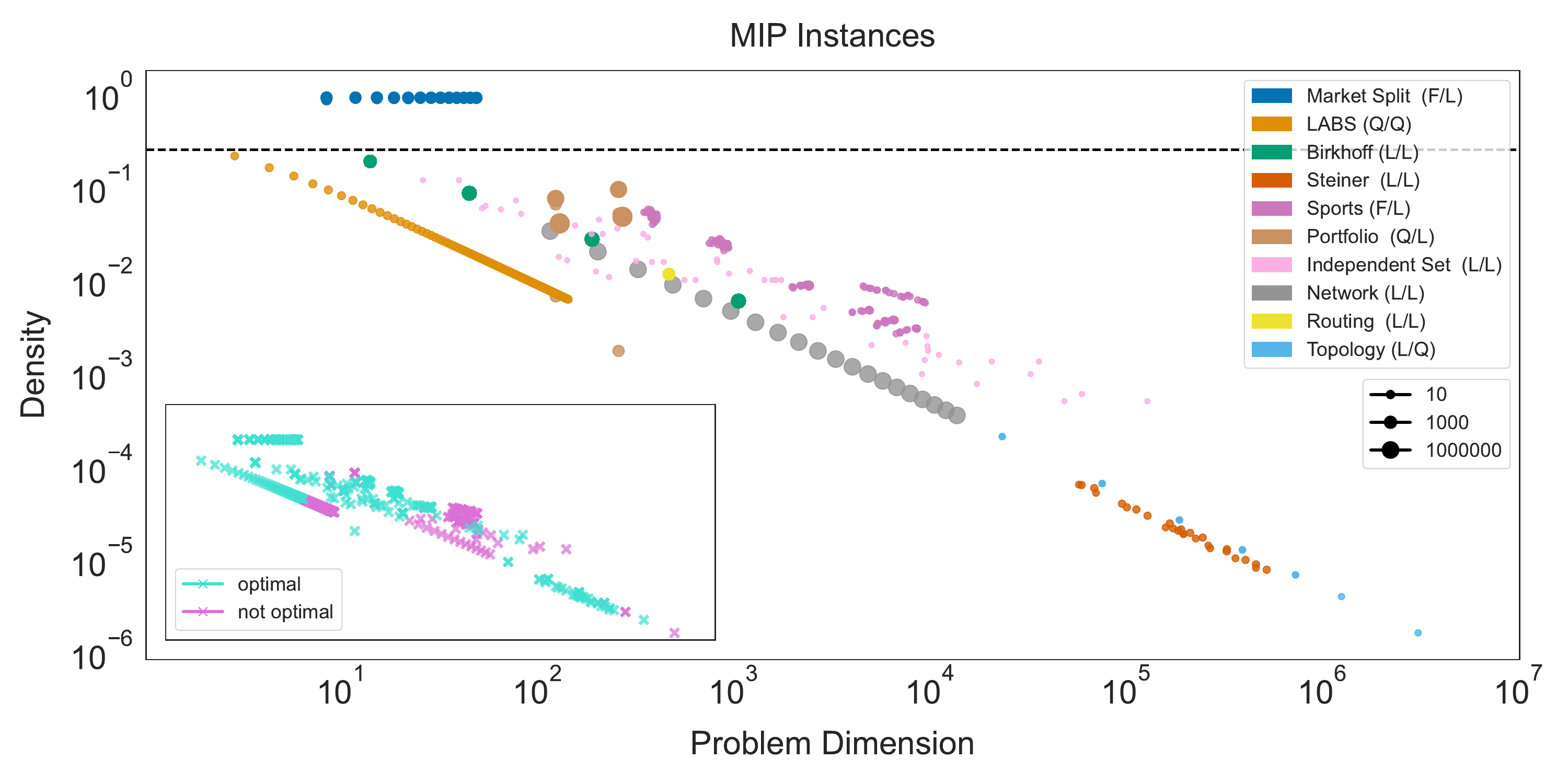}
\includegraphics[width=0.85\textwidth]{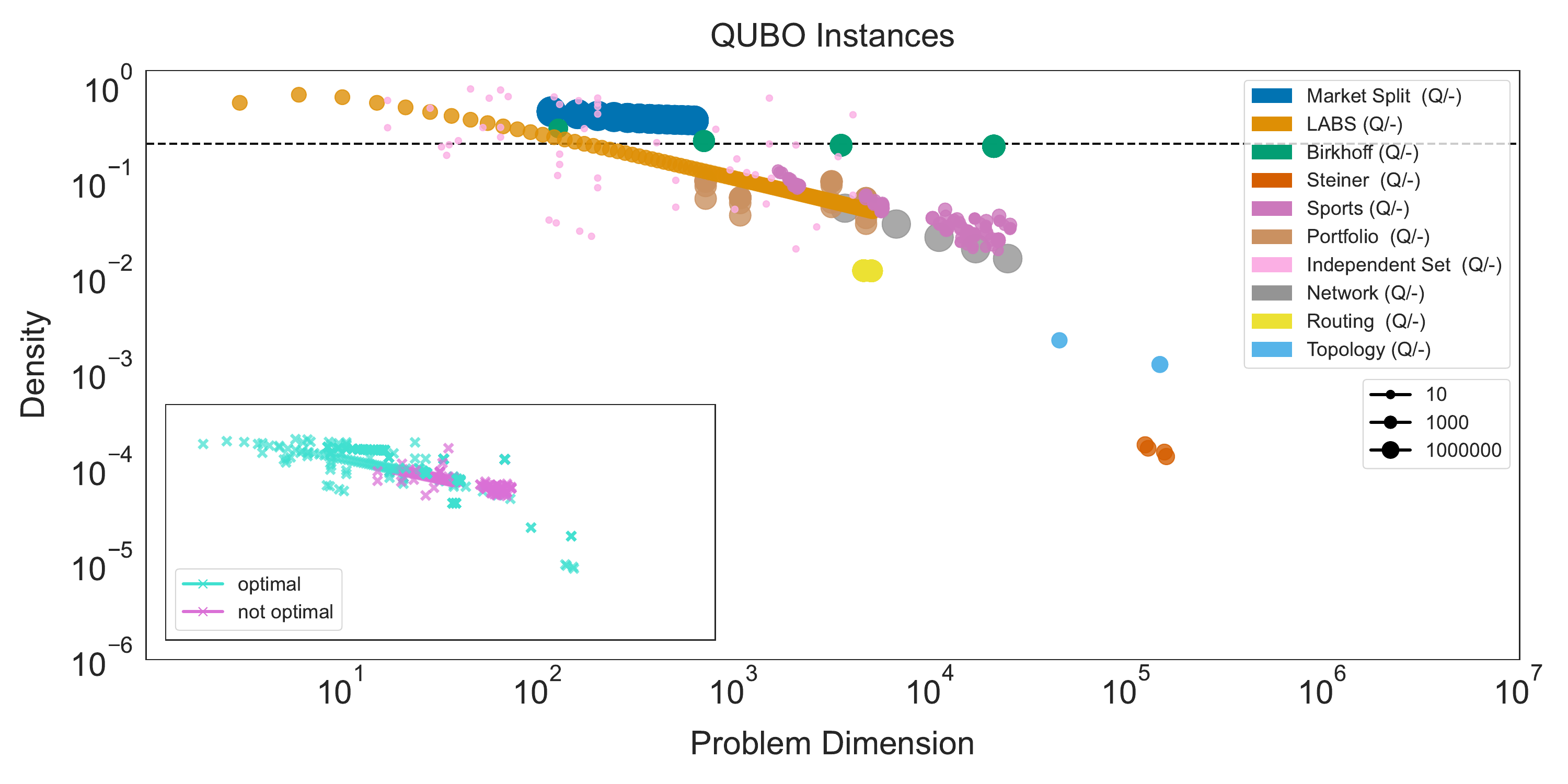}
\caption{ 
The figures plot the problem dimension (MIP: $\sqrt{\#variables \cdot \# constraints}$, QUBO: $\#variables$) against the density 
, i.e., the fraction of non-zero coefficients in the objective and, for the MIP instances, constraint functions, of the various problem instances provided in the \RN{} repository for MIP (top) and QUBO (bottom) formulations of the problem. The legend entries for the various problem classes name the class itself and, in brackets, give the form of the (objective/constraint) function with L and Q denoting linear and quadratic, respectively. Furthermore, the size of the individual points is directly related to the coefficient range of the respective problem instance per problem class--see legend in the center right. The dashed lines illustrate the separation between sparse and dense instances as described in Table \ref{tab:problem_classes}. Finally, the inset plots illustrate the instances in which the optimal solution is known.
It should be noted that while the MIP plot shows all problem instances, the QUBO plot only shows those instances that we could transform into QUBO models with less than $1.6\cdot10^5$ variables, cf.~Table~\ref{tab:problem_classes}. 
The respective model translation is done for Birkhoff, Steiner, Sports, Network, Routing, and Topology with Qiskit Optimization 0.6.1 \cite{QiskitOpt} using the default parameters. Marketsplit, LABS, Portfolio, and Independent Set can be trivially converted to QUBOs using ZIMPL 3.6.1 \cite{Koch2004}, since there are no slack variables to consider. Note that for Portfolio, this shows the density of the constraints matrix; the quadratic objective is always dense.
}
\label{fig:mip_qubo_plot}
\end{figure}

The number of decision variables for which MIP or QUBO formulations of the selected instances become difficult ranges from less than 100 to 45,000, and is therefore within reach for testing algorithms on near-term quantum hardware.
Depending on the problem, the ranges of the constraint coefficients lie between $1$ and $\approx 10^7$, assuming that the coefficients are scaled to integers, as we do throughout the paper. This can pose another challenge for quantum computers as they need to be able to represent the problem data in the necessary resolution.

Figure~\ref{fig:mip_qubo_plot} provides a more detailed overview of the instances provided for the various problem classes, complementing Table~\ref{tab:problem_classes}. For MIP and QUBO formulations of the considered problem instances, it illustrates the number of variables, the problem density, the coefficient range, and whether the optimal solution is known. Please note that if multiple MIP and QUBO models are available (see \cite{gitlab_repo}), we choose to represent the one resulting in the smallest number of variables.
We would further like to point out that the QUBO illustration only shows those instances that could be converted from an MIP model into a QUBO model with fewer than $1.6 \cdot 10^5$ variables.
As the plots show, the larger the number of variables, the more likely it is that we do not know the optimal solution. 
Finally, it can be seen that Market Split and LABS are the two problem classes for which difficult instances are found at the smallest sizes. Independent Set also has hard instances at relatively small sizes; these instances are also sparse and come with a small coefficient range. For these reasons, these three problem classes seem the most amenable for approaching with a quantum algorithm on near-term hardware.

In the remainder of this section, we discuss each problem class, describe the specific instances included, and present baseline results computed with classical optimization solvers. Of course, all instances, used models, and solutions are also made accessible in the \RN{} repository. 
We do not claim that our baseline results are the best solutions one can find with a classical computer, but they should provide an indication of what is possible using general solvers without extensive effort.
Moreover, the provided MIP and QUBO formulations were not subject to significant optimization efforts but are intended to serve as illustrations.
Our goal is to provide a framework to track the progress towards quantum advantage in optimization and have an initial baseline to start with. 
To build-up expertise and start tracking today's capabilities, we describe an informed approach to construct warm-up examples from the provided problem instances in  Appendix~\ref{app:instance_simplification}. 

\subsection{Market Split} \label{sec:problems:marketshare}
\renewcommand{\customcaptiontext}{(Market Split) }

\subsubsection{Background}
The market split instances are multi-dimensional subset-sum problems, which were explicitly designed to be hard to solve using classical methods. 
Since their introduction by \cite{CornuejolsDawande1999} in 1999, the progress in solving them has been moderate. Subset-sum is a classic \NP-hard problem \cite{GareyJohnson1979}. The time complexity for enumeration is $\mathcal{O}(2^{n/2} \cdot (n/4))$ \cite{SchroeppelShamir1981}. Several other methods have been tried, including smart enumeration, branch-and-cut, and lattice point enumeration with basis reduction \cite{AardalEtal2000,Wassermann2002,Vogel2012}.

\begin{newtextbox}[Why is it interesting?]
Market split problems are a type of multi-dimensional subset-sum problem. 
This problem class contains various instances for which it is challenging to find feasible solutions with classical digital systems--already for small system sizes, with little more than $100$ variables.
Applications of the market split problem could be interesting in various industries, such as the energy sector, e.g., to help derive optimal pricing strategies subject to customer acquisition, retention or competition constraints, and regional regulatory policies \cite{GRIMM2016493, FRAUNHOLZ2021111833}.
\end{newtextbox}

\subsubsection{Problem statement}
Given a matrix $A\in\NN^{m\times n}$ and a right-hand side $b\in\NN^m$, for $m,n\in\NN$, we want to find $x\in\{0,1\}^n$ such that
\begin{equation}
\label{eqn:marketsplit_feas} 
Ax=b .  
\end{equation}
This is a multi-dimensional subset-sum problem.
We can transform Eq.~\eqref{eqn:marketsplit_feas} trivially into unrestricted optimization problems: 
\[
\min_{x\in\{0,1\}^n} \left(b - Ax\right)^2
\qquad\text{or}\qquad
\min_{x\in\{0,1\}^n} \left\|b - Ax\right\|_{\infty},
\]
where a feasible solution exists if the global minimum is zero. 
With careful choices of the coefficients, this problem becomes hard to solve, e.g., for numbers $m \in \NN_+$, $D \in \NN$, set $n = 10 (m-1)$, $I = \{1, \ldots m\}, J = \{1, \ldots, n\}$ and 
$a_{ij} \in \{0, \ldots, D - 1 \}, b_i = \left\lfloor \frac{1}{2}\sum_{j \in J} a_{ij} \right\rfloor$ \cite{CornuejolsDawande1999}.

The best results so far have been achieved using forms of lattice enumeration with basis reduction. \cite{Wassermann2002,Wassermann2025} report solutions to selected instances up to $m=14$. Traditional branch-and-cut-based MIP solvers often already struggle before size $m\geq 7$. However, they are typically more effective on the lattice reformulation introduced in \cite{AardalEtal2000}. 

\subsubsection{Instances}
For this benchmark, we generated four random instances each for $m=\{3,\ldots,15\}$, and $D=\{50,100,200\}$ as described above. 
While not all procedures to generate Market Split instances result in feasible problems (e.g., \cite{CornuejolsDawande1999}, see the study in \cite{AardalEtal2000}), we generated them such that feasibility is guaranteed.
For $m\leq7$, we randomly generate problem instances and only include them if a MIP solver can find a feasible solution. For $m > 7$, we start with randomly generated feasible solutions and then generate problem instances that have that point as a solution, while resembling the characteristics of the instances in \cite{CornuejolsDawande1999};
details are given in Appendix~\ref{app:marketsplit}. In our tests, the resulting instances are still hard for classical solvers. 
Finally, we provide a routine in \RN{} that, given an instance and a vector $x\in\{0,1\}^n$, checks the feasibility of the vector.

\subsubsection{Classical Baseline}

We provide modeling examples for an integer linear program and a QUBO, which can be run with any MIP or QUBO solver. 
For the ILP, we use the constrained formulation with positive slack, minimizing the slack as objective. 
For the QUBO, we add all constraints as quadratic penalties to the objective.
Figure~\ref{fig:marketshare:baseline:plot:gurobi} in the Appendix
depicts detailed runtimes by \gurobi 11.0.0 \cite{Gurobi} and \abstwo~\cite{NakanoEtal2023} for instances with $3\leq m\leq 7$.

\begin{figure}[htb!]
\centering
\includegraphics[width=0.9\textwidth]{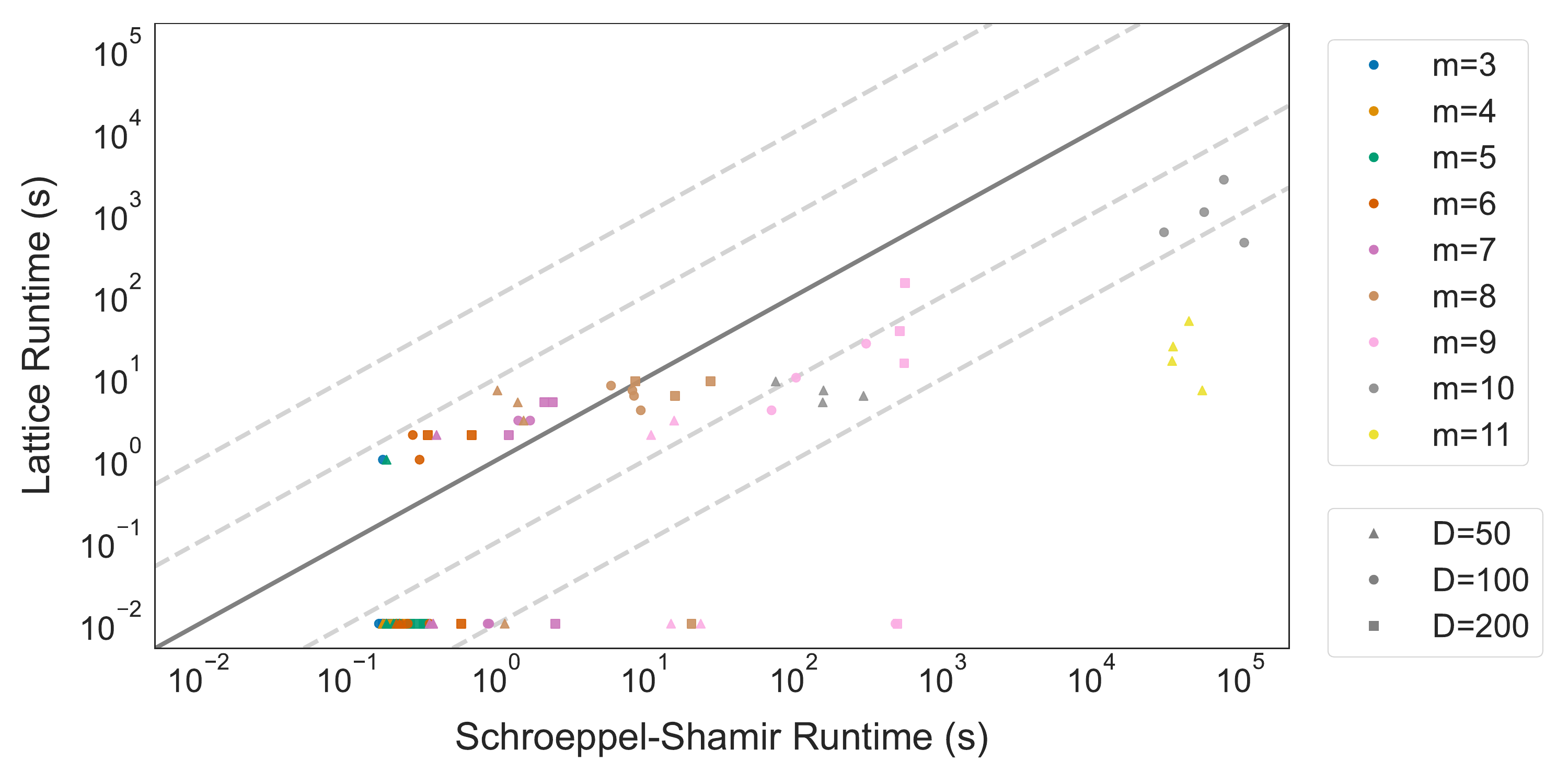}
\customcaption{ 
Comparison of runtimes using a lattice enumeration method and a GPU-accelerated Schroeppel-Shamir algorithm. 
The colors represent parameter $m$ and the shapes represent parameter $D$.}
\label{fig:marketshare:comparison}
\end{figure}

Enumeration methods, either Lattice-based according to \cite{Wassermann2002}, or Schroeppel-Shamir \cite{SchroeppelShamir1981}, show better performance and allow solving all instances up to size $m=11$ and partly up to $m=14$.
Figure~\ref{fig:marketshare:comparison} depicts a comparison of the runtimes of the algorithm from \cite{Wassermann2025} on an Intel Xeon E-2288G CPU @ 3.70~GHz and the GPU-accelerated Schroeppel-Shamir algorithm \cite{KempkeKoch2025} on an Nvidia GH200 system. Note that both methods are capable of computing all solutions to an instance. The time needed to find the ``first'' solution, however, varies easily by a factor of 10 between the instances. 

\subsection{Low Autocorrelation Binary Sequences} \label{sec:problems:labs}
\renewcommand{\customcaptiontext}{(LABS) }

\subsubsection{Background}

\begin{newtextbox}[Why is it interesting?]
Established classical algorithms struggle to find (good) solutions to LABS instances even at relatively small sizes. 
Moreover, the binary (spin-like) structure of LABS aligns naturally with Ising-type models, making it a suitable test case for quantum optimization techniques. Demonstrating improvements in scalability or solution quality on LABS would represent a significant step in addressing a known hard problem.
\end{newtextbox}

Low Autocorrelation Binary Sequences (LABS) were initially studied in the context of radar and sonar system design, where achieving minimal autocorrelation sidelobes is crucial for improving range resolution and target discrimination~\cite{boehmer1967binary, schroeder1970synthesis}. Over time, these sequences have also proven useful in digital communications, coding theory, and spread-spectrum systems, where low autocorrelation properties can enhance channel efficiency and signal clarity.
Unlike many hard optimization problems, each sequence length $N$ in LABS defines a single, unique problem instance. As a result, progress on LABS is easy to measure: pushing beyond known solutions or improving run-time for larger $N$ directly reflects meaningful advances in algorithmic capability. 

LABS also appear as ground states of the Bernasconi model in statistical physics, a system involving complex, long-range four-body spin interactions~\cite{mertens1998ground}. Despite significant research efforts, finding optimal LABS remains a challenging task. LABS instances can be solved using convex-relaxation-based branch-and-bound, see, e.g., \cite{elloumi2021solving}. In this line of research, LABS is formulated as a polynomial optimization problem over 0-1 variables and solved using branch-and-cut techniques. Both off-the-shelf solvers and specialized solvers can be used to deal with the corresponding formulation, but the exact solution of large instances remains difficult; see \cite{elloumi2021solving} as well as the benchmark results (with optimality gaps) in \cite{minlplib24}. For convex-relaxation-based branch-and-bound, the empirical performance on LABS instances varies greatly, and it is difficult to estimate. The empirically-estimated running time of a tailored classical branch-and-bound method that uses combinatorial bounds \cite{packebusch2016low} is approximately $1.73^N$, while the best performing heuristic discussed in the literature for this problem (Memetic Tabu Search) has an estimated running time of approximately $1.34^N$; see \cite{Shaydulin2024} and the references therein for a discussion of these running times and \cite{zhang2025newimprovementssolvinglarge} for recent results. The work presented in \cite{Shaydulin2024} also shows an empirical study, extrapolating from results on smaller instances, estimating the complexity of quantum optimization approaches, particularly those based on QAOA combined with amplitude amplification, to be around \(1.21^N\), considering idealized conditions. Although these figures are estimates for exponential-time algorithms, which are notoriously difficult to estimate correctly on unseen instances, they raise the possibility that quantum approaches could eventually outperform the fastest known classical algorithms.

\subsubsection{Problem Statement}

Given a sequence $S = (s_1, \ldots, s_n)$ of length $n$ with binary variables $s_j \in \{ -1, +1\}$, the minimum energy autocorrelations of the sequence $j \in \{ 0, \ldots, n-1\}$ corresponds to
\[
\min_{s\in\{-1, +1\}^n} \sum_{j=1}^{n-1}\left(\sum_{i=1}^{n-j} s_i s_{i+j}\right)^2.
\]
As LABS is an unconstrained problem, any sequence $S$ is a feasible solution.

\subsubsection{Instances}
Since there is exactly one instance per size $n$, all instances can easily be generated. 
We provide a checking routine that, given a sequence $S$, computes the autocorrelation energy and checks the result against known optima.

\subsubsection{Classical Baseline}

\begin{figure}[htb!]
\centering
\includegraphics[width=0.9\textwidth]{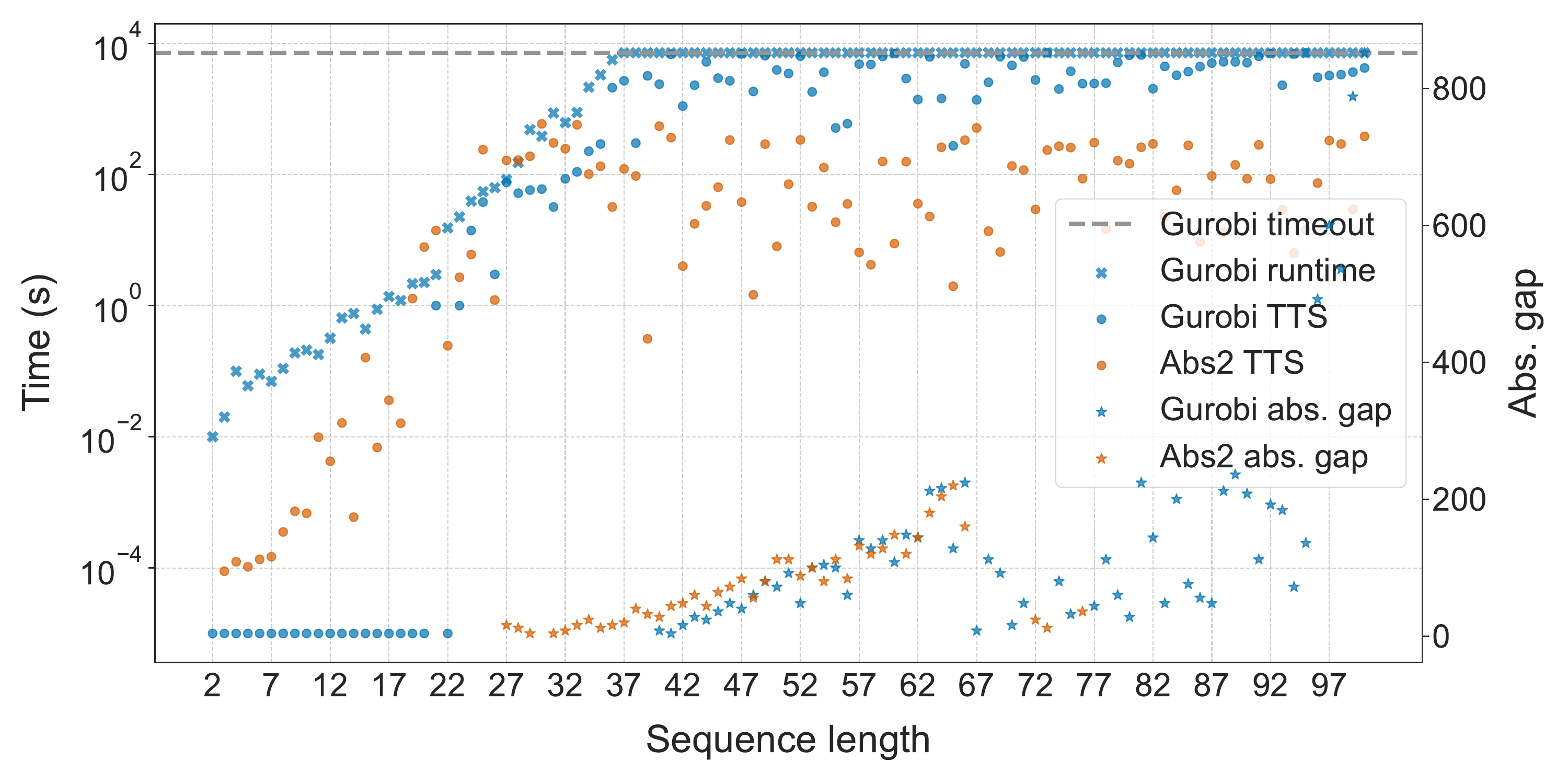}
\customcaption{Solution times using \gurobi and \abstwo for different LABS sequence lengths. 
\emph{Abs.~gap} denotes (the upper bound on) the absolute gap to the optimal (\gurobi)/best-known (\abstwo) solution.
Only gaps $>0$ are shown.}
\label{fig:labs:baseline:plot}
\end{figure}

\begin{table}[htb!]
    \centering
    \begin{minipage}[t]{0.45\textwidth}
        \centering
        \begin{tabular}{rlr}
        \toprule
        Length & Sequence & Obj \\
        \midrule
        2 & 2 & 1 \\
        3 & 12 & 1 \\
        4 & 112 & 2 \\
        5 & 113 & 2 \\
        6 & 1113 & 7 \\
        7 & 1123 & 3 \\
        8 & 1124 & 8 \\
        9 & 121113 & 12 \\
        10 & 111124 & 13 \\
        11 & 331211 & 5 \\
        12 & 111522 & 10 \\
        13 & 5221111 & 6 \\
        14 & 41112221 & 19 \\
        15 & 11213133 & 15 \\
        16 & 2112113131 & 24 \\
        17 & 11312144 & 32 \\
        18 & 1112115222 & 25 \\
        19 & 4111142212 & 29 \\
        20 & 11114142122 & 26 \\
        21 & 27221111121 & 26 \\
        \bottomrule
        \end{tabular}
    \end{minipage}
    \hfill
    \begin{minipage}[t]{0.45\textwidth}
        \centering
        \begin{tabular}{rlr}
        \toprule
        Length & Sequence & Obj \\
        \midrule
        22 & 11111212723 & 39 \\
        23 & 231131121413 & 47 \\
        24 & 122121111732 & 36 \\
        25 & 122121111733 & 36 \\
        26 & 3371111212211 & 45 \\
        27 & 34313131211211 & 37 \\
        28 & 112112131313431 & 50 \\
        29 & 112212111117323 & 62 \\
        30 & 132311111212164 & 59 \\
        31 & 1112111122122337 & 67 \\
        32 & 71112111133221221 & 64 \\
        33 & 742112111111122221 & 64 \\
        34 & 22229421121111111 & 65 \\
        35 & 11111132326612121 & 73 \\
        36 & 1121112121311132363 & 82 \\
        37 & 22228421121121111111 & 86 \\
        38 & 1222211111112112448 & 87 \\
        39 & 11111112343212112128 & 99 \\
        40 & 44412112131121313131 & 108 \\
        41 & 343111111222281211211 & 108\\
        \bottomrule
        \end{tabular}
    \end{minipage}
    \customcaption{Optimal solutions obtained by Gurobi where each sequence encodes the counts of consecutive identical tokens in the corresponding solution. The counts are listed in order--reflecting the length of each uninterrupted group of the same token.}
    \label{tbl:opt_labs_seq}
\end{table}

Currently, the best-known exact solutions are limited to $n=66$ and $n=127$ for skew-symmetric variants; see \cite{packebusch2016low}.
We provide an example where LABS is modeled as a binary quadratically constrained quadratic program (BQCQP), which can be fed to a variety of established solvers. 
Figure~\ref{fig:labs:baseline:plot} shows results obtained with \gurobi 11.0.0 on an AMD EPYC-7542 32-core processor using 64 threads and a time limit of 2 hours (we report the upper bound on the absolute optimality gap if optimality is not certified),
and the results of \abstwo running on a system with two Nvidia A100 SXM4 80GB GPUs using a time limit of 10 minutes (we report the absolute gap to the best found solution).  
The largest instance that could be solved to optimality using this approach--although exceeding the usually set time limit--was $n=41$. The respective execution time was 13 hours on a 72-core Nvidia GH200 system using \gurobi 11.0.2. An optimal solution was found in 6 hours and 47 minutes; the remaining time was used to prove optimality.
Note that problem sizes up to $n=19$ can be solved in less than one second and up to $n=25$ in less than one minute.
The optimal computed sequences are shown in Table~\ref{tbl:opt_labs_seq}.

       \subsection{Minimum Birkhoff Decomposition} \label{sec:problems:birkhoff}

\subsubsection{Background}
\label{sec:background_birkhoff}
\renewcommand{\customcaptiontext}{(Birkhoff) }

The minimum Birkhoff decomposition problem is a special case of the Birkhoff decomposition problem \cite{DUFOSSE2016108}. In 1946, Birkhoff showed that any $n\times n$ doubly stochastic matrix can be written as a convex combination of permutation matrices  \cite{birkhoff1946}. The method of proof used by Birkhoff inspired a class of algorithms for decomposing doubly stochastic matrices known as the \emph{Birkhoff algorithm} \cite[pp.~192]{brualdi1982notes}, which can decompose a doubly stochastic matrix with at most $(n-1)^2 + 1$ permutation matrices \cite{marcus1959diagonals}. It is possible to obtain tighter upper bounds on the number of permutations required depending on the number of zeros in the doubly stochastic matrix (see, e.g., \cite{Brualdi_1982}); however, the lower bound, i.e., the minimum number of permutation matrices required, is generally not known. In \cite{DUFOSSE2016108}, Dufoss\'e and U\c{c}ar showed that the problem of finding the minimum Birkhoff decomposition is \NP-hard despite the existence of algorithms that can obtain approximate decomposition where the error decreases exponentially with the number of permutations in the decomposition \cite{Kulkarni17}.

\begin{newtextbox}[Why is it interesting?]

The problem of finding the minimum Birkhoff decomposition is \NP-hard \cite{DUFOSSE2016108}, and it is hard to solve in practice, even for small instances.
Also, it is easy to craft problem instances and obtain non-trivial upper bounds on the optimal cost, which is generally challenging in \NP-hard problems. Another interesting aspect is that the problem of decomposing doubly stochastic matrices is well-studied, and many mathematical formulations and algorithms exist, including MIP and convex optimization. We also note that doubly stochastic matrices have a natural representation in terms of unitaries \cite{mariella2024quantumtheoryapplicationcontextual}, which are the backbone of quantum computation. 
\end{newtextbox}

Decomposing doubly stochastic matrices has applications in assignment problems, especially in networking \cite{832560, 7577883, liu2015scheduling, 9359108, bojja2016costly}. In those problems, a permutation matrix corresponds to a matching in a bipartite graph, which represents a schedule or system configuration used to exchange some resource (e.g., data packets, energy). The weight associated with a permutation matrix captures the time the system will spend in such a configuration. Sparse decompositions are important to reduce the time a system spends switching between configurations, which is non-negligible in practice. Studying this problem can also be of interest for more general graph scheduling problems where the underlying graph is not just bipartite, e.g., in peer-to-peer networks \cite{o2023quantum}. 

The algorithms for solving the (minimum) Birkhoff decomposition methods can be divided into three classes. Branch-and-bound algorithms for solving a MINLP \cite{hendrych2022convex}, Frank-Wolfe algorithms \cite{combettes2023revisiting, pmlr-v28-jaggi13}, and variants of the Birkhoff algorithms \cite{liu2015scheduling,9509349}. Each approach has different characteristics. Branch-and-bound algorithms are often the slowest ones but can obtain certifiable optimal solutions when the doubly stochastic matrices are not too large or too dense. In contrast, FW algorithms are very fast and can obtain coarse approximations with just a few iterations. However, they are often unable to obtain exact decompositions even when the number of permutations used exceeds $(n-1)^2+1$.  Birkhoff-type algorithms take an approach that can be considered between FW and branch-and-bound methods: they can obtain exact decompositions with at most $(n-1)^2+1$ permutations, and their speed is comparable to FW algorithms. Unlike branch-and-bound, Birkhoff-type algorithms do not guarantee that the solution obtained is optimal. 

\subsubsection{Problem statement}

Let $D$ be an $n\times n$ doubly stochastic matrix and $P_i$ the $i$-th $n \times n$ permutation matrix, $i \in  \{1,\dots, n! \}$. Recall that a matrix is doubly stochastic if its entries are non-negative and the rows and columns sum to one. Similarly, a permutation matrix is a doubly stochastic matrix with binary entries. 

For a given doubly stochatic matrix $D$, the \emph{minimum Birkhoff decomposition} problem is:

\begin{align}
\begin{tabular}{lll}
$\underset{c_i \in [0,s]}{\min}$ & $\sum_{i=1}^{n!} |c_i|^0$ \\
subject to & $sD = \sum_{i=1}^{n!} c_i P_i$ \\
& $\sum_{i=1}^{n!} c_i = s$
\end{tabular},
\label{eq:minimum_birkhoff_decomposition}
\end{align}
where we define $0^0 = 0$, i.e., $|c_i|^0$ counts the number of non-zero entries, and $s > 0 $ is a scalar, often equal to one. 
The goal is to find a subset of permutation matrices such that $D$ is in its convex hull and that this subset contains as few permutation matrices as possible. 
The following is an example of a decomposition of a $3 \times 3$ matrix:
\begin{align}
\begin{bmatrix}
0.2 & 0.3 & 0.5\\
0.6 & 0.2 & 0.2\\
0.2 & 0.5 & 0.3
\end{bmatrix} \label{eq:decomposition_example}
& =   0.2
\begin{bmatrix}
 0 & 1 & 0  \\
0  & 0  & 1\\
1 & 0 & 0
\end{bmatrix}
+ 0.2
\begin{bmatrix}
1 & 0 & 0 \\
0 & 1 & 0 \\
0 & 0 & 1 
\end{bmatrix}
+ 0.1
\begin{bmatrix}
0 & 1 & 0 \\
1 & 0 & 0 \\
0 & 0 & 1 
\end{bmatrix}
\\
& \ \! + 0.5
\begin{bmatrix}
0 & 0 & 1\\
1 & 0 & 0 \\
0 & 1 & 0  
\end{bmatrix}
+ 0.0
\begin{bmatrix}
0 & 0 & 1\\
0 & 1 & 0 \\
1 & 0 & 0  
\end{bmatrix}
 + 0.0
\begin{bmatrix}
1 & 0 & 0\\
0 & 0 & 1 \\
0 & 1 & 0  
\end{bmatrix}.\notag
\end{align}
Note that $\sum_{i=1}^{n!} |c_i|^0 = 4$ in the example above since the decomposition only uses four out of the $3! = 6$ possible permutation matrices. 

\subsubsection{Instances}
\label{subsub:birk_instances}
The instance set consists of doubly stochastic matrices of size $n\in \{3,\dots,16\}$ generated by sampling $n$ (sparse) and $n^2$ (dense) permutations uniformly at random. There are 10 instances for each matrix size and density. The weight associated with each permutation is also sampled uniformly at random with a fixed number of digits after the decimal points. The number of digits after the point depends on the size of the doubly stochastic matrix. We then scaled all coefficients to make them integers. See \cite{gitlab_repo} for a detailed description of how doubly stochastic matrices are generated. By construction, sparse doubly stochastic matrices can be decomposed with at most $n$ permutations, which is a non-trivial upper bound on the minimal decomposition compared to the upper bound $(n-1)^2+1$ that applies to all doubly stochastic matrices.
It should be noted that the instances were generated from feasible solutions to guarantee the feasibility of all given problem instances. 
However, it is not known whether these feasible solutions are optimal. 

\subsubsection{Classical Baseline}
\label{sec:Birkhoff_classicalbaseline}

The goal of this section is to illustrate the performance and limitations of different approaches to set a performance baseline with state-of-the-art classical solvers for the minimum Birkhoff decomposition problem instances. We compare two convex optimization algorithms, Blended FW \cite{braun2019blendedconditionalgradientsunconditioning, FrankWolfe.jl} and Birkhoff+ \cite{9509349, BirkhoffDecomposition.jl}, as well as CPLEX.

\begin{figure}
\centering
\includegraphics[width=0.48\textwidth]{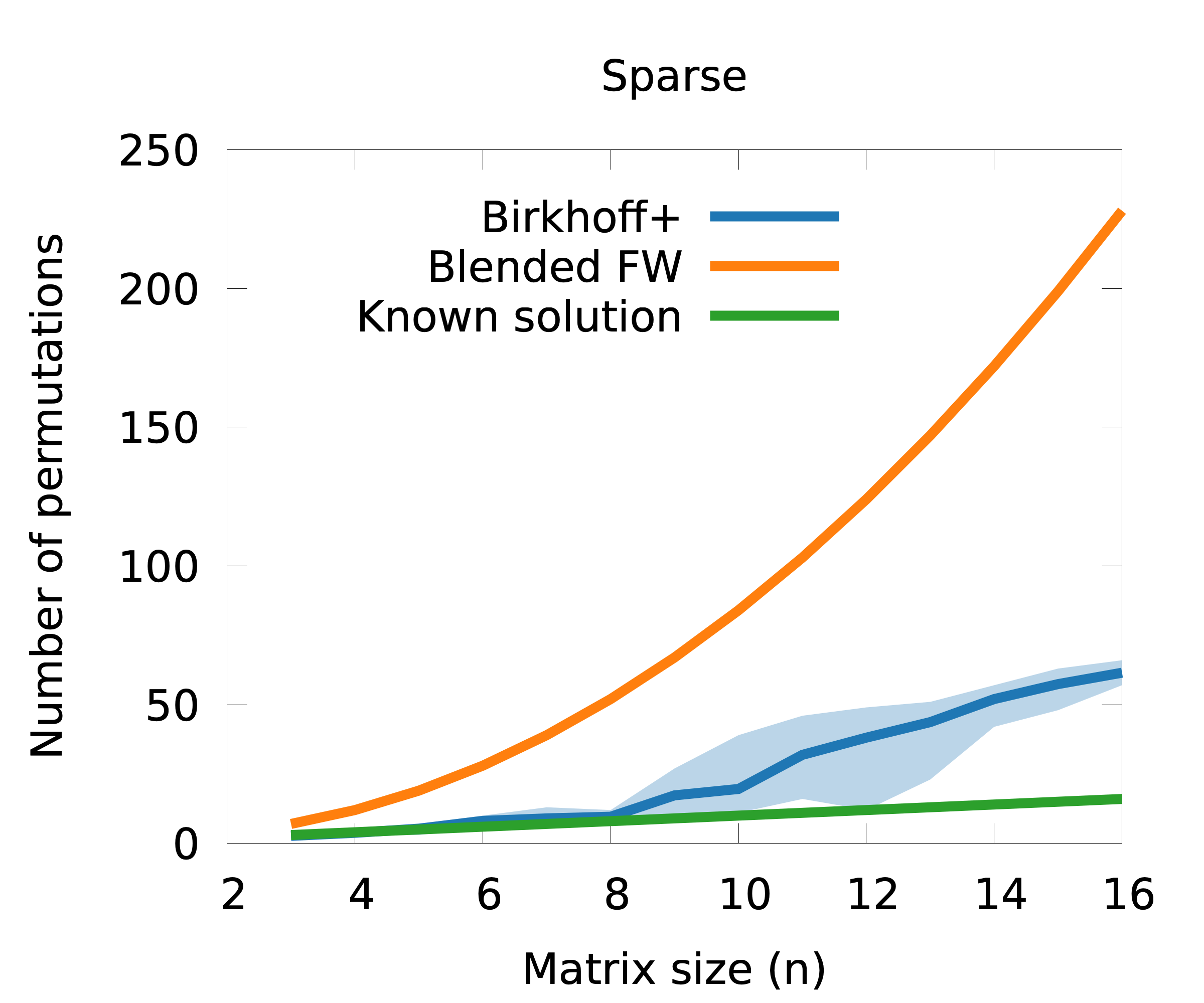}
\includegraphics[width=0.48\textwidth]{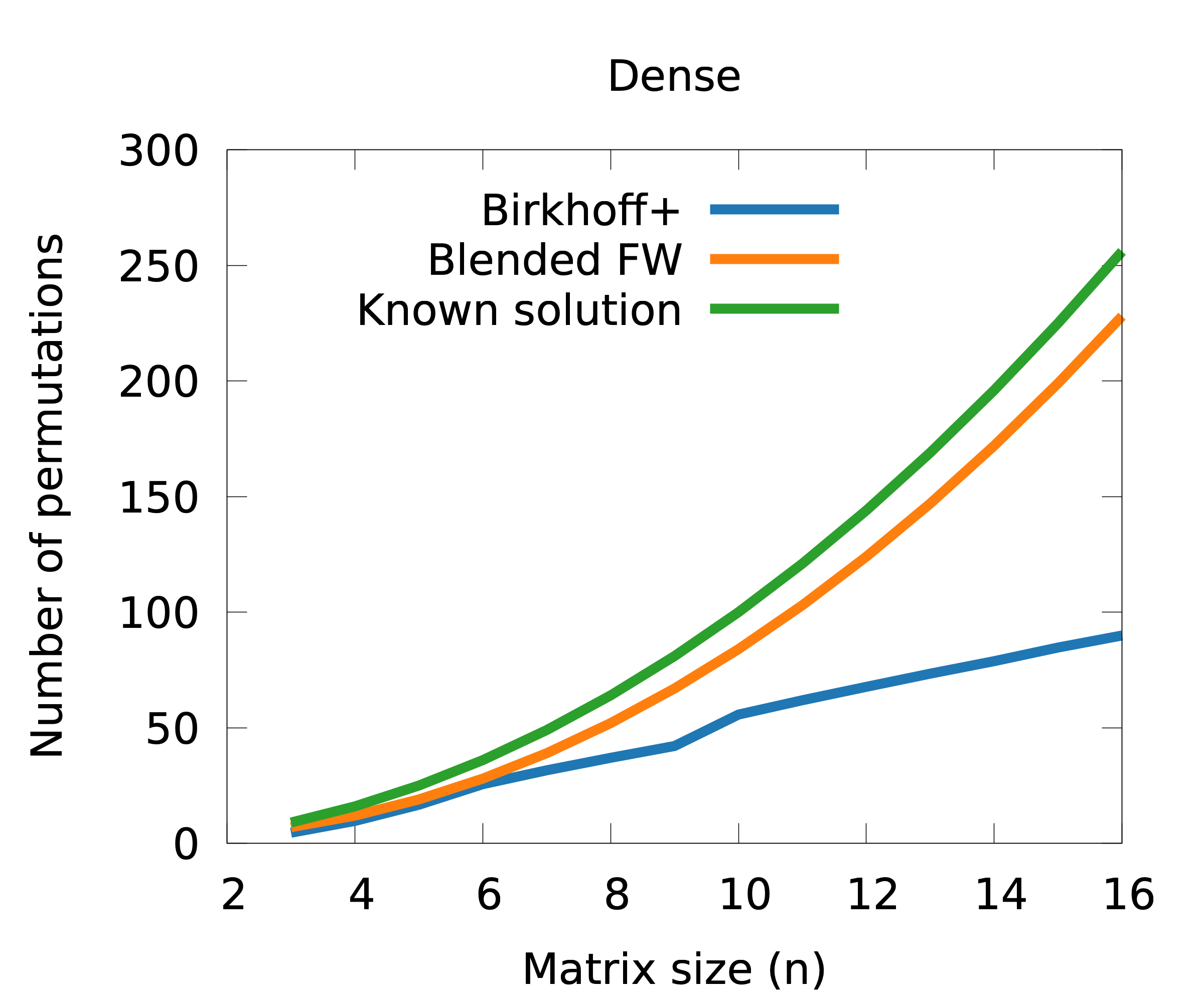}
\customcaption{ Illustration of the number of permutations required to decompose sparse (left) and dense (right) doubly stochastic matrices for varying matrix size. The shaded area--which is only visible in the sparse case--presents the maximum and minimum required number of
permutation matrices per problem size.
}
\label{fig:classical_benchmark}
\end{figure}

First, we focus on the two convex optimization algorithms devised to obtain sparse approximate decompositions, i.e., not a minimal and exact decomposition. 
Notably, classical heuristics are discussed for this problem because we also investigate quantum heuristics in Section~\ref{sec:min_birk_exp}.
We run the algorithms with their out-of-the-box
settings \cite{FrankWolfe.jl,BirkhoffDecomposition.jl} and show the results in Figure~\ref{fig:classical_benchmark}. The shaded area in Figure~\ref{fig:classical_benchmark} (only visible in the sparse case) indicates the maximum and minimum number of permutation matrices required for a given problem size, i.e., the best and worst case instance. The `known solution' corresponds to the number of permutation matrices used to generate the doubly stochastic matrix in the dataset, which is an upper bound to the minimum decomposition. Observe from the figure that Birkhoff+ obtains better performance than Blended FW in both cases. The reason for that is that Birhoff+ is guaranteed to obtain an exact decomposition with at most $(n-1)^2+1$ permutations, whereas Blended FW does not have such a property, even though it can reduce the approximation error exponentially fast with the number of permutations this adds to the decomposition. In the experiments, we limit the maximum number of permutations of Blended FW to $(n-1)^2 + 1$ and observe decomposition errors in between $10^{-2}$ and $10^{-6}$ depending on the instance. With Birkhoff+, all the decompositions are exact, and their performance varies depending on whether the matrix to decompose is sparse or dense. In the sparse case, Birkhoff+ obtains decompositions that match the number of permutations used to generate the matrices (i.e., $n$) when $n \le 8$. For $n \ge 9$, the number of permutations required varies significantly depending on the instance (shaded area), especially in the range $n \in [9,12]$. Finally, in the dense case, we obtain that the number of permutations obtained is smaller than the number of permutations used to generate the doubly stochastic matrix (i.e., $n^2$). There is no significant variability across instances, and the number of matrices required by Birkhoff+ to obtain an exact decomposition scales linearly up to the problem sizes investigated in this analysis.

\begin{figure}
\centering
\includegraphics[width=0.4\textwidth]{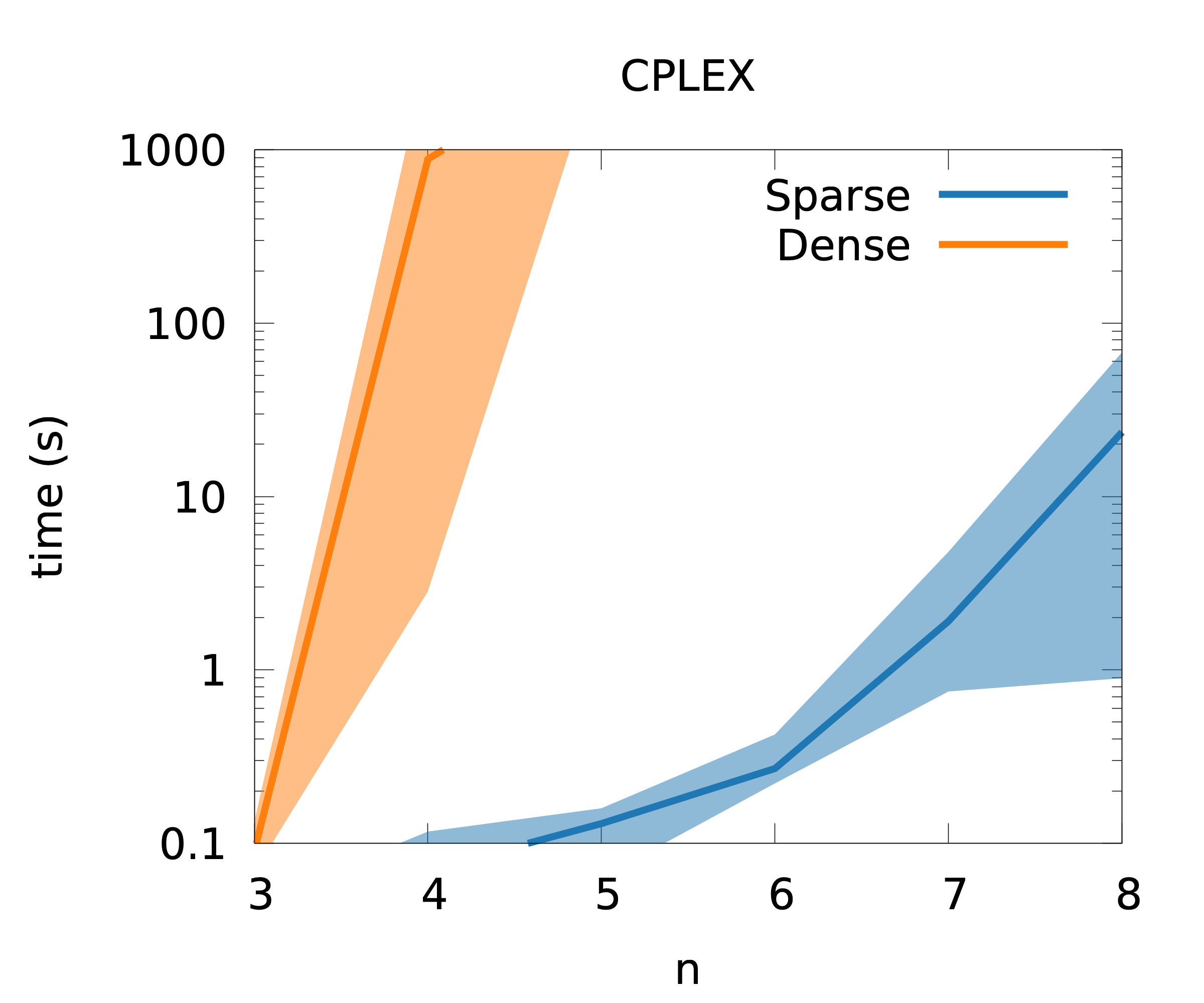}
\customcaption{Arithmetic mean of time required by CPLEX to decompose a doubly stochastic matrix with the minimum number of permutation matrices. The x-axis indicates the size of the doubly stochastic matrix, and the shaded area indicates the minimum and maximum time required to decompose an instance of such size.
}
\label{fig:cplex_benchmark}
\end{figure}

Next, we illustrate the performance of CPLEX when solving the following Integer program (IP): 
$\min_{c_i, P_i} \sum_{i=1}^{k} |c_i|^0$ s.t.\ $sD = \sum_{i=1}^{k} c_i P_i$, $\sum_{i=1}^{k} c_i = s$, $c_i \in \{0,1,\dots,s\}$, $P_i \in \{0,1\}^{n\times n}$, $P_i \mathbf 1 = P_i^T \mathbf 1 = \mathbf 1$ for all $i \in \{1,\dots,k\}$ where $\mathbf 1$ is the all-ones vector, and $s$ and $k$ integers. Note that the permutation matrices are formulated as decision variables, and all decision variables are represented by integers. Also, note that when the entries of $D$ have a fixed number of decimal points (see Section~\ref{subsub:birk_instances}), we can always select $s$ to be an integer and make $sD$ an integer-valued matrix.
For example, if we multiply Eq.~\eqref{eq:decomposition_example} by 10, we obtain $\left[ \begin{smallmatrix} 2 & 3 & 5 \\ 6 & 2 & 2 \\ 2 & 5 & 3 \end{smallmatrix} \right]$, which we can decompose with weights 2, 2, 1, and 5.
Parameter $k$ in the formulation controls the decomposition length--which is ideally as small as possible. Thus, to solve the problem, we start the optimization with $k=1$ and increase $k$ by one if CPLEX fails to return a feasible solution. Figure~\ref{fig:cplex_benchmark} illustrates the runtime of CPLEX as a function of the matrix size.
One can see from the figure that the runtime increases rapidly in both cases. 
With CPLEX on a laptop (M1 Max CPU and 32 GB memory) and a one-hour runtime limit, we can only solve instances up to $n=4$ when the matrix is dense and up to $n=8$ when the matrix is sparse.

\subsection{Steiner Tree Packing}
\label{sec:problems:steiner}
\renewcommand{\customcaptiontext}{(Steiner) }

\subsubsection{Background}
The Steiner Tree Packing Problem (STPP) describes the problem of placing $n$ individual Steiner Trees node-disjointly into a graph. For $n=1$ this is the Steiner tree problem, i.e., the combinatorial variant of the much older Euclidean Steiner problem. The Euclidean Steiner problem asks for the minimal tree that connects a given set of points in the plane: the special case with three points was discussed by Fermat before 1640, whereas the general version may have originated from Gauss; see \cite{arora1996polynomial}. The Steiner tree problem replaces points with vertices in a graph with given edges and weights, and we are asked to find a minimum cost tree that connects a given set of terminals \cite{hakimi1971steiner,levin1971algorithm}. There is abundant literature on the Steiner Tree problem, which is already \NP-hard: we refer to \cite{polzin2003algorithms,ljubic2021solving} for two excellent overviews, containing an extensive list of references and to \cite{RehfeldtKoch2023} regarding the current state-of-the-art in solving them.

STPP provides a natural mathematical optimization formulation for Very Large Scale Integration (VLSI) design \cite{grotschel1997steiner,HeldKorteRautenbachVygen2011}. Therefore, its efficient solution has tremendous practical interest. Early work on the STPP included polyhedral studies \cite{grotschel1996packing}, separation routines for valid inequalities \cite{grotschel1996packingsepa}, and approximation algorithms \cite{jain2003packing}. For a more recent discussion of computational results on difficult STPP instances, see \cite{hoang2012steiner}.

\begin{newtextbox}[Why is it interesting?]
The minimization problem underlying the Steiner Tree Packing Problem comes with a linear objective function; the density is relatively sparse, the coefficients are all one, and the decision variables are binary.
Computing optimal solutions, especially for dense routing areas, proved to be very hard, and little progress--apart from improvements in hardware and general solvers--has been achieved in the last two decades. This problem is also notoriously difficult for heuristics since, in a dense solution, one wrong decision can render the problem infeasible.
In practice, this problem class can be used to model network design problems while considering cost minimization and reliability constraints. 
\end{newtextbox}

\subsubsection{Problem statement}
For $n\in\NN_+$, we write $[n]=\{1,\ldots,n\}$. In the STPP, we are given an undirected simple graph $G=(V,E)$, and $n\in\NN_+$ disjunct terminal sets $T_1,\ldots,T_n$  with $T_i\subseteq V$ for all $i\in[n]$ and $T_i \cap T_j = \emptyset$ for $i,j \in [n]$, $i \neq j$.
The goal is to find an edge set $S^*\subseteq E$ of minimal cardinality such that $(V(S^*), S^*)$ is a forest, where $\bigcup_{i\in[n]} T_i\subseteq V(S^*)$ and no vertex from $T_i$ is connected to any vertex of $T_j$ for all $i\neq j$.
The general STPP with $n>1$, as defined above, requires the inclusion of several disjoint Steiner trees into the same graph. A solution is a forest, and each terminal set $T_i$ is connected by a single tree in the forest. 

\begin{figure}[htb!]
    \centering
    \includegraphics[width=0.65\linewidth]{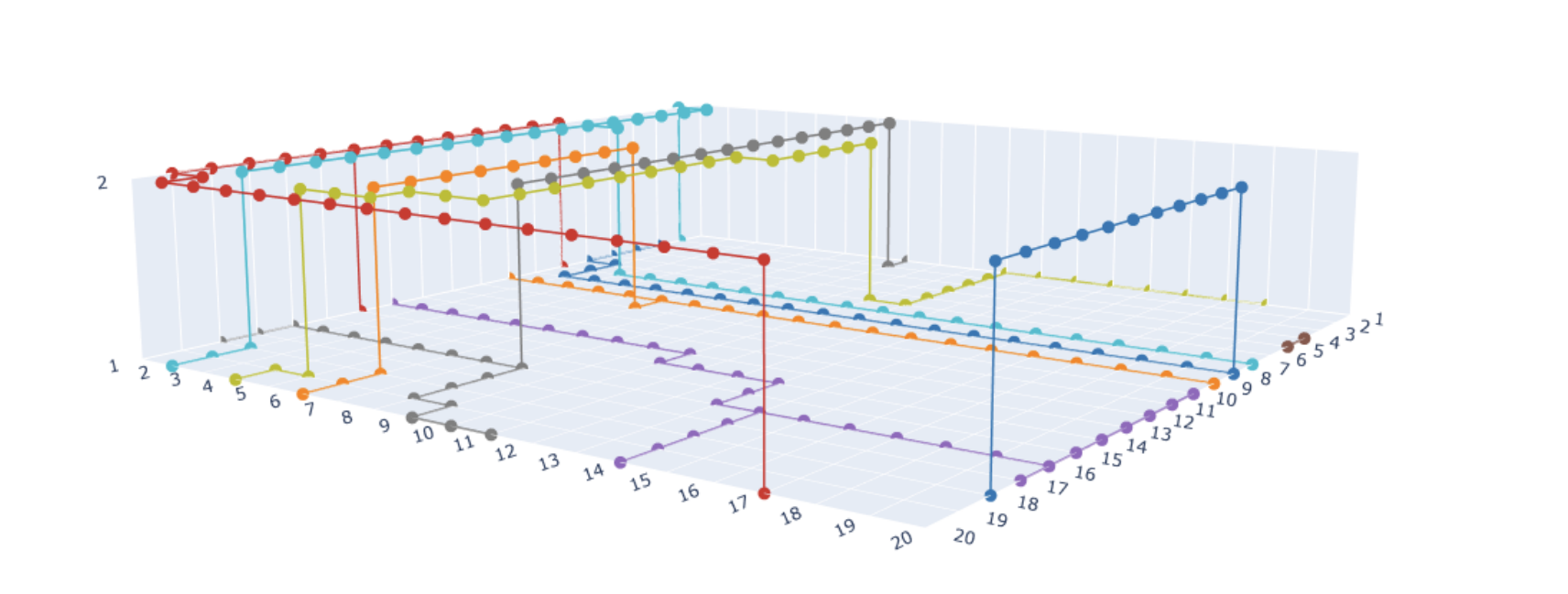}
    \customcaption{ Representation of Steiner tree packing instance of size $20 \times 20 \times 2$ with the corresponding optimal solution, packing 8 Steiner trees, each one depicted in a different color. Here, for visualization purposes, we invert the $z$ axis compared with the usual convention, i.e., we put terminals in the bottom layer.}
    \label{fig:stp_example}
\end{figure}
\subsubsection{Instances}

We generated 3-dimensional $S\times S\times L$ grid graphs with holes.
All terminals are placed at the border of the top layer. We require all trees to be node-disjoint.
The parameters for each problem are the size of the grid graph $S$, the number of layers $L$, the maximum number of terminals per terminal set $T$, and the number of holes in the grid $H$. The holes can extend to multiple layers. An example of a $20 \times 20 \times 2$ instance with the corresponding optimal solution is given in Figure~\ref{fig:stp_example}.
In practice, the problem can also be defined with edge weights. For example, changing from one layer to another could be more expensive. However, since this does not make the problem more difficult in general, we kept it simple and set all edge weights to one.
To ensure feasibility, the instances were reverse-engineered from feasible solutions, which may not be optimal. We observed that Gurobi started having difficulties with instances of size larger than $S=20$.

Finally, we provide a program that checks if a given solution to the STPP is feasible for a given instance, i.e., that each network's terminals are connected by a Steiner Tree that is node-disjoint from all other Steiner Trees. 

\begin{figure}[htb!]
    \centering
    \includegraphics[width=0.45\linewidth]{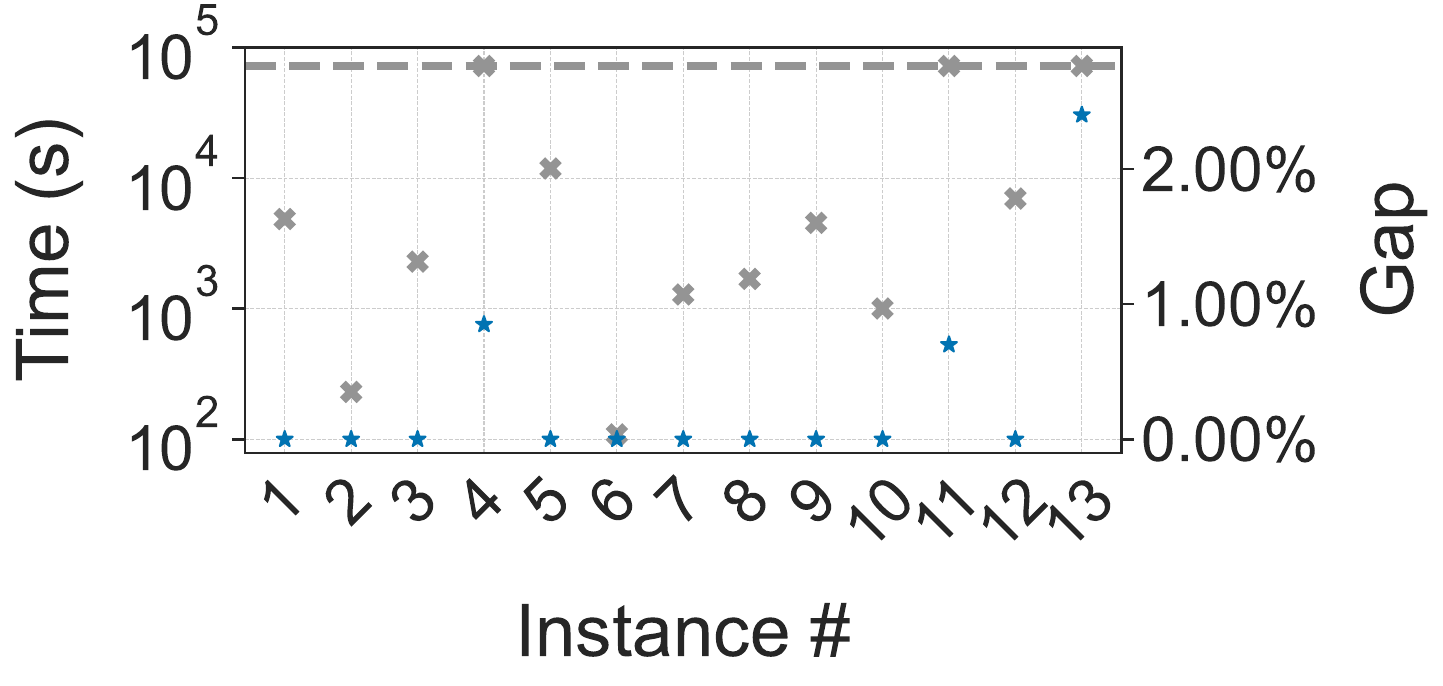}
    \includegraphics[width=0.45\linewidth]{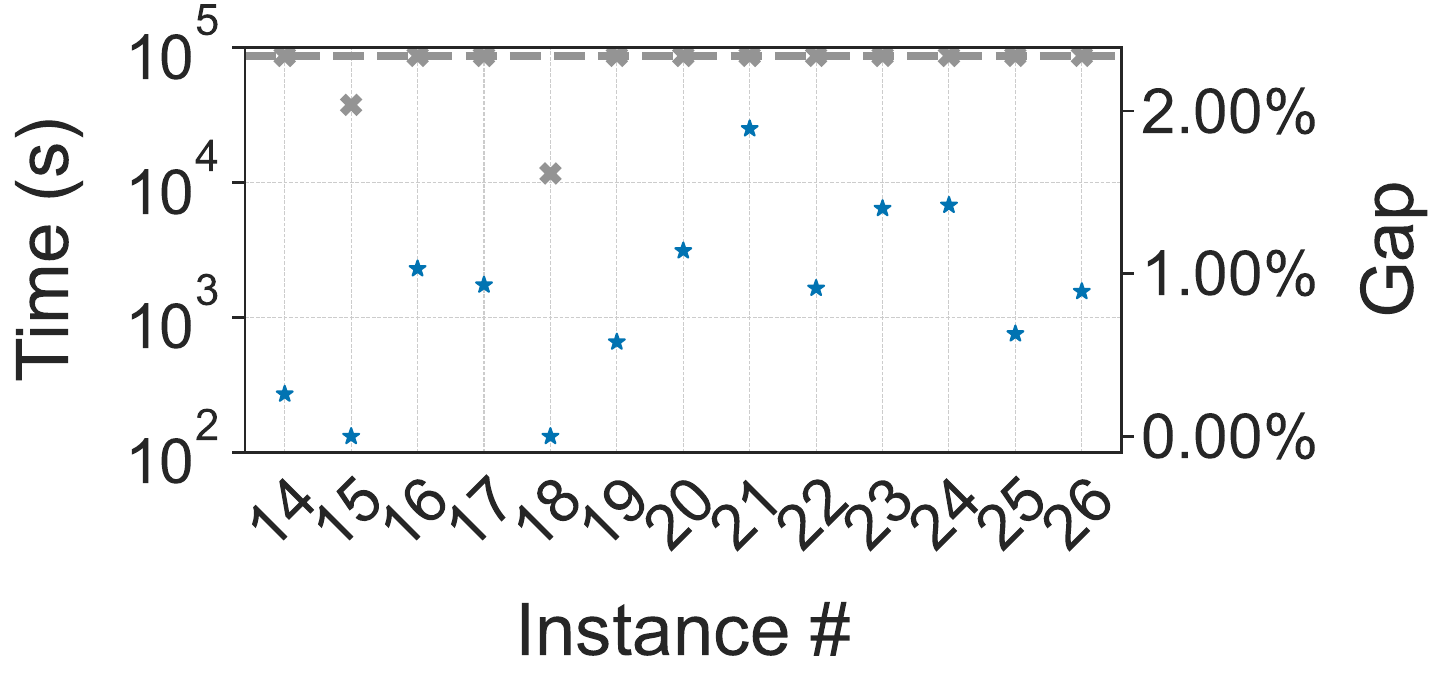}
    
    \customcaption{Total runtime and gap to the best-known (or even provably optimal) solution, solved with \gurobi for different size $20$ instances with a timeout of 72000s (left) and different size $30$ instances with a timeout of 86400s (right). The ticked line represents the timeout, the crosses indicate the runtime, and the stars show the gap.}
    \label{fig:steiner:runtime_gap_size20}
\end{figure}

\begin{table}[htb!]
\centering
\footnotesize
\setlength{\tabcolsep}{2.5pt}
\begin{tabular}{@{}r l r r r r | r l r r r r@{}}
\toprule
\textbf{\#} & \textbf{Instance (size 20)} & \textbf{Sol.} & \textbf{Bd.} & \textbf{Gap} & \textbf{Time} &
\textbf{\#} & \textbf{Instance (size 30)} & \textbf{Sol.} & \textbf{Bd.} & \textbf{Gap} & \textbf{Time} \\
\midrule
 1  & stp\_s020\_l2\_t3\_h2\_rs24098 & 228 & 228 & 0.00 & 4852.66 
 & 14 & stp\_s030\_l2\_t3\_h1\_rs97531 & 389 & 388 & 0.26 & -- \\
 2  & stp\_s020\_l2\_t3\_h3\_sd97531 & 219 & 219 & 0.00 & 231.00
 & 15 & stp\_s030\_l2\_t3\_h3\_rs97531 & 388 & 388 & 0.00 & 37495.02 \\
 3  & stp\_s020\_l2\_t4\_h0\_rs24098 & 242 & 242 & 0.00 & 2292.39
 & 16 & stp\_s030\_l2\_t4\_h2\_rs97531 & 387 & 383 & 1.03 & -- \\
 4  & stp\_s020\_l3\_t3\_h2\_rs24098 & 355 & 352 & 0.85 & --
 & 17 & stp\_s030\_l2\_t5\_h1\_rs24098 & 430 & 426 & 0.93 & -- \\
 5  & stp\_s020\_l3\_t4\_h0\_rs24098 & 230 & 230 & 0.00 & 47150.22
 & 18 & stp\_s030\_l3\_t3\_h1\_rs97531 & 327 & 327 & 0.00 & 11652.86 \\
 6  & stp\_s020\_l3\_t4\_h2\_rs97531 & 163 & 163 & 0.00 & 109.33
 & 19 & stp\_s030\_l3\_t3\_h2\_rs97531 & 343 & 341 & 0.58 & -- \\
 7  & stp\_s020\_l3\_t4\_h3\_rs97531 & 217 & 217 & 0.00 & 1286.76
 & 20 & stp\_s030\_l3\_t4\_h0\_rs97531 & 439 & 434 & 1.14 & -- \\
 8  & stp\_s020\_l4\_t3\_h3\_rs97531 & 268 & 268 & 0.00 & 1692.55
 & 21 & stp\_s030\_l3\_t5\_h1\_rs24098 & 476 & 467 & 1.89 & -- \\
 9  & stp\_s020\_l4\_t4\_h0\_rs24098 & 281 & 281 & 0.00 & 4543.95
 & 22 & stp\_s030\_l4\_t3\_h1\_rs97531 & 328 & 325 & 0.91 & -- \\
10  & stp\_s020\_l4\_t4\_h3\_rs37235 & 189 & 189 & 0.00 & 1005.19
 & 23 & stp\_s030\_l4\_t4\_h0\_rs97531 & 429 & 423 & 1.40 & -- \\
11  & stp\_s020\_l5\_t3\_h3\_rs24098 & 284 & 282 & 0.70 & --
 & 24 & stp\_s030\_l4\_t4\_h2\_rs97531 & 424 & 418 & 1.42 & -- \\
12  & stp\_s020\_l5\_t4\_h0\_rs24098 & 281 & 281 & 0.00 & 6973.54
 & 25 & stp\_s030\_l5\_t3\_h1\_rs97531 & 320 & 318 & 0.63 & -- \\
13  & stp\_s020\_l5\_t4\_h3\_rs24098 & 333 & 325 & 2.40 & --
 & 26 & stp\_s030\_l5\_t4\_h0\_rs24098 & 447 & 443 & 0.89 & -- \\
\bottomrule
\end{tabular}
\customcaption{Results for instances of size 20 (left) and 30 (right). Sol. is the best solution objective found, and Bd. is the best lower bound found. 
Gap is the gap to the optimal solution, and Time is the total runtime in seconds, where -- indicates a timeout (20 hours for size 20, 24 hours for size 30).}
\label{tab:steiner_gurobi_results}
\end{table}

\subsubsection{Classical Baseline}

Using the multicommodity flow ILP formulation for the node-disjoint STPP given in \cite{hoang2012steiner}, we ran \gurobi 11.0.2 on an AMD EPYC-7513 32-core processor with 64 threads to attempt to solve the smallest instances in our benchmark set (gridsize 20 or 30). 
We used the model ``as is'', without additional valid inequalities or decomposition strategies (the model is also provided in the repository \cite{gitlab_repo}). Hardware information and results are summarized in Figure~\ref{fig:steiner:runtime_gap_size20}, whereas solution values and bounds are given in Table~\ref{tab:steiner_gurobi_results}. With this ILP model, we could solve most of the size 20 instances (with an average solution time of 53 minutes) within 20 hours, but only two of the size 30 instances within 24 hours (the solution times were 194 minutes and 624 minutes). Reimplementing the techniques of \cite{hoang2012steiner} using current solver technology, one may be able to solve a few more instances, but the largest grid sizes in the benchmark set remain out of reach for current exact methods.

\subsection{Sports Tournament Scheduling} \label{sec:problems:sports}
\renewcommand{\customcaptiontext}{(Sports) }

\subsubsection{Background}

\begin{newtextbox}[Why is it interesting?]
	As soon as any side constraints are considered, 
the complex combinatorial structure makes it very difficult for established heuristic methods to find feasible solutions. 
For several medium-sized instances, no known (and investigated) method was able to find any feasible solution \cite{VANBULCK20231249ITC, vanbulck2023algorithm}--despite the fact that the generation process ensures the existence of feasible solutions.
Furthermore, the problem is well aligned with real-world applications, which typically `only' differ in terms of a few additional side constraints.
\end{newtextbox}

The design of algorithms to construct sports timetables goes back to at least the 1950s (see, e.g., \cite{Freund1956}).
{Particularly well researched are so-called single round-robin sports timetables where each team plays every other team once.}
Even though a {single} round-robin timetable can be constructed in time polynomial in the number of teams (see, e.g., \cite{Werra1981}), the problem becomes {\NP-hard} as soon as some elementary constraints are added (see, e.g., \cite{Briskorn2010b}).
One of the prime difficulties in sports timetabling is the vast size of the solution space: every single round-robin timetable corresponds to an {(oriented)} 1-factorization of $K_n$ and the other way around {(see Fig.~\ref{fig:onefactor})}, where $K_n$ denotes the complete graph on $n$ nodes. 
Considering just 14 teams and even ignoring the order of rounds, {the permutation of team names}, and the home-away status of games, more than 1,132,835,421,602,062,347 unique, corresponding timetables already exist \cite{Kaski2009}.

\begin{figure}[htb!]
	\hfill
	\begin{minipage}{0.32\linewidth}
		\begin{tikzpicture}
        \node (1) at (0,0) {};			
			\graph[circular placement, radius=0.30\linewidth,
			nodes={circle,draw, minimum size=14pt, inner sep=1pt}, clockwise, n=6] {
				\foreach \x in {1,...,6} {
					\foreach \y in {\x,...,6} {
						\x --[dashed,line width=0.3pt] \y;
					};
				};
			};
			\draw (1) edge[->, thick] node[black,left,pos=.2] {} (2);
			\draw (3) edge[->, thick] node[black,left,pos=.2] {} (5);
			\draw (4) edge[->, thick] node[black,left,pos=.2] {} (6);
			\draw (5) edge[->, double, dashed] node[black,left,pos=.2] {} (1);
			\draw (2) edge[->, double, dashed] node[black,left,pos=.2] {} (4);
			\draw (3) edge[->, double, dashed] node[black,left,pos=.2] {} (6);
			\draw (4) edge[->, dashed, thick] node[black,left,pos=.2] {} (5);
			\draw (1) edge[->, dashed, thick] node[black,left,pos=.2] {} (3);
			\draw (6) edge[->, dashed, thick] node[black,left,pos=.2] {} (2);
			\draw (3) edge[->, ultra thick] node[black,left,pos=.2] {} (4);
			\draw (2) edge[->, ultra thick] node[black,left,pos=.2] {} (5);
			\draw (6) edge[->, ultra thick] node[black,left,pos=.2] {} (1);
			\draw (2) edge[->, double] node[black,left,pos=.2] {} (3);
			\draw (1) edge[->, double] node[black,left,pos=.2] {} (4);
			\draw (5) edge[->, double] node[black,left,pos=.2] {} (6);
		\end{tikzpicture}
	\end{minipage}
	\hfill
	\begin{minipage}{0.32\linewidth}
		\begin{tabular}{lllll}
			\toprule
			S1 & S2 & S3 & S4 & S5\\
			\midrule
			1-2 & 1-3 & 1-4 & 5-1 & 6-1 \\
			3-5 & 4-5 & 2-3 & 2-4 & 2-5 \\
			4-6 & 6-2 & 5-6 & 3-6 & 3-4 \\
			\bottomrule
		\end{tabular}
	\end{minipage}
	\hfill
	\,
	\caption{Oriented one-factorization of $K_6$ (left) and its associated single round-robin timetable (right), where each $i-j$ represents a home game of team $i$ against team $j$ for time slots $1$ to $5$.}
	\label{fig:onefactor}
\end{figure}

Numerous IP-based methods have previously been proposed to construct sports timetables.
See, e.g., \cite{vanDoornmalen2023} for a discussion on IP formulations, \cite{Nemhauser1998} for a popular decomposition approach known as first-break-then-schedule, and \cite{VanBulck2023} for an application of Benders' decomposition.

At the same time, several QUBO-inspired algorithms have been proposed for a related sports tournament problem known as the break-minimization problem.
In this problem, the objective is to minimize the overall number of breaks (i.e., consecutive games at home or away), given that the pairing of opponents is fixed in each round (e.g., \cite{Fujii2023,Kuramata2021,Urdaneta2018}).

\subsubsection{Problem statement}

Given a set of teams $T=\{1,\dots, n\}$ and time slots $S=\{1,\dots, 2(n-1)\}$, $n$ even, a double round robin (2RR) corresponds to a perfect matching of teams in each time slot such that over all time slots each team plays precisely once at home against every other team.
We call a 2RR `phased' if each team meets every other team exactly once during the first and last $n-1$ time slots.

The feasibility version of the ITC2021 sports scheduling problem considered in this benchmark is to find a phased 2RR that additionally respects the following hard constraints: 
\begin{itemize}
    \item {\bf Capacity constraints.} Regulate when teams can play home or away.
    \begin{itemize}
    \item CA1: team $i\in T$ plays at least or no more than $k$ home games in time slots $P\subseteq S$;
    \item CA2: same as CA1 but considering opponents as well;
    \item CA3: no more than two consecutive home or two consecutive away games; 
    \item CA4: teams in $T_1\subseteq T$ play no more than $k$ home games against teams in $T_2\subseteq T$ during time slots in $P\subseteq S$.
\end{itemize}
    \item {\bf Break constraints.} A team has a break if it plays consecutively at home or away.
  \begin{itemize}
    \item BR1: team $i\in T$ has no more than $k$ breaks during time slots in $P\subseteq S$;
    \item BR2: no more than $k$ breaks in total.
\end{itemize} 
\item {\bf Game constraints.} Forbid some games during specific time slots.
\begin{itemize}
    \item GA1: at most $k$ games from $G\subseteq T\times T$ during time slots in $P\subseteq S$.
\end{itemize} 
\end{itemize}

The original version of the ITC2021 sports scheduling problem also allowed the constraints above to be soft, penalizing violations of the soft constraints while strictly adhering to the hard ones.
Additionally, the original version of the problem considered two more soft constraints:
\begin{itemize}
\item {\bf Fairness and separation constraints.} Increase the attractiveness and fairness of the tournament.
\begin{itemize}
    \item FA2: at any point in time, the difference in the number of home games played between any two teams does not exceed two.
    \item SE1: there are at least 10 time slots between each pair of games involving the same teams.
\end{itemize} 
\end{itemize}

\subsubsection{Instances}

The problem instances stem from the International Timetabling Competition 2021 \cite{VanBulck2022b} and a follow-up study that generated additional problem instances using the same format \cite{VanBulck2024}.
These problem instances were generated using `Instance Space Analysis' \cite{Smith-Miles2023}. 
They feature a realistic set of constraints for real-world applications. 
Furthermore, they are challenging for state-of-the-art MIP and even dedicated sports tournament solvers. Finally, they feature a wide variety of characteristics that can expose different strengths and weaknesses of algorithms; see, e.g., \cite{VanBulck2024}.
Moreover, the instances were constructed so that a feasible solution is known to exist.
More specifically, we made use of the fact that generating a basic sports schedule is easy, even when some of the constraints are considered. 
Suppose a problem instance should suffice $p$ `easy' constraints from type A and $q$ `difficult' constraints from type B.
Then, we started by generating a solution that satisfies constraints A and added constraints of type B to this solution--such that the respective solution is compatible with constraints of type B.

The problem instances in this benchmark can be divided into four subclasses: ITC2021, Large, Medium, and Small.
The ITC2021 class corresponds to the 45 original problem instances from the International Timetabling Competition on Sports Timetabling 2021 \cite{ITC}, considering all hard and soft constraints.
Within the ITC2021 class are multiple instances for which no feasible solution was reported within a reasonable amount of time.
The Large, Medium, and Small classes represent feasibility versions of the ITC2021 problem instances. 
More specifically, to lower the entrance barrier and simplify the problem, we ignore all soft constraints and, therefore, reduce the optimization problem to its feasibility version.
For further experimentation, we also generated an additional set of problem instances having exactly the same number of constraints as the original ones; however, with 6, 8, and 12 teams only. 
These instances are contained in Medium and Small.

\subsubsection{Classical Baseline}

We provide feasible solutions to all problem instances--which were obtained during the instance generation.
An online solution checker is available in the validator section of \cite{ITC}.
Furthermore, the repository also includes a MIP formulation modeling example proposed in \cite{Berthold2021}.

\begin{figure}
	\centering
	\includegraphics[width=\linewidth]{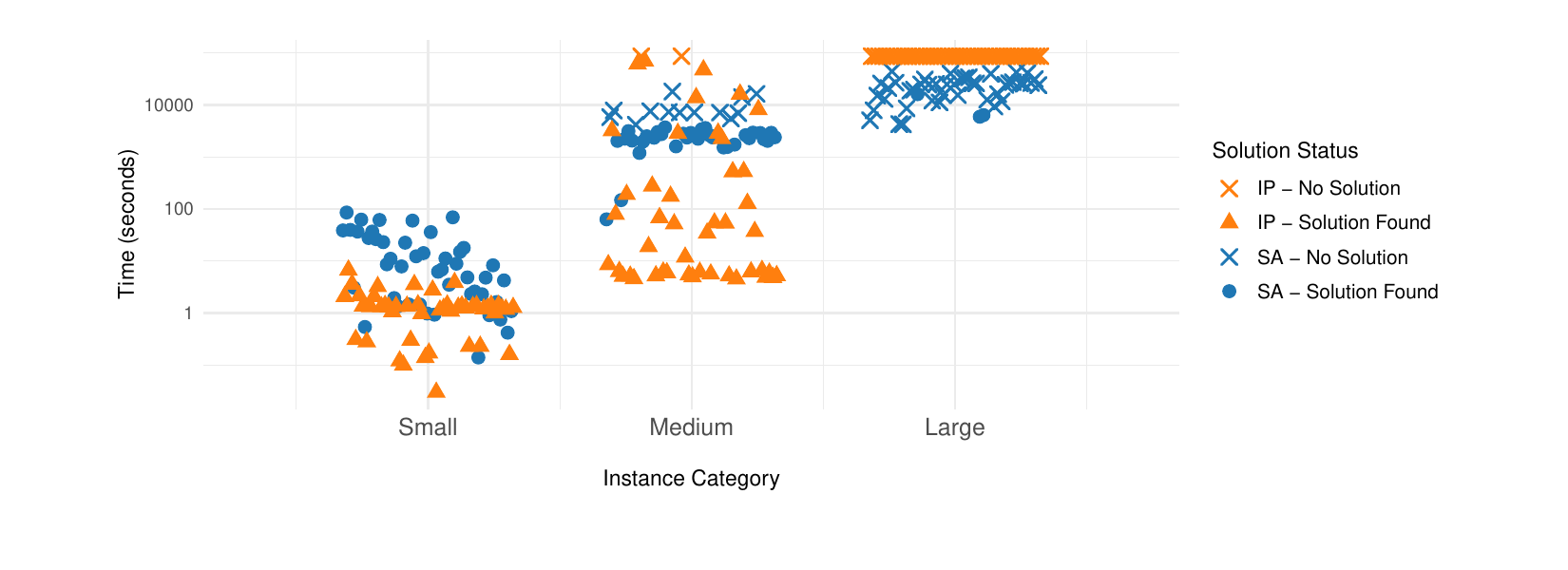}

    \caption{(Sports Scheduling) Solution times with Simulated Annealing (SA) and CPLEX (ILP) for the feasibility version of the problem. The time limit for SA was iteration-based, while the ILP solver timed out after 24 hours.}
	\label{fig:solTimeITC2021}
\end{figure}

Figure~\ref{fig:solTimeITC2021} shows the average solution times for a state-of-the-art Simulated Annealing (SA) algorithm (see \cite{Rosati2022}) and for a MIP formulation similar to the one presented in \cite{Berthold2021}. Both algorithms were run on a machine equipped with an AMD EPYC-7532 32-core processor, using 8 cores for solving the MIP (CPLEX 12.10) and a single core for the SA algorithm.
The MIP solver was given a time limit of 24 hours, while the SA algorithm was stopped after a fixed number of iterations. 
The MIP performed reasonably well on small and most medium-sized instances but failed to solve any of the large instances.
Also, the SA algorithm managed to find only solutions to three large instances, even though it was given considerable computational resources.
As a reminder, the instances were selected in such a way that it is known that existing state-of-the-art solvers heavily struggle to find feasible solutions.

\subsection{Portfolio Optimization} \label{sec:problems:portfolio}
\renewcommand{\customcaptiontext}{(Portfolio) }
\subsubsection{Background}

The problem of quantitative investment and risk diversification in portfolio optimization spans a broad spectrum of methodologies, each incorporating different simplifications, investor priorities, and trade-offs between risk exposure, returns, computational complexity, robustness, and out-of-sample performance. A foundational approach is Markowitz’s mean-variance optimization, formulated as a convex quadratic program~\cite{Markowitz1952}. This framework has been extended to address real-world constraints~\cite{Kolm2014}, estimation errors~\cite{Haff1980,ChamberlainRothschild1983,LalouxCizeauBouchaud1999}, sensitivity to input variability~\cite{MichaudR2008,Prado2019,BENTAL1999,Tuetuencue2004}, and Bayesian techniques that integrate expert judgment with statistical models~\cite{BlackLitterman1992,AvramovZhou2010,CorPeTue2018}. 
Computational limitations of the methodologies for practically relevant cases~\cite{Michaud2015} have led to the adoption of, e.g., fixed-mix strategies that improve model transparency and stability~\cite{Merton1969,DeMiguel2007}, hierarchical risk parity and risk budgeting techniques~\cite{Prado2016,GavaTurc2022}, and multiperiod approaches which help to capture intertemporal effects and hedging demands~\cite{Almgren2000,Garleanu2013}. 
In fact, portfolio optimization models can be defined with varying complexity \cite{Jan2015,Kolm2019,Xiaoyue2022,Camilo2023} depending on whether one aims to also jointly model risk, return predictability, and impact costs over time while addressing the exponential growth of time complexity with respect to the number of assets and rebalancing periods~\cite{Brown2011}. Portfolio optimization can transform into a stochastic dynamic programming problem, which, under realistic conditions that account for, e.g., transaction costs or short selling, has been proven to be computationally intractable for portfolios with more than two risky assets~\cite{Constantinides1979,Brown2010op}.

Given its relevance and complexity, portfolio optimization has attracted considerable interest from the quantum computing community. Theoretical research suggests that quantum algorithms may offer speedups under specific conditions that may, however, not hold in real-world instances; see, e.g., \cite{kerenidis2019quantum,rosenberg2016solving,lim2024quantum}. 
Another line of research has investigated how one may approach simple portfolio optimization instances with near-term compatible quantum algorithms \cite{brandhofer2022benchmarking, slate2021quantum, mugel2022dynamic}.
The simplified version of the problem derived here focuses on the computational complexity in the multiperiod setting, using a binary quadratic formulation of the mean-variance model with linear constraints, and perfect knowledge of expected future returns and covariances. 
This allows for the creation of compact yet computationally challenging problem instances, as the problem is \NP-hard \cite{Bienstock1996}, and the systematic benchmarking of current quantum algorithmic capabilities.

\begin{newtextbox}[Why is it interesting?]
A vast array of industrial and scientific domains are interested in solving problems where one aims to find an optimal composition of a portfolio of coupled objects with respective observables that typically share a common probability space. 
(Financial) Portfolio management is one of the hardest variants of this problem. It concerns the risk-sensitive allocation of a time-dependent investment budget for a number of diversified financial assets whose behavior varies with time according to unknown stochastic processes, which are driven by external factors. 
The hardness of this problem can be explored at different levels of complexity, even at a small scale, and, as such, allows systematic tracking of algorithmic and hardware progress.
\end{newtextbox}

\subsubsection{Problem statement}

Let \(\mathcal{T} := \{ t_1, \dots, t_m \}\) be an ordered set of discrete time points such that \(0 \leq t_1 < t_2 < \cdots < t_m\), representing a finite investment horizon in the future. 
Consider an artificial deterministic market that is fully described by a closed system of $n \in \mathbb{Z}^+$ financial assets $\{a_i \mid i=1\dots,n \}$, where each asset $a_i$ has known expected market prices $\{\bar{p}_{it}\}$ at each time $t \in \mathcal{T}$.
Assuming a myopic investor in a mean-variance framework and no access to additional information, the objective is to find a multiperiod trading strategy $\mathcal{S}_1 := \{x_{it}\}$ at $t_1$, where portfolio allocations $x_{it} \in \{0, 1\}$ represent binary investment decisions of up to $C$ normalized capital units $u \in \mathbb{R}^+$ into assets $\{a_i\}$.
We define the price $p_{it}$ as follows: Let $e_i := u/\bar{p}_{it_0}$, then $p_{it}=e_i \bar{p}_{it}$, i.e., we compute how many assets of type $i$ we can buy for one unit of cash at time $t_1$. 
The amount of stocks $e_i$ we can get at time $t_1$ for one unit of cash $u$ is then used for all later time steps, i.e., 
$p_{it}$ tells us what has become out of one cash unit invested into asset $i$ at time $t$. 
The investor seeks to balance returns and risk at terminal time $t_m$, where the temporal returns are given by $(p_{it} - p_{i, t-1})x_{it}$ and the temporal risk is measured through the risk preference  $\lambda$-weighted positive semidefinite covariance matrix $\Sigma$ as $\lambda (px)^T \Sigma xp$, while obeying a set of regularizing trading constraints. 

These constraints aim to incorporate more realistic real-world conditions, although here simplified into a 
quadratic optimization problem:
\begin{itemize}
    \item Transaction cost: Rebalancing of $x$ over $\mathcal{T}$ incurs costs due to fees, taxation, etc.~that negatively offset the returns by a factor of transaction cost $\delta \in \mathbb{R}^+$ for each change $p_{it}\lvert x_{it-1} - x_{it} \rvert = p_{it}(x_{it-1} + x_{it} - 2 x_{it-1} x_{it})$ and a liquidation cost $\delta p_i x_i$ at terminal time $t_m$. This is a strong simplification, as cost factors are neither global nor time-invariant scalars, but rather stochastic functions driven by market uncertainties.
    \item Short selling cost: Given a subset $f\in\mathcal{F} \subseteq \{a_i\}$ of assets permissible for short selling positions over $\mathcal{T}$, a fixed loan rate for short positions $\rho \in \mathbb{R}^+$ is applied to each allocation $\rho p_{ft} x_{ft}$ with $f \in \mathcal{F}$.
    \item Cash flow limits: Let \( u \in \mathbb{R}^+ \) denote the unit value of cash, and \( C \) the total available units of normalized capital. Each asset is assigned an indicator \(\tau_i \in \{-1, +1\}\), representing a short (\(-1\)) or long (\(+1\)) position. At any time, the total net investment $\sum_i \tau_i x_{it}$ must not exceed the capital limit $C$, and the total number of assets held at each time step must be bounded by $B$, which serves as an upper limit on the number of acquired positions. 
    \item Cash interest: Temporarily unused capital earns risk-free interest 
    $\nu u (C-\sum_i\tau_i x_{it})$
    with a risk-free rate $\nu \in \mathbb{R}^+$. 
    \item Multiple shares: We also allow that each asset \(i\) can be selected up to \(k\) times. We use a unary encoding where the Hamming weight of \(k\) binary variables encodes the number of shares, i.e., we essentially consider \(k\) copies of the same asset that can be selected. For the small number of shares considered here, this is as efficient as other encodings, but it has the advantage that we can easily model the transaction costs. Thus, we expand the vector of portfolio allocations to a \(2kn\)-dimensional binary vector \(\mathbf{x}\) for the \(n\) assets at each time \(t\), where each block of \(k\) binary variables represents a single asset type. The inclusion of a factor of 2 accounts for the allowance of short selling, with an equal number of vector elements allocated to represent both long and short positions. 
\end{itemize} 

The problem can then be formulated as a Binary Quadratic Program (BQP), where the objective is a quadratic function subject to linear constraints, and the decision variables are restricted to be binary:
\begin{align}
\min_{{x \in \{0,1\}^{n \times m}, y \in \{0,1\}^{|\mathcal{C}| \times m}}} \quad 
& \sum^m_{t=1} \Bigg(
\lambda \overbrace{ \sum^n_{i=1} \sum^n_{j=1} \tau_i p_{it} x_{it} \sigma_{ijt} \tau_j p_{jt} x_{jt}}^{\text{risk}} 
\\ 
& - \sum^{n-1}_{i=1}  
    \underbrace{\tau_i (p_{i,t+1} - p_{it}) x_{it}}_{\text{return}} 
    - \sum^n_{i=2}\underbrace{\delta p_{it} (x_{it-1} + x_{it} - 2 x_{it-1} x_{it})}_{\text{transaction cost}} \notag \\
& \qquad - \underbrace{\nu u \sum_{c\in\mathcal{C}}2^c y_{ct}}_{\text{cash interest}} 
+ \underbrace{\rho \sum_{f \in \mathcal{F}} p_{ft} x_{ft}}_{\text{short selling cost}} 
\Bigg) 
+ \underbrace{\delta \sum^n_{i=1} (p_{i1} x_{i1}+p_{i m} x_{i m})}_{t_1\text{ transaction $+$ liquidation cost}}
\label{eq:po:main}
\end{align}
subject to
\begin{align}
\sum^n_{i=1} \tau_i x_{it} 
+ \sum_{c\in\mathcal{C}}2^c y_{ct} &=  
C &&\text{for all }t \in \mathcal{T}\label{eq:po:cash}
\\
\sum^n_{i=1} x_{it} 
&\leq
B &&\text{for all }t \in \mathcal{T}\label{eq:po:assets}
\end{align}
where \eqref{eq:po:cash} describes the capital limit and \eqref{eq:po:assets} limits the number of assets that can be included in the portfolio. 
Since we restricted the problem to binary variables, we modeled the amount of available cash units as binary variables  $y_{ct}\in
\{0,1\}$ with a suitable $c\in\mathcal{C}:=\{0,\ldots,\lfloor\log_2(C)\rfloor\}$ such that $\sum_{c\in\mathcal{C}} 2^c\ge C$.
It is worth noting that alternative formulations may also be possible. 

\subsubsection{Instances}

\begin{table}[htb!]
\centering
\begin{minipage}[]{0.52\linewidth}
\begin{tabular}{cll}
\toprule
&\textbf{Parameter} & \textbf{Value} \\
\hline
$\nu$ & risk-free rate & 0.01\% \\
$\delta$ & transaction cost rate & 0.1\% \\
$\rho$ & loan rate on short positions & 0.0025\%  \\
$C$ & available \# of cash units & 10 units \\
$k$ & Max. \# units of each asset & 3 \\
\bottomrule\\\\\\
\end{tabular}
\end{minipage}
\begin{minipage}[]{0.45\linewidth}
\begin{tabular}{cccccr}
\toprule
\textbf{\# inst}&\textbf{$n$} & \textbf{$m$} & \textbf{$u$} & \textbf{$B$} & \textbf{\# vars}\\
\hline
4 & 10 & 10 & 3 & 4 & 710\\
4 & 10 & 15 & 3 & 4 & 1065\\
4 & 50 & 10 & 3 & 20 & 3110\\
4 & 50 & 15 & 3 & 20 & 4665\\
4 & 200 & 10 & 3 & 50 & 12110\\
4 & 200 & 15 & 3 & 50 & 18165\\
4 & 400 & 10 & 3 & 100 & 24110\\
4 & 400 & 15 & 3 & 100 & 36165\\
\bottomrule
\end{tabular}
\end{minipage}
\customcaption{(left) Fixed parameters used to generate the portfolio optimization problem instances. The annual rates of $\nu$ and $\rho$ correspond to 2.55\% and 0.92\%, respectively. The total amount of available cash equals \$1,000,000. (right) Overview of the different sizes and parameters of the Portfolio Optimization instances for binary variables. We generated four instances for each setting.}
\label{tab:experiment_parameters}
\end{table}

Table~\ref{tab:experiment_parameters} (left) outlines the parameters that are used in our experiments.
The parameter choices are based on established literature
\cite{gaivoronski2005optimal,d2002market,tepla2000optimal}. 
Using these parameters, we generated 4 instances each (1 containing the original data and 3 with randomly perturbed data) for 10, 50, 200, and 400 possible assets with two different time horizons: 10 days and 15 days.
We used real S\&P 500 data from January 1, 2024, to May 31, 2024, for the asset prices $p_{it}$ and the corresponding covariances $\sigma_{ijt}$.  
The size of the instances is defined by 
the number of assets \(n\), the number of considered time steps \(m\), and the number of chosen asset units \(k\). 
The risk-affinity/aversion parameter \(\lambda\), the unit limit on asset selection \(B\), and the capital bound \(C\) have a strong influence on the solvability of the instances. 
We would like to highlight that the repository includes instances with $\lambda=0$, which represents an edge-case setting that lacks operational meaning but serves as a simplified mathematical representation.

Moreover, we generated randomly perturbed instances based on real-world S\&P 500 data from January 1, 2024, to May 31, 2024.
The generated data is sampled independently at each time step using the $t$-distribution with $\nu = 5$, scaled to the mean and covariance of the stocks as given by the original data, with a rolling window of 30 days. 
Outliers are randomly added to the dataset.
We always consider the $n$ largest companies, ranked by their market capitalization as of January 1, 2025.
Table~\ref{tab:experiment_parameters}(right) gives an overview of the instances in the problem class, with the instances on top being randomly generated and the instances on the bottom representing real-world data.

\subsubsection{Classical Baseline}

\begin{figure}[htb!]
    \centering
    \includegraphics[width=
    0.7\linewidth]{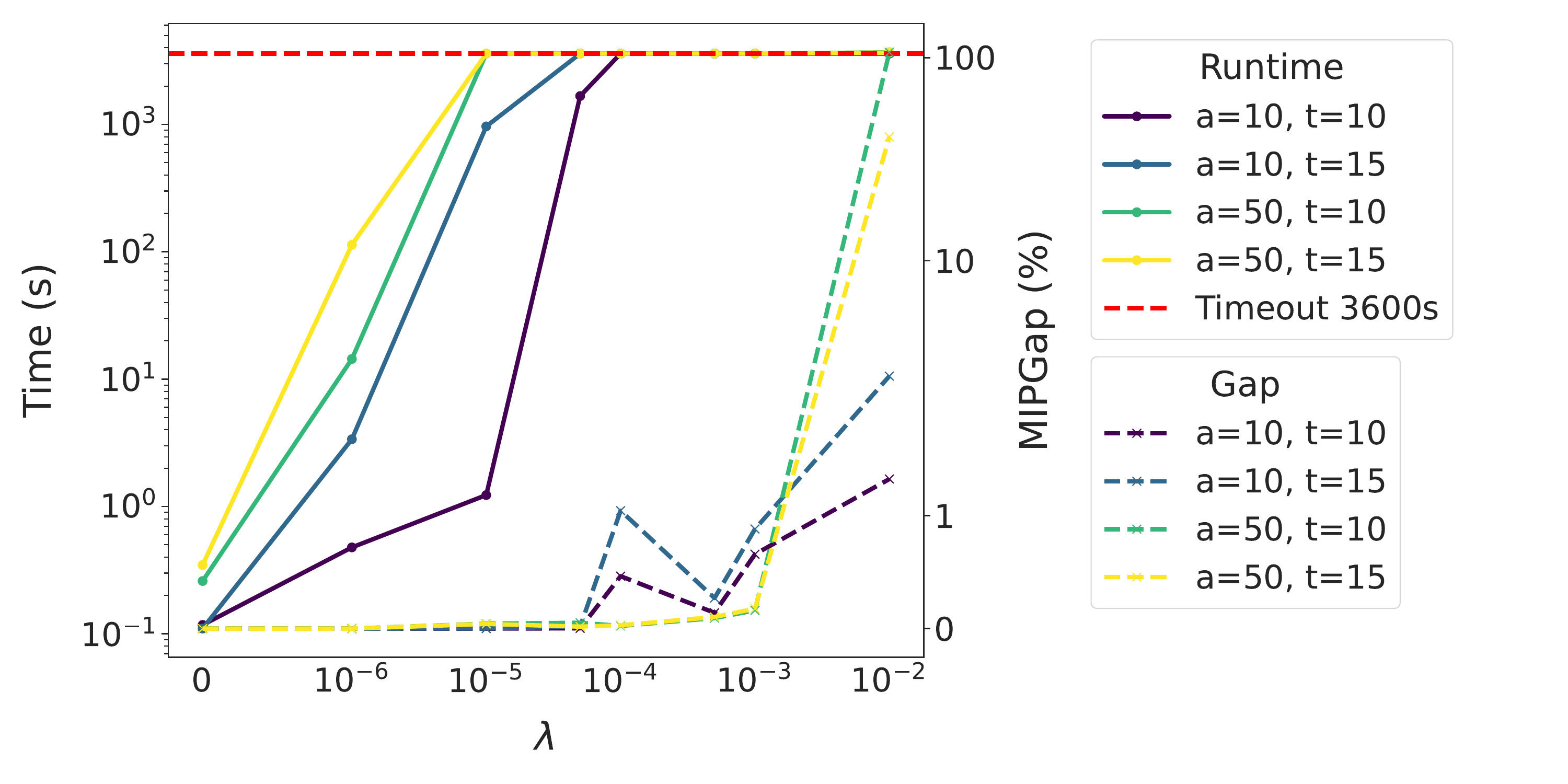}
    \customcaption{ Runtimes (solid lines) and MIP Gap (dashed lines) for \gurobi with a time limit of 3600s for different values of $\lambda$ averaged over three randomly generated similar instances.}
    \label{fig:portfolio:gurobi}
\end{figure}

\begin{table}[htb!]
\centering

\begin{tabular}{r|rrrr|rrr}
\toprule
 & \multicolumn{4}{c}{Gurobi} & \multicolumn{3}{c}{Abs2} \\
$\lambda$ & LB & Incumbent & Gap (\%) & RT (s) & Incumbent & TTS (s) & Gap (\%) \\
\midrule
0 & -879572 & -879572 & 0.00 & 0.4 & -879572 & 90.0 & 0.00 \\
0.000001 & -872141 & -872055 & 0.01 & 398.7 & -872055 & 79.1 & 0.00 \\
0.00001 & -887265.36 & -819899 & 8.21 & 3600.1 & -819899 & 30.2 & 0.00 \\
0.00005 & -692341.57 & -680124 & 1.80 & 3600.2 & -680124 & 50.7 & 0.00 \\
0.0001 & -604495.30 & -586823 & 3.01 & 3600.2 & -586823 & 206.7 & 0.00 \\
0.0005 & -424863.29 & -386990 & 9.80 & 3600.1 & -384862 & 1497.6 & 0.55 \\
0.001 & -367893.74 & -314736 & 16.90 & 3600.2 & -299298 & 741.9 & 4.91 \\
0.01 & -141194.79 & -437920 & 3124.21 & 3661.8 & -2825 & 1510.8 & 99.35 \\
\bottomrule
\end{tabular}

\customcaption{Comparison of (classical) BQP using \gurobi 12.0.1 with a tolerance of 1e-4 vs.(quantum-ready) QUBO formulation of a portfolio optimization problem--a050\_t15\_s00 instance. The \abstwo Gap is the relative gap between the \abstwo incumbent and the gurobi incumbent.}
\label{tab:po:a200_t10}

\end{table}

The tests are conducted by running the BQP model with \gurobi 12.0.1 on an Intel Xeon Gold 6338 CPU @ 2.00GHz processor using 128 threads with a timeout of 3600s and using the QUBO model with \abstwo on a system with four A40 SXM4 48 GB GPUs with a timeout of 3600s.
An overview of different parameters up to $n=50$ assets can be seen in Figure~\ref{fig:portfolio:gurobi}. With increasing $\lambda$, the running time and the upper bound on the optimality gap (MIP Gap) also increase.
Table~\ref{tab:po:a200_t10} shows execution details on the instance for $n=50$, $m=15$, $k=3$, $B=20$, $C=10$: 
All the objective function values shown are negative because we are minimizing the risk, and therefore, the profit is subtracted in the objective function.
As may be expected, when the risk adversity increases, the profit decreases, as well as the gap between the lower and upper bounds of the solution.
The computational performance varies depending on the risk aversion parameter $\lambda$. 
For $\lambda=0$, the risk term is ignored, and only the profit is optimized. This can be quickly solved to optimality. In the risk-averse case with a high $\lambda=0.01$, the opposite happens. 
The lowest risk strategy is to take the risk-free interest rate $\nu=0.01$ and receive a profit of $C\cdot m\cdot\nu=1000$. 

\begin{figure}[htb!]
    \centering
    \includegraphics[width=0.9\textwidth]{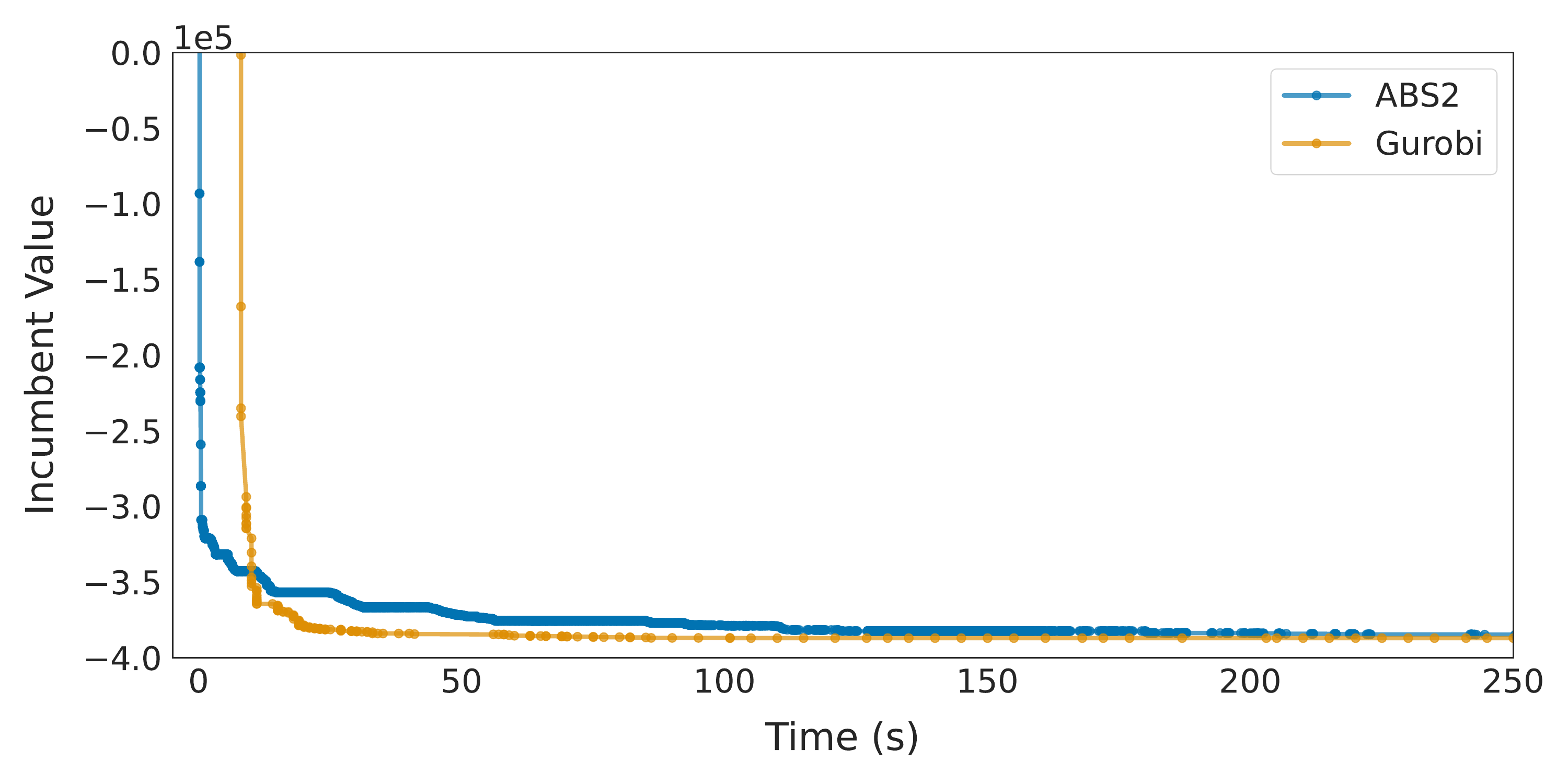}        
    \customcaption{The plot illustrates the best found objective values at every step of the solving for different risk aversion parameters with Abs2 and Gurobi solvers for one instance with $\lambda = 0.0005$ and 50 assets. While \abstwo finds good solutions faster, Gurobi eventually achieves better results.\label{fig:a050_t15_q0.0005}}
\end{figure}

Figure~\ref{fig:a050_t15_q0.0005} shows an 
in-depth analysis for a problem instance with $n=50$, $m=15$, $k=3$, $B=20$, $C=10$. More specifically, the Figure shows how the objective value changes over time for $\lambda = 0.0005$. The same system was used for \gurobi and \abstwo.
Although BQP achieves a better final objective value, the QUBO approach reaches good solutions in a substantially shorter time--without developing specialized heuristics. 

\subsection{Maximum Independent Set} \label{sec:problems:stable-set}
\renewcommand{\customcaptiontext}{(Independent Set) }

\subsubsection{Background}
An independent (or stable) set is a set of vertices not adjacent to each other in a graph, i.e., no edge connects any two vertices of the set.
The maximum independent set (MIS) problem (or maximum stable set problem) is to find the largest independent set for a given graph. It is deeply rooted in graph theory and combinatorial optimization and has a long history of research. Its decision problem is one of the first problems shown to be \NP-complete~\cite{karp1972reducibility}. It is known to be strongly \NP-hard for general graphs, while it can be solved in polynomial time for certain classes of graphs, e.g., $P_5$-free graphs and perfect graphs.
The independent set is a fundamental feature of a graph and relates to various other graph parameters. The MIS for a graph is equivalent to the maximum clique problem for its complement graph. A set of independent vertices is equivalent to its complement set being a vertex cover. Thus, the sum of the size of the MIS and the size of the minimum vertex cover is equal to the number of vertices in the graph.
The MIS is also useful for modeling industry problems,
e.g., non-intersecting label placement on a map~\cite{agarwal1998label},
discovering stable genetic components for designing engineered genetic systems~\cite{hossain2020automated}, and strategic planning, such as choosing locations for stores~\cite{wurtz2022industry}.

\begin{newtextbox}[Why is it interesting?]
The unweighted maximum independent set problem is a classic \NP-hard mathematical problem. Starting at problem sizes of several hundred variables, there are known instances that are hard to solve to proven optimality with existing methods \cite{OEIS_A265032}, and which are difficult to tackle even heuristically. 
An interesting property of this problem class is that it is well-suited for translation into relatively sparse QUBOs--often with small coefficients.
\end{newtextbox}

There have been extensive studies both on exact and approximate algorithms.
As far as we know, the best-known exact algorithm can compute the MIS of an $n$-vertex graph in $1.1996^n n^{O(1)}$ time and polynomial space~\cite{xiao2017exact}.
As for approximate algorithms, most studies (too many to list here) consider a particular class of graphs and develop algorithms with constant approximation ratios since the maximum independent set problem, in general, is Poly-APX-complete, meaning it is as hard as any problem that can be approximated to a polynomial factor~\cite{bazgan2005completeness}.
Quantum algorithms have also been investigated recently.
For example, a method is proposed for solving the maximum independent set problem on unit disk graphs~\cite{CLARK1990165} using a quantum simulator with Rydberg atomic arrays~\cite{ebadi2022quantum}. It should be noted that for unit disk graphs, better approximation results are available than for the general case; see, e.g., \cite{Marathe1995,DAS202063}.

\subsubsection{Problem statement}
Let $G = (V, E)$ be a graph with vertices $V=\{1,\ldots,n\}$ and edges $E \subset V \times V$.
The maximum independent set problem for the graph $G$ is 
\[
\max_{S \subseteq V} |S| \text{ with } i,j \in S \Rightarrow (i,j) \notin E.
\]
It can be represented as a Binary Linear Program (BLP):
\[
\max_{x \in \{0,1\}^n} \sum_{i \in V} x_i
\text{ subject to } x_i + x_j \leq 1 \text{ for all } (i, j) \in E, 
\]
or as a Binary Quadratic Program (BQP):
\[
\max_{x \in \{0,1\}^n} \sum_{i \in V} x_i^2
\text{ subject to } x_i x_j =0 \text{ for all } (i, j) \in E. 
\]
Note that $x_i = x_i^2$ holds for binary variables.
The constraint in each formulation forbids $x_i = x_j = 1$, for all $(i, j)\in E$, i.e., both ends of each edge cannot be in an independent set.
For the weighted version, some weights $w_i > 0$, $i \in V$,  can be added to the vertices.

\subsubsection{Instances}
\label{sec:stable_set_instances}
Any collection of graphs might be a suitable benchmark set for MIS. 
We aimed to collect some instances known to be challenging and added several smaller ones to facilitate progress tracking.
We collected instances from various sources, in particular:
Network Repository \cite{nr}, SteinLib \cite{KMV00}, OESIS Implementation Challenge \cite{OEIS_A265032}, and self-generated graphs.
Furthermore, we provide a program that takes a graph and an independent set and verifies that it is indeed a stable set in the graph.

\subsubsection{Classical Baseline}
\begin{figure}[htb!]
    \centering
    \includegraphics[width=0.9\linewidth]{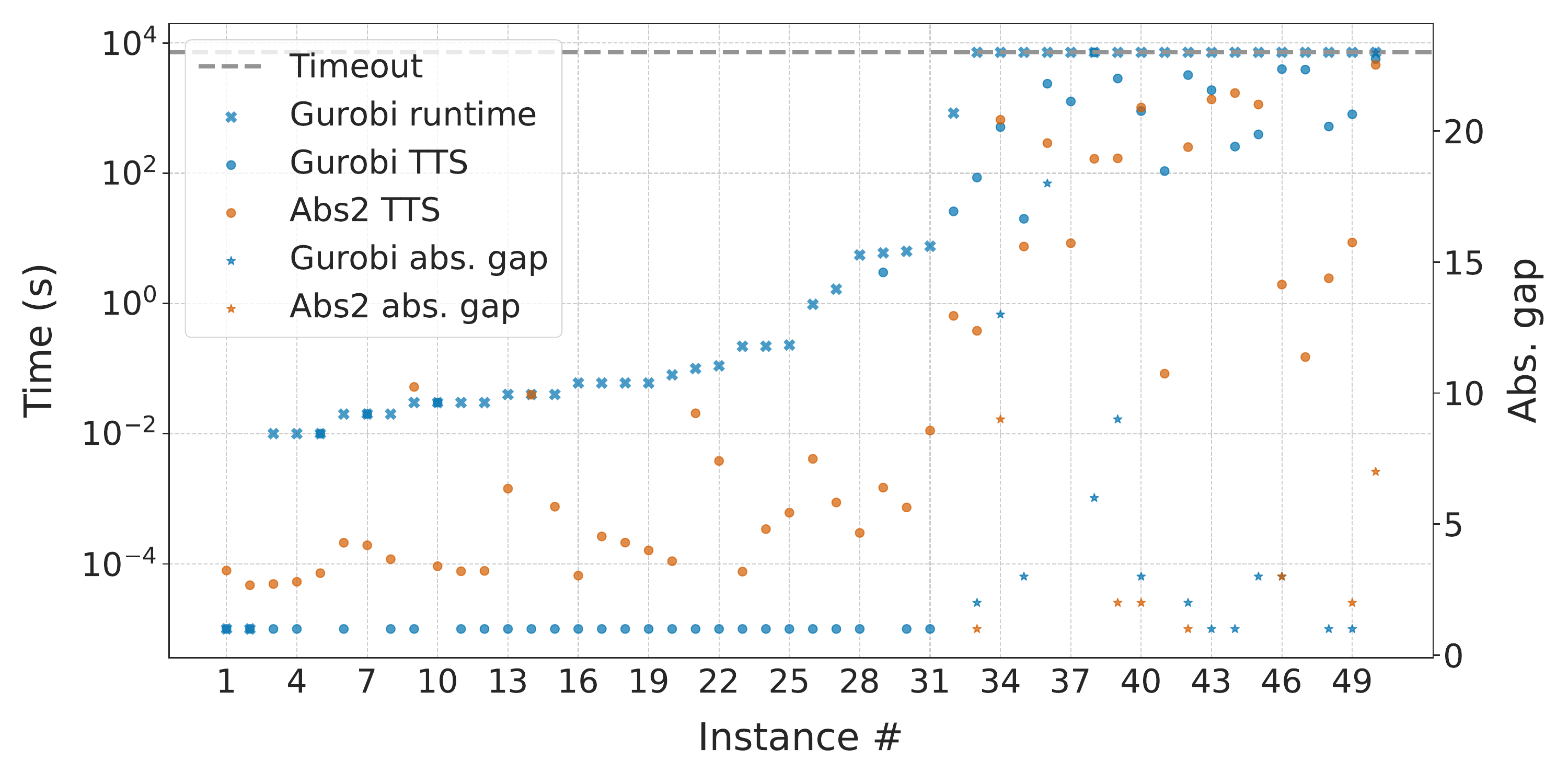}
    \customcaption{ Runtimes, time-to-solution, and absolute gap to optimal/best-known solution for the problem sorted by \gurobi's runtime. Only gaps $> 0$ are plotted.}
    \label{fig:independent_set:runtime_and_tts}
\end{figure}

We provide formulations as binary linear programs and as QUBO.
These can be solved by any suitable solver.
We provide baseline results for the binary linear program using \gurobi 12.0.1 run on an Intel Xeon Gold 6338 CPU @ 2.00~GHz processor using 128 threads with a timeout of 7200 seconds. 
The baseline results for the QUBO formulation are computed using \abstwo as a heuristic run on four Nvidia A40 48GB with a timeout of 7200 seconds.
Corresponding runtime, TTS, and absolute gaps to the optimal/best-known solutions can be found in Figure~\ref{fig:independent_set:runtime_and_tts}.
Exact data and instance names can be obtained from Table~\ref{tab:independent_set:_gurobi_and_abs2_results}.

\begin{table}[htb!]
\centering
\resizebox{\textwidth}{!}{%
\begin{tabular}{r|l|rr|r|rrr|rr|l}
\toprule
\multicolumn{5}{r}{} & \multicolumn{3}{r}{Gurobi} & \multicolumn{2}{r}{Abs2} &  \\
\# & Name & Nodes & Edges & Best & RT (s) & TTS (s) & Gap & TTS (s) & Gap & Source \\
\midrule
4 & farm & 17 & 39 & 10 & 0.0 & 0.0 & 0 & 0.0 & 0 & \cite{nr} \\
2 & mammalia-kangaroo-interactions & 17 & 91 & 4 & 0.0 & 0.0 & 0 & 0.0 & 0 & \cite{nr} \\
3 & johnson8-2-4 & 28 & 210 & 7 & 0.0 & 0.0 & 0 & 0.0 & 0 & \cite{nr} \\
8 & ibm32 & 32 & 90 & 13 & 0.0 & 0.0 & 0 & 0.0 & 0 & \cite{nr} \\
5 & karate & 34 & 78 & 20 & 0.0 & 0.0 & 0 & 0.0 & 0 & \cite{nr} \\
1 & football & 35 & 118 & 16 & 0.0 & 0.0 & 0 & 0.0 & 0 & \cite{nr} \\
7 & chesapeake & 39 & 170 & 17 & 0.0 & 0.0 & 0 & 0.0 & 0 & \cite{nr} \\
16 & MANN-a9 & 45 & 918 & 3 & 0.1 & 0.0 & 0 & 0.0 & 0 & \cite{nr} \\
10 & aves-sparrow-social & 52 & 454 & 13 & 0.0 & 0.0 & 0 & 0.0 & 0 & \cite{nr} \\
11 & insecta-ant-colony1-day38 & 56 & 1134 & 6 & 0.0 & 0.0 & 0 & 0.0 & 0 & \cite{nr} \\
6 & sloane\_1dc\_64 & 64 & 543 & 10 & 0.0 & 0.0 & 0 & 0.0 & 0 & \cite{nr} \\
23 & hamming6-4 & 64 & 704 & 12 & 0.2 & 0.0 & 0 & 0.0 & 0 & \cite{OEIS_A265032} \\
12 & hamming6-2 & 64 & 1824 & 2 & 0.0 & 0.0 & 0 & 0.0 & 0 & \cite{nr} \\
20 & johnson8-4-4 & 70 & 1855 & 5 & 0.1 & 0.0 & 0 & 0.0 & 0 & \cite{KMV00} \\
13 & es60fst03 & 113 & 142 & 55 & 0.0 & 0.0 & 0 & 0.0 & 0 & \cite{nr} \\
19 & johnson16-2-4 & 120 & 5460 & 15 & 0.1 & 0.0 & 0 & 0.0 & 0 & \cite{nr} \\
15 & es60fst01 & 123 & 159 & 60 & 0.0 & 0.0 & 0 & 0.0 & 0 & \cite{nr} \\
26 & C125-9 & 125 & 787 & 34 & 1.0 & 0.0 & 0 & 0.0 & 0 & \cite{OEIS_A265032} \\
24 & sloane\_1zc\_128 & 128 & 1120 & 18 & 0.2 & 0.0 & 0 & 0.0 & 0 & \cite{OEIS_A265032} \\
22 & sloane\_1dc\_128 & 128 & 1471 & 16 & 0.1 & 0.0 & 0 & 0.0 & 0 & \cite{KMV00} \\
18 & sloane\_2dc\_128 & 128 & 5173 & 5 & 0.1 & 0.0 & 0 & 0.0 & 0 & \cite{KMV00} \\
25 & insecta-ant-colony3-day09 & 160 & 8883 & 9 & 0.2 & 0.0 & 0 & 0.0 & 0 & \cite{OEIS_A265032} \\
9 & es60fst04 & 162 & 238 & 78 & 0.0 & 0.0 & 0 & 0.1 & 0 & \cite{nr} \\
27 & keller4 & 171 & 5100 & 11 & 1.7 & 0.0 & 0 & 0.0 & 0 & \cite{nr} \\
14 & es60fst02 & 186 & 280 & 88 & 0.0 & 0.0 & 0 & 0.0 & 0 & \cite{nr} \\
17 & c-fat200-1 & 200 & 1534 & 18 & 0.1 & 0.0 & 0 & 0.0 & 0 & \cite{nr} \\
21 & gen200\_p0-9\_44 & 200 & 1990 & 44 & 0.1 & 0.0 & 0 & 0.0 & 0 & \cite{KMV00} \\
29 & brock200-2 & 200 & 10024 & 12 & 6.0 & 3.0 & 0 & 0.0 & 0 & \cite{nr} \\
31 & brock200-3 & 200 & 12048 & 9 & 7.6 & 0.0 & 0 & 0.0 & 0 & \cite{nr} \\
30 & brock200-4 & 200 & 13089 & 8 & 6.3 & 0.0 & 0 & 0.0 & 0 & \cite{nr} \\
28 & brock200-1 & 200 & 14834 & 6 & 5.6 & 0.0 & 0 & 0.0 & 0 & \cite{nr} \\
37 & brock400-1 & 400 & 20077 & 27 & 7200.1 & 1269.0 & 0 & 8.4 & 0 & \cite{nr} \\
33 & R\_500\_005\_1 & 500 & 6256 & 91 & 7200.0 & 86.0 & 2 & 0.4 & 1 & \cite{nr} \\
35 & C500-9 & 500 & 12418 & 57 & 7200.1 & 20.0 & 3 & 7.5 & 0 & \cite{Rivetta2024} \\
45 & brock800-1 & 800 & 112095 & 23 & 7200.3 & 395.0 & 3 & 1139.3 & 0 & \cite{nr} \\
49 & frb45-21-3 & 945 & 58245 & 45 & 7201.5 & 806.0 & 1 & 8.7 & 2 & \cite{nr} \\
39 & R\_1000\_005\_1 & 1000 & 24670 & 117 & 7200.1 & 2859.0 & 9 & 169.6 & 2 & \cite{nr} \\
41 & hamming10-4 & 1024 & 89600 & 40 & 7200.2 & 108.0 & 0 & 0.1 & 0 & \cite{GleixnerHendelGamrathetal2021} \\
42 & frb50-23-3 & 1150 & 81068 & 50 & 7200.2 & 3226.0 & 2 & 252.3 & 1 & \cite{nr} \\
40 & frb53-24-1 & 1272 & 94227 & 53 & 7200.1 & 903.0 & 3 & 1023.5 & 2 & \cite{nr} \\
32 & socfb-haverford76 & 1446 & 59589 & 282 & 838.7 & 26.0 & 0 & 0.6 & 0 & \cite{nr} \\
47 & p\_hat1500-3 & 1500 & 277006 & 94 & 7200.8 & 3915.0 & 0 & 0.2 & 0 & \cite{nr} \\
48 & p\_hat1500-1 & 1500 & 839327 & 12 & 7201.4 & 523.0 & 1 & 2.4 & 0 & \cite{nr} \\
46 & frb59-26-2 & 1534 & 126163 & 59 & 7200.8 & 3980.0 & 3 & 1.9 & 3 & \cite{nr} \\
34 & sorrell7 & 2048 & 39424 & 198 & 7200.1 & 513.0 & 13 & 663.5 & 9 & \cite{nr} \\
44 & sorrell4 & 2048 & 504451 & 24 & 7200.2 & 257.0 & 1 & 1708.8 & 0 & \cite{nr} \\
43 & socfb-trinity100 & 2613 & 111996 & 499 & 7200.2 & 1891.0 & 1 & 1361.5 & 0 & \cite{nr} \\
36 & keller6 & 3361 & 1026582 & 59 & 7200.1 & 2377.0 & 18 & 290.9 & 0 & \cite{Rivetta2024} \\
50 & frb100-40 & 4000 & 572774 & 100 & 7205.1 & 5695.0 & 23 & 4634.8 & 7 & \cite{nr} \\
38 & C4000-5 & 4000 & 3997732 & 18 & 7200.1 & 7200.1 & 6 & 167.0 & 0 & \cite{nr} \\
\bottomrule
\end{tabular}

}
\customcaption{Best solution, Runtimes, TTS, and absolute gaps to best found (some even optimal) solutions for Gurobi 12.0.1 and Abs2. Notably, only Gurobi checks optimality. Abs2 runs as a heuristic. Please note that we do not report any runtimes for \abstwo, since the stopping criteria for these heuristic solvers vary. 
}
\label{tab:independent_set:_gurobi_and_abs2_results}
\end{table}

\subsection{Network Design} \label{sec:problems:network}
\renewcommand{\customcaptiontext}{(Network) }

\subsubsection{Background}

Network design models have wide applications in telecommunications, transportation planning \cite{gavish1991topological, magnanti1984network}, and power networks \cite{SaberiPowerNetworks23}. One common formulation for this problem is to use an MIP formulation \cite{bienstock1995computational}. Given an \(n \times n\) matrix \(T\), where each entry \(t_{ij}\) represents the traffic to be routed between vertex \(i\) and \(j\), and an integer \(p > 0\), construct a simple directed graph \(D = (V, A)\) with vertex set \(V = \{1, \ldots, n\}\) and a subset of selected edges \(S^* \subseteq A\), where each vertex has indegree and outdegree equal to $p$. Furthermore, all the demands are routed such that the maximum load on any single edge of the network is minimized. The load on an edge is the sum of all the traffic that passes through that edge. The problem is \NP-hard \cite{bienstock1995computational}.

\begin{newtextbox}[Why is it interesting?]
The min-max objective function underlying network design models is linear, and the problem often has a sparse structure. 
While it is trivial to construct a feasible solution, finding optimal solutions is very difficult. The particular problem presented in this section has been withstanding all attempts to solve it to proven optimality for the last 30 years in its 24-node size. 
Furthermore, network design models have wide applications in telecommunications and transportation planning \cite{gavish1991topological, magnanti1984network}.
\end{newtextbox}

Although we will focus on the mathematical formulation as a multicommodity-directed model, many other formulations have
been used in network design, depending on the problem context. In \cite{gendron1999multicommodity,armacost2002composite}, the authors provide surveys of alternative formulations. 
In the standard formulation, both flow decision variables 
and design decision variables 
are modeled explicitly \cite{bienstock1995computational, bienstock1998minimum, gendron1999multicommodity}. 
In this case, the linear programming relaxations of multicommodity flow formulations provide lower bounds that are normally weak \cite{gendron1999multicommodity,armacost2002composite}. 
To overcome this issue, in \cite{armacost2002composite}, the authors remove the flow variables as explicit decisions and embed them within the design variables. 
Furthermore, in \cite{gendron1999multicommodity,armacost2002composite}
the authors also provide a survey of different methods that have been used for network design.  For example, in \cite{magnanti1986tailoring}, the authors apply Benders decomposition and demonstrate the effectiveness of variable elimination preprocessing and a dual ascent procedure to accelerate the decomposition algorithm.    
In \cite{bienstock1995computational}, the authors report on computational experience with a cutting plane algorithm for the above formulation. In \cite{barnhart1996air}, they use the column generation technique to obtain near-optimal solutions.

\subsubsection{Problem statement}

Given a matrix $T\in\NN^{n\times n}$, and an integer $p>0$, 
construct a simple directed graph $D=(V,A)$ with vertex set $V=\{1,\ldots,n\}$, where each vertex has indegree and outdegree equal to $p$, and simultaneously, in $D$, route $t_{ij}$ units of flow from vertex $i$ to $j$, for all $1\leq i\neq j\leq n$, such as to minimize the maximum aggregated flow on any arc in $D$. 

We can formulate this network design problem as a multicommodity design problem as follows:
For $k, i, j \in \{1, \ldots, n\}$ we define binary variables $x_{i j}\in\{0,1\}$ which are $1$ iff an arc is established between vertex $i$ and $j$ and $0$, otherwise. Further, we define non-negative integer variables $f_{kij}\in\ZZ_+$ that represent the flow of commodity $k$ between vertices $i$ and $j$, and a suitable big constant $M$. Then we get:

\begin{align}
   \underset{x, f, z}{\min}\:z\qquad\quad&&&k, i, j \in \{1, \ldots, n\}\\  
   \text{subject to} \sum\limits_{j \neq i} x_{ij} & = p && \text{for all } i\:, \label{eq:design:con1a}\\
    \sum\limits_{j \neq i} x_{ji} & = p && \text{for all } i \:,\label{eq:design:con1b}\\
    f_{kij} & \leq M x_{ij} && \text{for all } k, i \neq j \:,\label{eq:design:con2}\\
    \sum\limits_{j \neq i} f_{kij} - \sum\limits_{j \neq i} f_{kji} & = t_{ki} && \text{for all } i,k \:,\label{eq:design:con3}\\
    \sum\limits_{k} f_{kij} & \leq z && \text{for all } i \neq j\:.\label{eq:design:con4}\\
\end{align}

Constraints \eqref{eq:design:con1a} and \eqref{eq:design:con1b} set the outdegree and indegree of each vertex to $p$. Constraint \eqref{eq:design:con2} ensures that flow is only allowed on arcs in $A$. Constraint \eqref{eq:design:con3} ensures the flow conservation for each commodity $k$. 
\eqref{eq:design:con4} ensures $z$ is greater than the aggregated flow on any arc. Since the objective is to minimize $z$, \eqref{eq:design:con4} will be met with equality at the arc with the highest aggregated flow. 

\subsubsection{Instances}
The original problem by \cite{bienstock1995computational} set $p=2$, $n=24$, $T\in\{0,\ldots,100\}^{n\times n}$, and had continuous flow variables. We have multiplied all demands by 1,000 and require an integral flow, making the problem completely integer. To provide smaller versions of the problem, the benchmark instances use the data of the first $n$ vertices from $n=5$ to $24$. 

We provide a checking routine that, given a sequence $n$, a file with the demands, and a file with a solution, checks whether the solution is feasible and computes the objective value.

\subsubsection{Classical Baseline}

\begin{figure}[htb]
    \centering
    \includegraphics[width=0.9\linewidth]{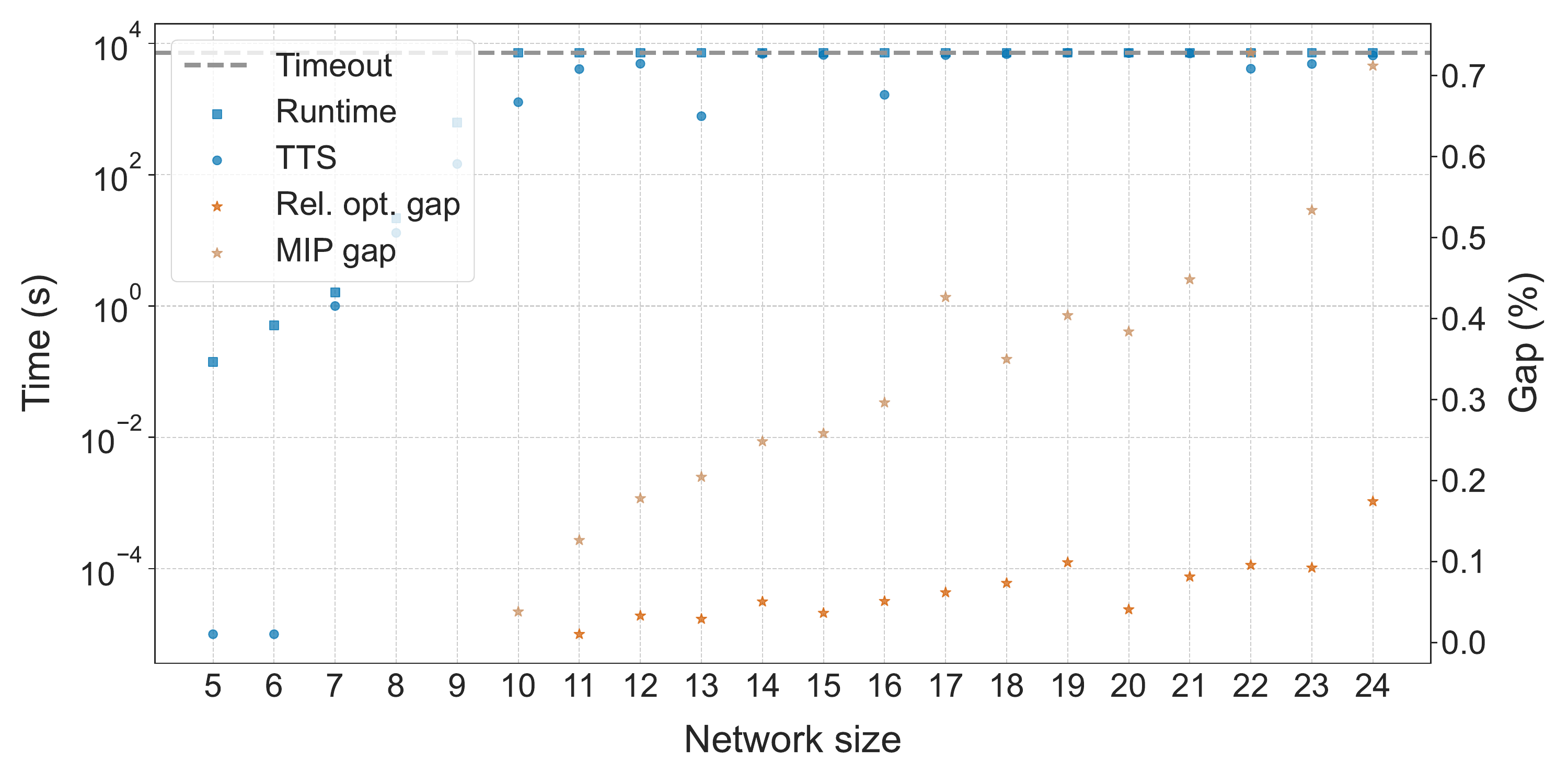}
    \customcaption{ \emph{Runtime} is the time to compute a proven optimal solution. \emph{Time-to-solution} is the time until the best solution was found. \emph{Rel.\ opt.\ Gap} is the gap to the optimal or best-known solution. \emph{MIP Gap} is the gap to the lower bound. Only gaps $>0$ are plotted. }
    \label{fig:network:runtime_gaps}
\end{figure}

We provide a basic IP formulation of the problem. 
More sophisticated cutting planes are described in \cite{bienstock1995computational}. 
Results with the basic IP formulation found using \gurobi 11.0.0 on an AMD EPYC-7542 32-core processor with 64 threads and a time limit of 2 hours are summarized in Figure~\ref{fig:network:runtime_gaps}. 
We were able to solve the problem up to size $n=9$ to proven optimality within the time limit. 
For larger sizes, heuristic, but not necessarily optimal, solutions are provided. The instance with 24 vertices and continuous flow variables is also part of MIPLIB2017 \cite{GleixnerHendelGamrathetal2021}, the latest results of which can be found at \cite{dano3mip}. Note that those results correspond to the unscaled model version with continuous variables, which, therefore, has a slightly different objective value than the scaled one. However, the solutions are compatible.

\subsection{Vehicle Routing Problem} \label{sec:problems:routing}
\renewcommand{\customcaptiontext}{(Routing) }

\subsubsection{Background}

The Vehicle Routing Problem (VRP) is the problem of determining the most efficient routes for a fleet of vehicles to deliver goods or services to a set of customers, typically starting and ending at a central depot. This problem is highly relevant for many real-world applications in logistics, transportation, and supply chain management, see, e.g., \cite{cook2022constrainedlocalsearchlastmile}. Originally proposed in \cite{dantzig1959truck} as `the truck dispatching problem', many different variants of VRP have emerged since then \cite{helsgaun2017,vidal2020}.
Providing a collection of relevant and difficult VRP instances is a complex task because there are numerous variants. 
Although classical heuristics for VRP are practically very successful, many papers on quantum approaches for VRP problems have been published \cite{xie2024feasibility, mohanty2023analysis, leonidas2023qubit, Fitzek2021ApplyingQA}.
We include the VRP problem class in this collection to facilitate a comparison with existing work and to explore the capabilities of quantum optimization algorithms for a practically relevant problem. 

\begin{figure}[htb!]
\centering\includegraphics[width=0.75\linewidth]{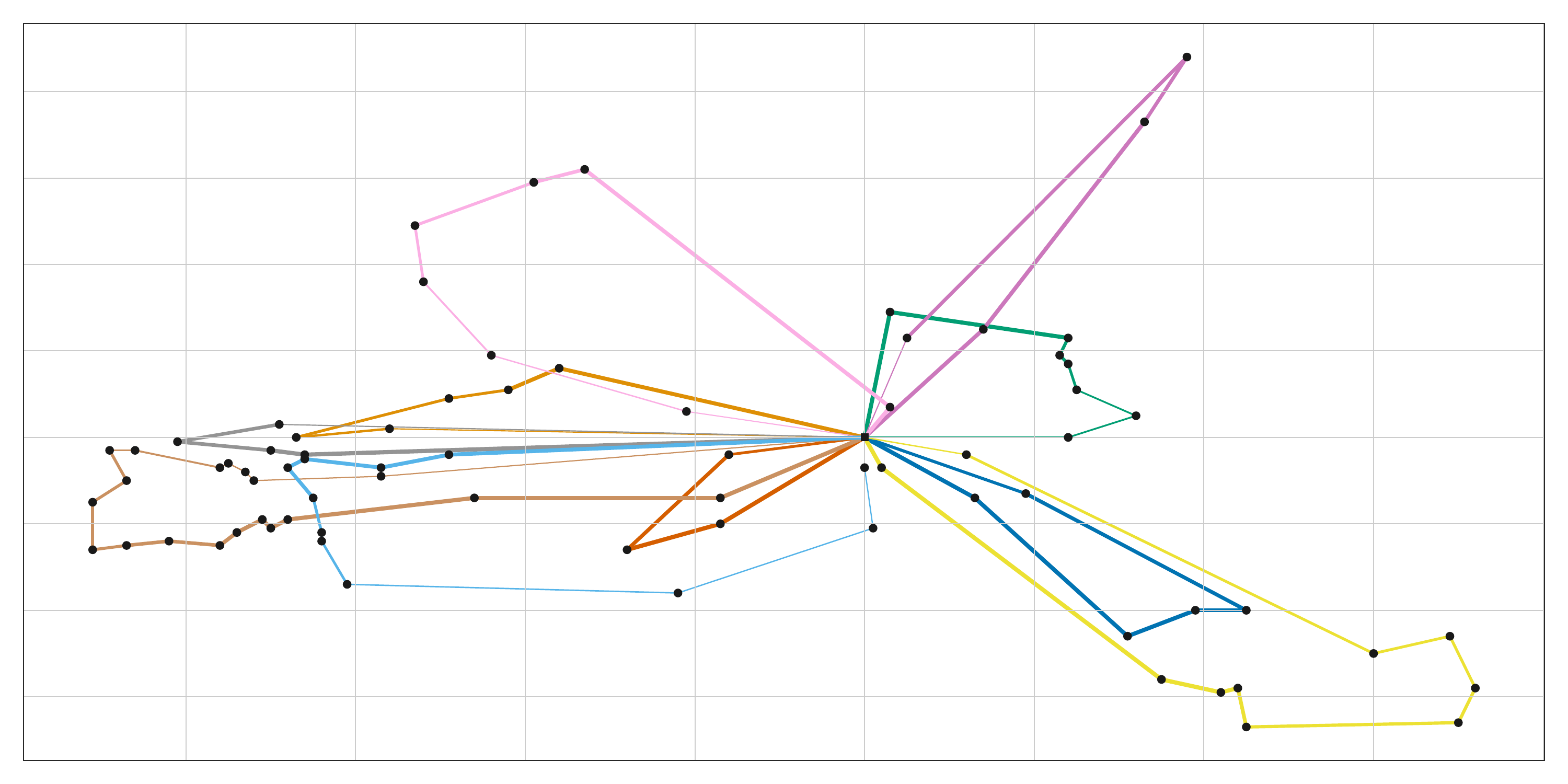}
\customcaption{ Example instance \emph{tai75a} (75 customers, 10 vehicles) from the Rochat and Taillard dataset \cite{rochat1995} and optimal solution as obtained from CVRPLIB \cite{cvrplib}. Each line color represents the route of one vehicle between the customers (\protect\tikz[baseline=-.55ex]{\protect\draw[fill=black] (0,0) circle (1.5pt);}) starting and ending at the central depot (\protect\tikz[baseline=-0.1ex]{\protect\draw[fill=black] (0,0) rectangle (4pt,4pt);}). The thickness of a line indicates the remaining capacity of the corresponding vehicle along the route. This example demonstrates the combinatorial complexity of finding the best route among all possible routes while fulfilling capacity constraints.}
\label{fig:cvrp-example}
\end{figure}

We limit ourselves to a specific variant of VRP, the Capacitated Vehicle Routing Problem (CVRP), where the objective is to minimize the aggregated costs across all routes while satisfying the vehicle capacity, effectively limiting the number of goods that each vehicle can carry. This variant, in fact, corresponds to the original problem formulation from \cite{dantzig1959truck}, where it was presented as a generalization of the TSP \cite{mtz1960}. Compared to TSP, VRP involves an additional level of decision-making, as customers not only need to be served but also assigned to vehicles. An example CVRP instance and its optimal solution are visualized in Figure~\ref{fig:cvrp-example}.

\begin{newtextbox}[Why is it interesting?]
Determining the optimal solution to a Vehicle Routing Problem (VRP) is \NP-hard \cite{lenstra1981} and of high practical relevance as it has many direct industrial applications and, hence, the potential for a significant business value impact. 
Similar to the network problem from the previous section, it is trivial to find feasible (but not necessarily optimal) solutions. 
Due to the complexity of the problem, mostly small instances can be solved exactly, but heuristics and metaheuristics are often well-suited to tackle real-world instances.
This problem class has already received significant attention from the quantum community.
\end{newtextbox}

Due to the practical and academic relevance of the CVRP, numerous optimization approaches have been developed, and research continues \cite{sidorov2021}. Exact methods include branch-and-bound and branch-and-cut strategies. A well-known heuristic is the \emph{savings algorithm} \cite{clarke1964}, which has been modified and improved in various ways. CVRPs can be further divided into symmetric CVRPs and asymmetric CVRPs, depending on whether the direction of a tour affects its cost. Some solution approaches can only be applied to symmetric CVRPs. A review of established exact and heuristic methods can be found in \cite{toth2002a,toth2002b}. Metaheuristics include genetic algorithms, simulated annealing, and Tabu search. A recent comparison of different heuristics and metaheuristics can be found in \cite{muriyatmoko2024}. Alternative research directions are approaches based on machine learning \cite{eljaouhari2024}. More recently, quantum optimization has been explored as a potential strategy. Since naive QUBO formulations for CVRP require many variables \cite{harikrishnakumar2020}, hybrid strategies have been used, for example, by separating the problem into a clustering and routing phase \cite{feld2019hybrid, holliday2024}. Many hybrid approaches rely on solving the vehicle-customer assignment classically and then solving TSPs using quantum optimization. Successful applications will need to overcome this simplification, as solving TSPs exactly, e.g., via Concorde \cite{applegate2003implementing}, is unlikely to be outperformed by quantum methods anytime soon.

\subsubsection{Problem statement}
Consider a CVRP with $n \geq 1$ customers, where each customer $i$ has a demand $d_i \in \mathbb{N}$ with $i \in \{1,\dots,n\}$. Furthermore, we assume that there is a homogeneous fleet of $K \geq 1$ vehicles, where each vehicle has a total capacity of $Q \in \mathbb{N}$. Each vehicle starts and ends its tour at the depot. The cost of moving a vehicle from customer $i$ to $j$ is given by $c_{ij} \in \mathbb{N}$ with $i,j \in \{0,\dots,n\}$ and $i \neq j$, where ``customer $0$'' is used to represent the depot. With this definition, we presume all-to-all connectivity between the customers (and the depot). While the cost matrix is generally arbitrary, we focus here on a special case relevant to the benchmarking instances we provide, where the costs are based on graph vertex distances in a metric space that satisfies non-negativity (as given by the cost domain) and symmetry (i.e., $c_{ij} = c_{ji} \,\forall\, i,j \in \{0,\dots,n\}$, the condition for the aforementioned symmetric CVRP), which means they also obey the triangle inequality (i.e., $c_{ij} \leq c_{ik}+c_{kj} \,\forall\, i,j,k \in \{0,\dots,n\}$). 
Each customer must be served exactly once by a vehicle, while the total customer demand on a given route must not exceed the capacity limit of the vehicle. The optimization goal is to determine a feasible solution that minimizes the aggregated costs across all routes. This does not necessarily require the use of all $K$ vehicles, i.e., some may remain in the depot and, hence, do not contribute to the costs. 

A CVRP can be modeled in many different ways, one of which is the two-index vehicle flow formulation with Miller-Tucker-Zemlin (MTZ) constraints \cite{sidorov2021}. Following \cite{munari2017}, this formulation reads:

\begin{align}
\min_{x,y} & \sum_{\substack{i,j=0 \\ i \neq j}}^{n+1} c_{ij} x_{ij} &  \label{eqn:cvrp} \\
\text{subject to} \sum_{\substack{j=1 \\ j \neq i}}^{n+1}  x_{ij} &= 1&& \text{for all }i \in \{ 1 \ldots, n\} \label{eqn:cvrp:c1} \:,\\
\sum_{\substack{i=0 \\ i \neq h}}^{n} x_{ih} &= \sum_{\substack{j=1 \\ j \neq h}}^{n+1} x_{hj} &&\text{for all } h \in \{1, \ldots, n \}\label{eqn:cvrp:c2}\:,\\
\sum_{j=1}^{n} x_{0j} &\leq K \label{eqn:cvrp:c3} \:,\\
y_{j} &\geq y_{i} + d_j x_{ij} - Q ( 1 - x_{ij} ) && \text{for all }i\neq j \in \{ 0, \ldots, n+1\} \label{eqn:cvrp:c4} \:,\\
d_{i} &\leq y_{i} \leq Q, && \text{for all }i \in \{0, \ldots, n+1\} \label{eqn:cvrp:c5} \:,\\
x_{ij} &\in \{0,1\} && \text{for all }i,j \in\{0, \ldots, n+1\} \label{eqn:cvrp:c6} \:,\\
y_{i} &\in \mathbb{N}_0 &&\text{for all } i \in\{ 0, \ldots, n+1\}\:. \label{eqn:cvrp:c7}
\end{align}

In this IP, the optimization is done with respect to two types of decision variables. First, the binary variables $x \in \{0,1\}^{(n+2)\times(n+2)}$ and second, the integer variables $y \in \mathbb{N}_0^{n+2}$. The binary variable $x_{ij}$ takes the value 1 if and only if there is a direct route between customers $i$ and $j$ for $i,j \in \{0,\dots,n+1\}$, where ``customer $0$'' is the depot, as introduced above. To simplify the notation, we also symbolically introduce ``customer $n+1$'' to represent the same depot, which allows us to distinguish between starting at the depot ($i=0$) and ending at the depot ($i=n+1$), where $c_{j,n+1}:=c_{0j}$ for all $j \in \{1,\dots,n\}$. Furthermore, the integer variable $y_i$ represents the cumulated demand on the route that visits customer $i$ for $i \in \{0,\dots,n+1\}$, where $d_0:=d_{n+1}:=0$.

The objective, Eq.~\eqref{eqn:cvrp}, represents the aggregated costs of all routes that need to be minimized. The first set of constraints, Eq.~\eqref{eqn:cvrp:c1}, ensures that each customer is only visited once. The second set of constraints, Eq.~\eqref{eqn:cvrp:c2}, ensures a correct flow of vehicles. The third constraint, Ineq.~\eqref{eqn:cvrp:c3}, ensures that only up to $K$ vehicles can leave the depot. The fourth and fifth set of constraints, Ineq.~\eqref{eqn:cvrp:c4}) and \eqref{eqn:cvrp:c5}, follow the MTZ modeling of a TSP \cite{mtz1960} and are responsible for ensuring compliance with the vehicle capacities. Using this formulation, a polynomial number of capacity constraints is sufficient (in contrast to an exponential number of constraints with alternative models). Finally, Eq.~\eqref{eqn:cvrp:c6} and \eqref{eqn:cvrp:c7} are the integrality constraints.

The problem is often also formulated with real-valued costs and real-valued demands, which only requires the change of $y$ from an integer variable to a continuous variable $y \in \mathbb{R}$, effectively leading to a mixed-integer program. For a more convenient benchmarking setup, we only provide the integer-based formulation, since it lets us avoid having to consider machine precision for the instance data, while simultaneously keeping the problem complexity the same. 

\subsubsection{Instances}

A survey on benchmark data sets can be found in \cite{gunawan2021}. 
All instances in the repository are given for $k=4$ vehicles and $n=20$ customers and were created to be 100\% tight, i.e., the sum of the demands equals the vehicle capacity $Q$. The instances follow the TSPLIB/CVRPLIB format \cite{cvrplib}.
We randomly generated and solved several thousand instances, selecting those that took the longest to solve, which is still only seconds on modern digital systems. 
Out of the set of problem sizes considered, the chosen instances are among the most difficult.
It has proven unexpectedly hard to find small-to-medium instances that are primal difficult. Classical methods solve these sizes quickly, see~\cite{uchoa2017new,gunawan2021}. 
This indicates that quantum approaches would have to be extremely fast to be competitive on these problems in the near term. We plan to add and investigate larger instances in the future.

\subsubsection{Classical Baseline}

It should be noted that numerous open-source codes for VRP are available, e.g., \cite{LKH3,googleor,HGSCRVP}, with some even utilizing GPU acceleration \cite{cuOPT}. 
Additionally, several companies offer specialized, highly optimized solvers for this problem; however, it is difficult to assess their performance. Furthermore, they are tailored to real-world scenarios, which, for example, include multiple vehicle classes and driver break times.

Table~\ref{tab:cvrp:instances} shows the optimal cost and the average time taken by HGS-CVRP \cite{HGSCRVP,Vidal2022} to find this solution, measured over 10 runs with different seeds and a time limit of 10 seconds. Notably, the optimal solution could not be found within this limit, for instance, 21. However, given that HGS-CVRP takes only a few milliseconds to find an optimal solution for typical instances with $n=20$, we expect it to be found within a marginally longer time limit. The optimal solutions were certified using VRPSolver~\cite{ErramiEtal2023,PessoaEtal2020,VRPSsolver,VRPSolverEasy}.

\begin{table}[htb!]
    \begin{minipage}[t]{0.3\textwidth}
\begin{tabular}{rrr}
\toprule
No  & Cost  & Time (s)\\
\midrule
01   &          646       &       1.74\\
02     &          650     &       0.33\\
03     &          508     &       1.30\\
04     &          776     &       1.64\\
05     &          702     &       1.01\\
06     &          690     &       0.85\\
07     &          730     &       0.82\\
08     &          718     &       0.65\\
09     &          707     &       0.88\\
10     &          737     &       0.61\\
11    &          914      &      4.48\\
12    &          709      &      1.18\\
13    &          628      &      0.47\\
14    &          696      &      1.44\\
15    &          780      &      0.94\\
16    &          830      &      3.55\\
17    &          605      &      1.03\\
18    &          997      &      1.07\\
19    &          976      &      2.10\\
20 &          648         &   8.71\\
\bottomrule      
\end{tabular}      
    \end{minipage}\hfill
    \begin{minipage}[t]{0.3\textwidth}
\begin{tabular}{rrr}
\toprule
No  & Cost  & Time (s)\\
\midrule
 21 &          842        &   $>$10.00\\
 22 &          760        &    0.98\\
 23 &          851        &    2.84\\
 24 &          725        &    0.82\\
 25 &          462        &    0.49\\
 26 &          688        &    1.01\\
 27 &          540        &    1.40\\
 28 &          669        &    2.27\\
 29 &          658        &    0.94\\
 30 &          685        &    0.78\\
 31 &          445        &    7.16\\
 32 &          882        &    2.38\\
 33 &          568        &    0.60\\
 34 &          891        &    8.51\\
 35 &          707        &    1.18\\
 36 &          610        &    2.03\\
 37 &          650        &    0.81\\
 38 &          991        &    0.98\\
 39 &          717        &    5.23\\
 40 &          700        &    2.16\\

\bottomrule      
\end{tabular}      
    \end{minipage}\hfill
        \begin{minipage}[t]{0.3\textwidth}
\begin{tabular}{rrr}
\toprule
No  & Cost  & Time (s)\\
\midrule

 41 &          736        &    1.33\\
 42 &          998        &    1.30\\
 43 &          465        &    1.95\\
 44 &          973        &    1.11\\
 45 &         1071        &    5.10\\
 46 &          812        &    0.79\\
 47 &          583        &    0.54\\
 48 &          544        &    1.81\\
 49 &         1045        &    0.56\\
 50 &          627        &    0.89\\
 51 &          491        &    0.91\\
 52 &          622        &    2.35\\
 53 &          967        &    7.12\\
 54 &          791        &    2.12\\
 55 &          617        &    3.73\\
 \bottomrule   
 \\
 \\
 \\
 \\
 \\
   
\end{tabular}      
    \end{minipage}
\caption{CVRP: Optimal cost values and running times.}
\label{tab:cvrp:instances}
\end{table}

\subsection{Topology Design} \label{sec:problems:topology}
\renewcommand{\customcaptiontext}{(Topology) }

\subsubsection{Background}

\begin{newtextbox}[Why is it interesting?]
Several GraphGolf competitions \cite{graphgolf} over the years have shown that specifically the Order Degree Problem (ODP) is difficult to solve in practice--although the resulting problem formulation can be relatively sparse, and a problem instance is specified by providing just three integers. 
Furthermore, the variables are typically binary, which facilitates the encoding of the problem in a quantum system. The problem finds application in computer network design \cite{shin2011small}.
\end{newtextbox}

Given a graph $G = (V,E)$, we define three possibly competing objectives.
First, one may want to focus on \textit{order} such that the number of vertices $n = |V|$ is maximized. Second, one can target the \textit{degree}, which relates to the minimization of the maximum vertex degree $d = \max_{v \in V}\delta (v)$. Lastly, one may want to minimize the \textit{diameter} $k = \max_{u,v \in V}$ length\_of\_shortest\_path$(u,v)$.
Fixing two parameters within the set $\{$order, degree, diameter$\}$, and optimizing the third, results in the following optimization problems:
\begin{itemize}
  \item Order Degree Problem (ODP): minimize the diameter;
  \item Degree Diameter Problem (DDP): maximize the number of vertices;
  \item Order Diameter Problem (ODiP): minimize the maximum vertex degree.
\end{itemize}
Notably, the name of the problem is given by the two fixed parameters.
The ODiP has received little attention to date. The DDP
has been considered extensively, with the foundations discussed already in the literature from the 1960s \cite{hoffman1960moore,cerf1974lower}, and further explored in the 1980s, e.g., \cite{chung1986diameters,chung1987diameters}. For a comprehensive survey, we refer the reader to \cite{miller2012moore}. In this work, we consider the ODP, for which--to our knowledge--an \NP-hardness proof is not available in the open literature. The ODP arises in the search for network topologies with low latency \cite{shin2011small}, and a variety of methodologies for constructing networks with low latency are discussed in the literature \cite{besta2014slim,singla2012jellyfish}. We use the ODP variant employed in the GraphGolf competitions \cite{graphgolf}: the goal is to solve the ODP, using the average shortest path length (ASPL) as a tie-breaker for graphs with the same diameter, with lower ASPL being preferred. This problem is difficult to solve in practice \cite{graphgolf}.

\subsubsection{Problem statement}
Let $\Delta(s,t)$, $s,t\in V$ be the shortest distance of two vertices in a simple unweighted graph $G=(V,E)$.  
Given a number $n\in\NN$ of vertices, and a maximum vertex degree $d\in\NN$.
We are looking for a simple graph $G=(V,E)$, $V=\{1,\ldots,n\}$, $E\subseteq\{(u,v)\in V\times V|u\neq v\}$ with $\delta(v)\leq d$, for all $v\in V$, that minimizes $\max_{s,t\in V} \Delta(s,t)$.
Multiple formulations for the ODP are possible; for an overview, we refer to \cite{waniek2021graphgolf}. There is considerable freedom in the choice of how the lengths of the all-pairs shortest paths are computed. 

\begin{table}[htb!]
    \centering
    \begin{minipage}{0.24\linewidth}
        \centering
        \begin{tabular}{r|r|r}
            \toprule
            $n$ & $d$ & $k$\\
            \hline
            15 & 4 & 2*\\
            15 & 3 & 3*\\
            20 & 3 & 3*\\
            20 & 4 & 3*\\
            20 & 5 & 2*\\
            25 & 3 & 4*\\
            25 & 4 & 3*\\
            25 & 5 & 3*\\
            \bottomrule
        \end{tabular}
    \end{minipage}
        \begin{minipage}{0.24\linewidth}
        \centering
        \begin{tabular}{r|r|r}
            \toprule
            $n$ & $d$ & $k$\\
            \hline
            25 & 6 & 2*\\
            30 & 4 & 3*\\
            30 & 5 & 3*\\
            30 & 6 & 3\\
            35 & 5 & 4\\
            35 & 6 & 3\\
            40 & 6 & 3*\\
            50 & 4 & 5\\
            \bottomrule
        \end{tabular}
    \end{minipage}
    \hfill
    \begin{minipage}{0.48\linewidth}
        \centering
        \begin{tabular}{r|r|r|l}
            \toprule
            $n$ & $d$ & $k$ & \textbf{Submitter} \\
            \hline
            512 & 4 & 6 & 
            {EvbCFfp1XB} \\
            512 & 6 & 5 & {MasakiChujo} \\
            1024 & 4 & 7 & {EvbCFfp1XB} \\
            1726 & 30 & 3 & {(random)} \\
            4855 & 15 & 4 & {Teruaki Kitasuka} \\
            9344 & 6 & 7 &{EvbCFfp1XB} \\
            65536 & 6 & 9 & {(random)} \\
            100000 & 8 & 7 & {Hajime Terao} \\
            1000000 & 16 & 6 & {Teruaki Kitasuka} \\
            1000000 & 32 & 5 & {(random)} \\
            \bottomrule
        \end{tabular}
    \end{minipage}
    \customcaption{Overview of the provided instances and their optimal/best-known solutions. Proven optimality is indicated by $*$. The solutions shown in the left and center tables were computed by us using Gurobi 11.0.0. and result submissions shown in the right table can be found at \cite{graphgolf}.}
    \label{tab:topology_instances}
\end{table}

\subsubsection{Instances}

The instances range from $15$ vertices and degree $3$ to $10^6$ vertices and degree $32$.
A detailed presentation of the provided instances can be found in Table~\ref{tab:topology_instances}.
We provided a program to check solutions, i.e., given a triple $(n,d,k)$ and a graph, it verifies that the graph has $n$ nodes, a maximum degree of $d$, and a diameter of at most $k$.

\begin{figure}[htb!]
\centering
\includegraphics[width=0.7\textwidth]{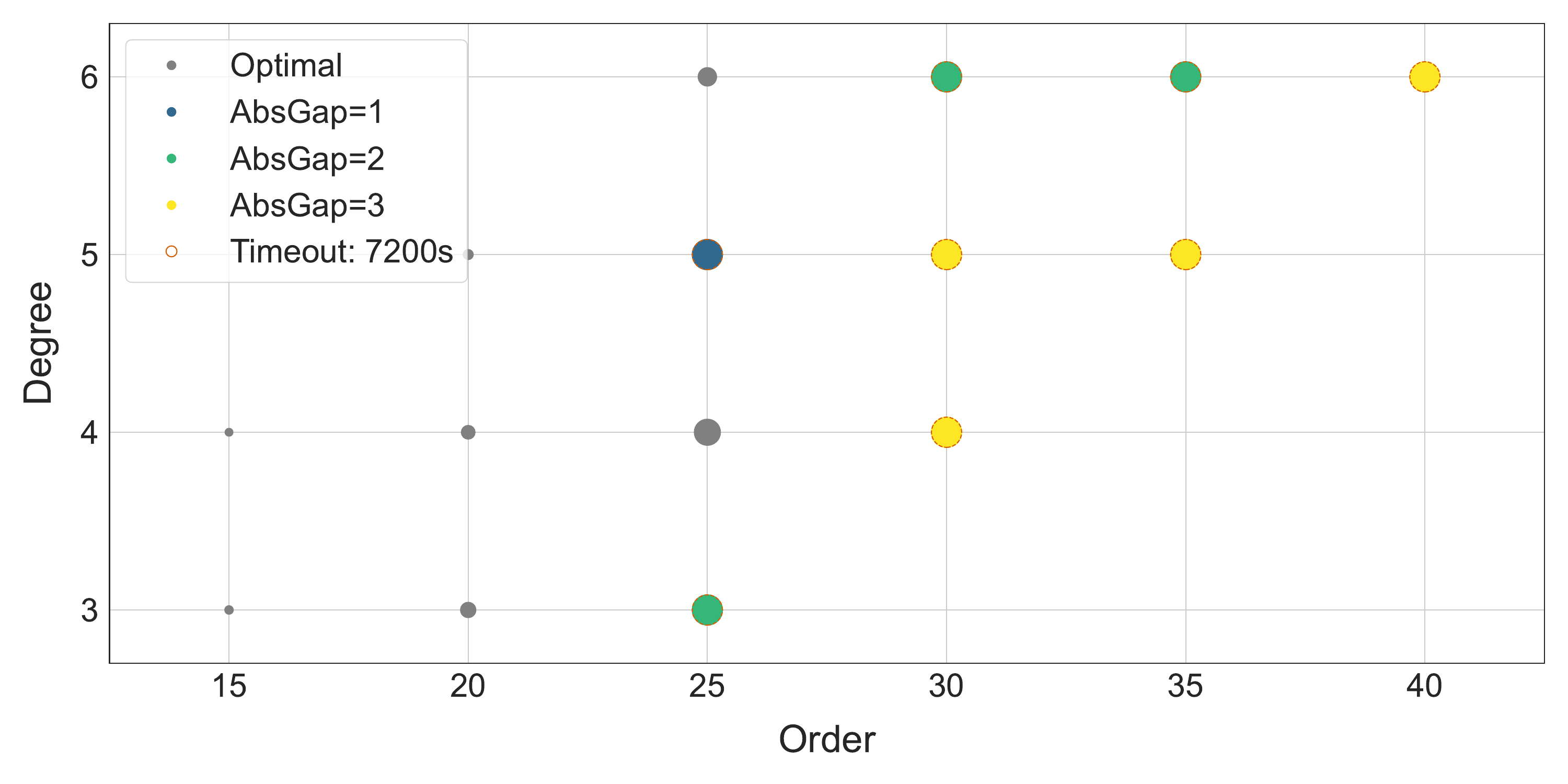}
\includegraphics[width=0.45\textwidth]{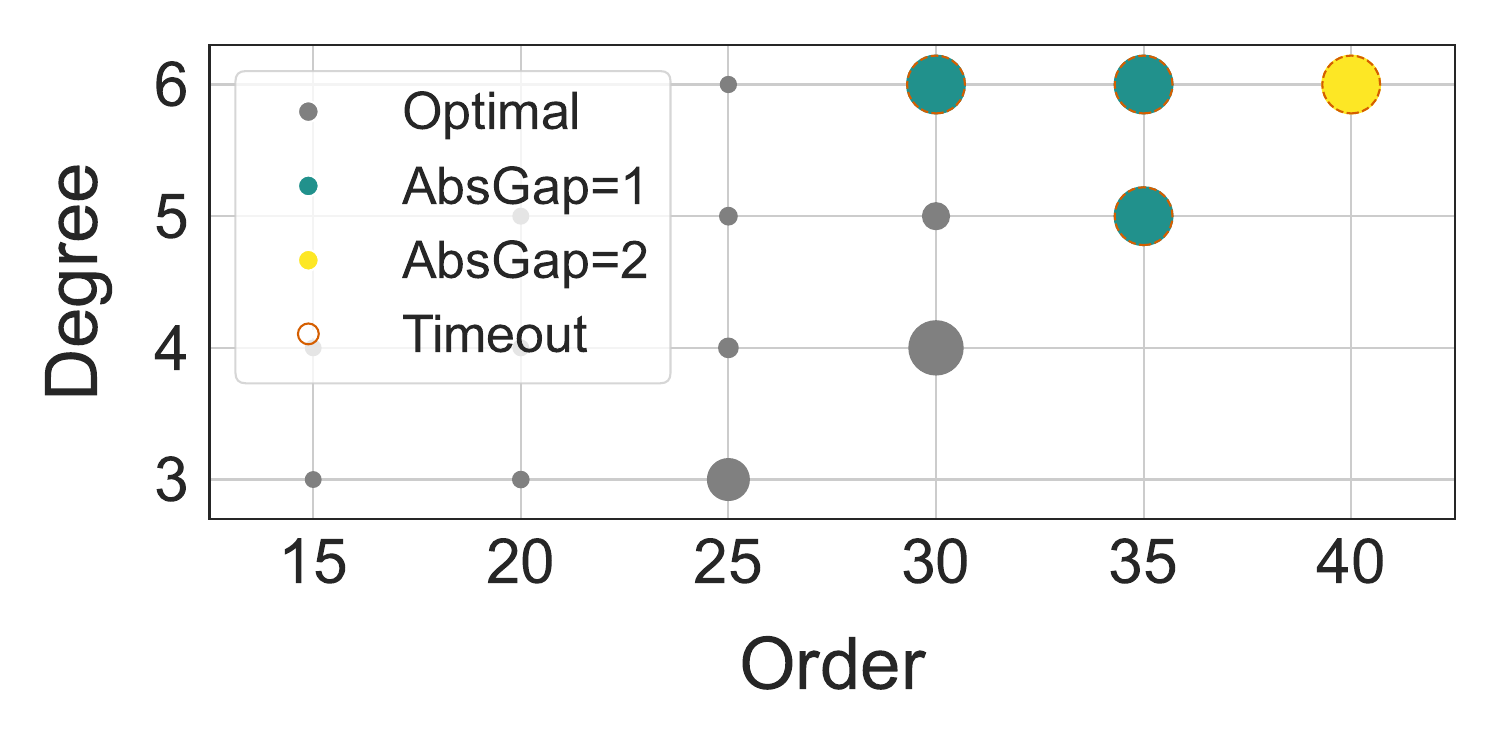}
\includegraphics[width=0.45\textwidth]{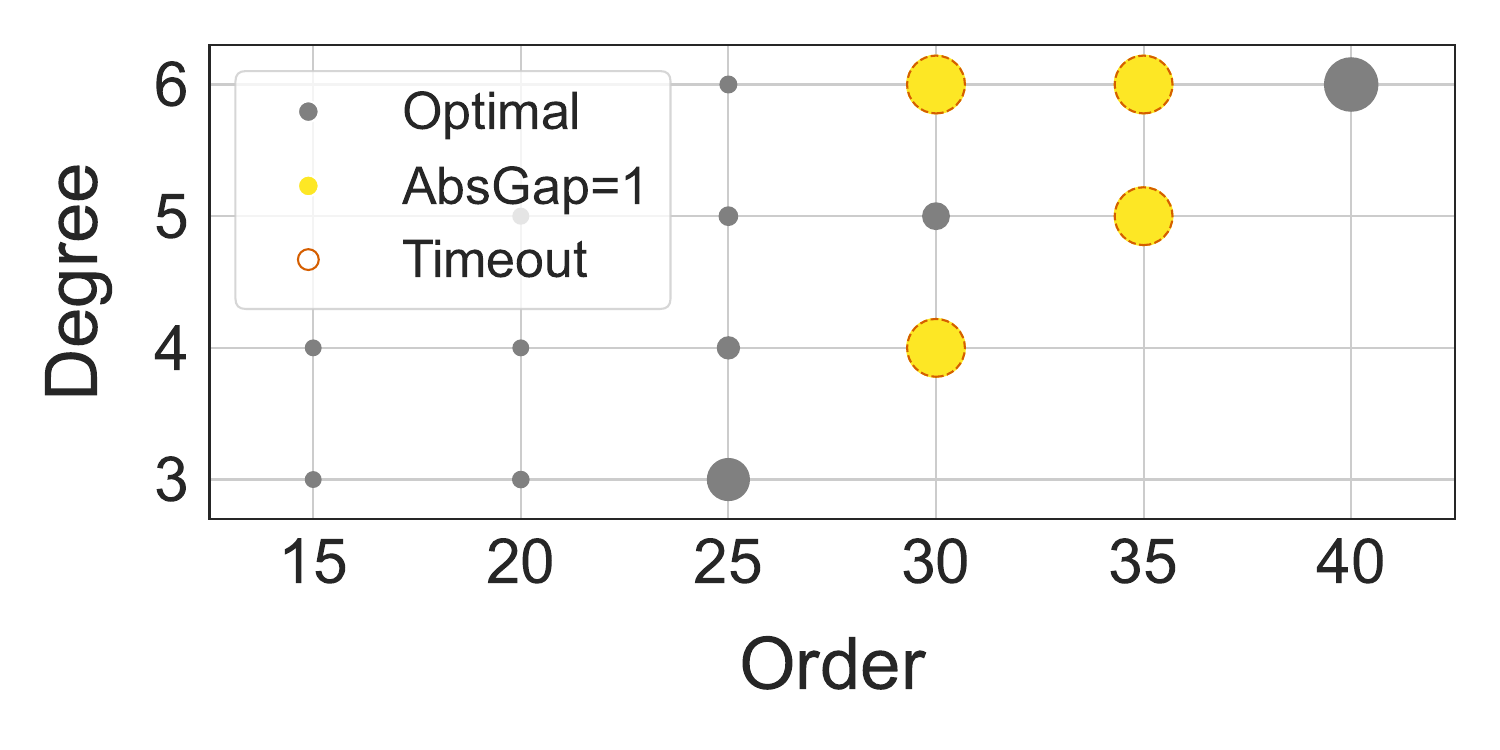}
\customcaption{ Absolute MIP gap corresponding to order and degree of the \textit{flow} (top) / \textit{quadratic} (bottom left) and \textit{linearized} (bottom right) Seidel model. The size of the points indicates the running time--the bigger, the longer. Furthermore, the timeout is set to 7200s.}
\label{fig:topology:plot}
\end{figure}

\subsubsection{Classical Baseline}

Our baseline was computed using \gurobi 11.0.0 on an AMD EPYC-7542 32-core processor using 64 threads with a time limit of 7200s.
In the repository, we provide three different MIP formulations. 
The first model uses a standard s-t flow model to compute shortest path values.
The second model uses Seidel's quadratic all-pairs shortest path formulation.
The third model is a linearization of Seidel's quadratic model. This problem is hard to model as a compact integer program.
Gaps and runtimes can be found in Figure \ref{fig:topology:plot}.
The results indicate that we can prove optimality only on rather small instances if we use an MIP solver. 
With the provided models, even solving $n=30, d=6$ takes too long. 
However, as can be seen on \cite{graphgolf}, using heuristic methods, it is possible to compute solutions to larger instances.

\renewcommand{\customcaptiontext}{}

\section{Submission Guidelines and Illustrative Results}
\label{sec:illustrative_quantum_baselines}

The following section presents guidelines for submitting benchmark results to \RN{} and provides exemplary baseline results for three selected problem classes--\textit{Low Autocorrelation Binary Sequences}, \textit{Minimum Birkhoff Decomposition} and \textit{Independent Set} problem--each solved with another quantum optimization algorithm.
These results can also be found in the \RN{} repository.

We provide a submission template in Table~\ref{tbl:metrics} which should be submitted together with the discovered solution. The metrics were chosen to enable a fair and systematic comparison of solutions and algorithms.
More specifically, we ask for the identification of the concrete problem instance and the submitter(s).
Given that not every detail can be given in a compact form, submissions ideally refer to a paper or a code repository with further information, such as hyperparameters, additional hardware specifications, software versions, etc.
Furthermore, the best objective value found (for optimization problems) as well as the corresponding solution (in addition to the template) should be given.
If supported by an algorithm, an a posteriori bound on the optimal objective value can be provided.
Next, the chosen modeling approach should be described (e.g.,~QUBO, ILP, etc.) as well as the resulting number and type of decision variables and (non-zero) coefficients needed to represent the considered problem instance.
Furthermore, the submission should provide a brief summary of the complete optimization workflow to facilitate reproducibility. This includes pre-processing, pre-solver, main optimization algorithm, and post-processing, as well as an indication if the algorithm is deterministic or stochastic.
Stochastic algorithms are generally recommended to be repeated multiple times. 
In this case, the number of successful runs that result in feasible solutions or return solutions close to the best found solutions (cf.~Table~\ref{tbl:metrics} for more details) should be reported.
Lastly, the overall runtime and the runtime spent on the various hardware platforms should be provided.
In case of multiple repetitions, the average runtime over all repetitions may be reported. 
Additional information, such as the distribution of the runtimes or potential correlations with solution quality, is encouraged to be described in a corresponding publication or reference. 
As discussed in Section~\ref{sec:methods}, the runtimes should not include potential queuing times for hardware access.
For most problems, we provide solution checkers in \cite{gitlab_repo}. These should be used to verify proposed solutions before a benchmark result is submitted to \RN.

\begin{table}[!ht]
\small{
\begin{center}
\begin{tabular*}{\textwidth}{l|l}

\toprule
\textbf{Problem Identifier} & Identifier of the considered problem instance. \\
\textbf{Submitter}          & Name(s) of the submitter(s) and affiliation(s). \\
\textbf{Date}               & Date of submission. \\

\midrule
\textbf{Reference} & Reference to a paper/repository with more details. \\

\midrule
\textbf{Best Objective Value}  & The best objective value found by the algorithm across all repetitions. \\
\makecell[l]
{\textbf{Optimality Bound}\\~} & \makecell[l]{Lower bound (minimization) or upper bound (maximization) for the\\
                                              optimal objective value, if supported; otherwise, set to N/A.} \\ 

\midrule
\textbf{Modeling Approach}        & Describe how the considered problem instance is modeled.\\
\textbf{\#Decision Variables}     & Total number of decision variables.\\ 
\textbf{\#Binary Variables}       & Number of binary decision variables.\\ 
\textbf{\#Integer Variables}      & Number of integer decision variables.\\ 
\textbf{\#Continuous Variables}   & Number of continuous decision variables.\\ 
\textbf{\#Non-Zero Coefficients}  & Number of non-zero coefficients in objective function and constraints.\\ 
\textbf{Coefficients Type}        & Type of coefficients, such as integer, binary, continuous.\\ 
\textbf{Coefficients Range}       & Range of non-zero coefficients, i.e., min/max values.\\ 

\midrule
\makecell[l]
{\textbf{Workflow}\\~}             & \makecell[l]{Description of the optimization workflow: pre-processing, pre-solvers,\\
                                                   machine learning training, optimization algorithms, post-processing, etc.} \\
\textbf{Algorithm Type}            & Indicate whether the algorithm is deterministic or stochastic. \\ 
\textbf{\#Runs}                    & The number of times the experiment has been repeated. \\
\textbf{\#Feasible Runs}           & The number of times a run found a feasible solution. \\
\makecell[l]
{\textbf{\#Successful Runs}\\~\\~} & \makecell[l]{Number of runs that found a feasible solution with objective value\\
                                                  $\leq (1+\epsilon)*f_{\text{min}}$ (minimization) or 
                                                  $\geq(1-\epsilon)*f_{\text{max}}$ (maximization), \\
                                                  where $f_{\text{min/max}}$ is the best solution found by the algorithm.} \\
\textbf{Success Threshold}         & The threshold $\epsilon$ to define a successful run.
\\\midrule
\textbf{Hardware Specifications}      & Specifications of hardware used to run the workflow.
\\\midrule
\textbf{Total Runtime}  (Wall-clock) & Total runtime to run the complete workflow.\\ 
\textbf{CPU Runtime}      & CPU runtime to run the workflow. \\
\textbf{GPU Runtime}      & GPU runtime to run the workflow. \\ 
\textbf{QPU Runtime}      & QPU runtime to run the workflow.  \\
\textbf{Other HW Runtime} & Runtime on other hardware to run the workflow.
\\
\bottomrule
\end{tabular*}
\end{center}
\customcaption{\label{tbl:metrics} The table summarizes the metrics that should be reported when submitting benchmark results to \RN. All \textit{runtimes should be reported as average} if multiple algorithm runs were executed.}
} 
\end{table}

\subsection{Low Autocorrelation Binary Sequences}
\renewcommand{\customcaptiontext}{(LABS) }

In the following, we focus on tackling the LABS problem using the bias-field digitized counterdiabatic quantum optimization (BF-DCQO) algorithm \cite{iskay}. The respective results are compared with QAOA and various classical optimization algorithms.

\begin{figure}[htb!]
    \centering
    \includegraphics[width=0.9\linewidth]{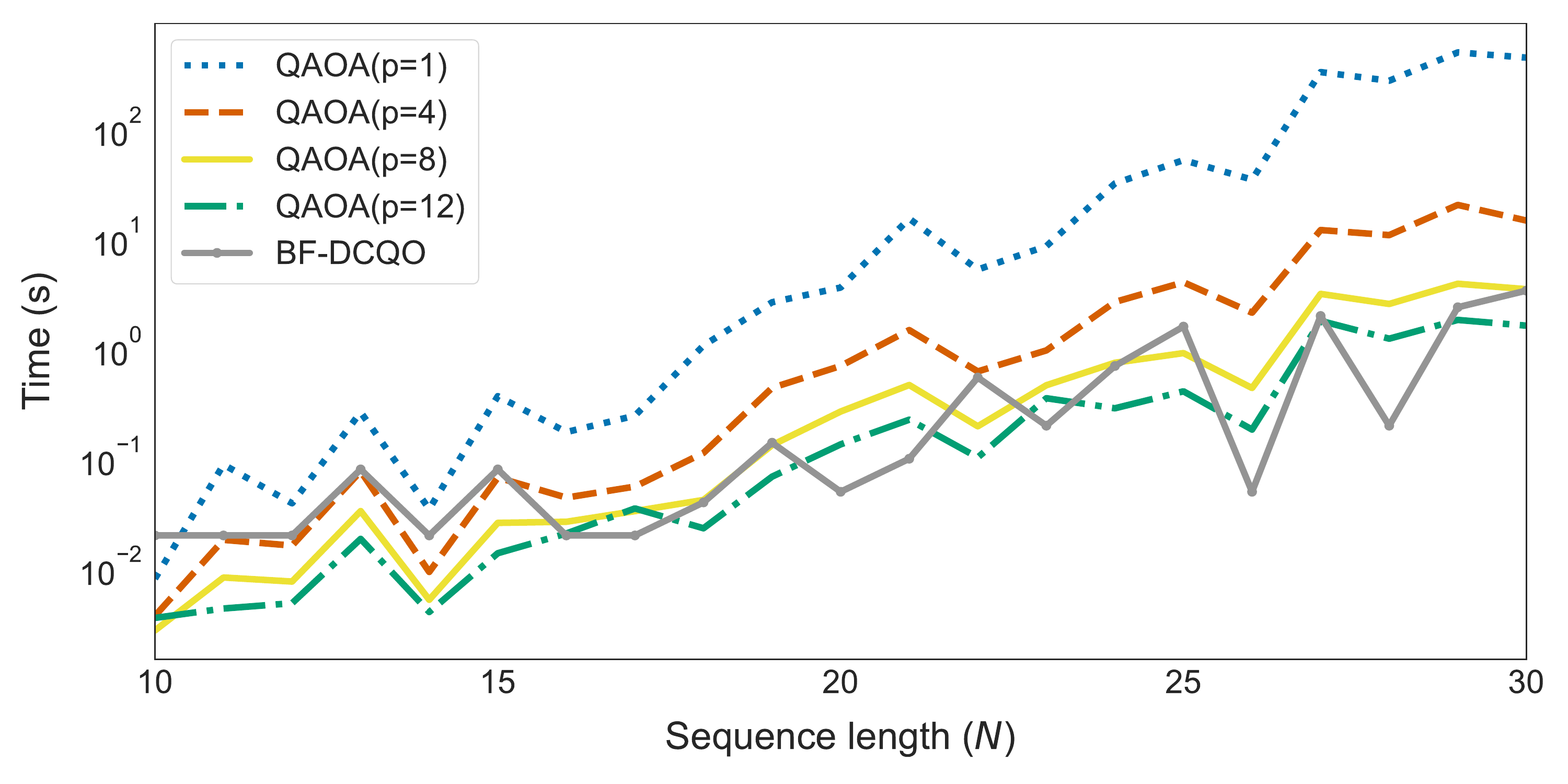}
    \customcaption{Comparison of the time-to-solution for LABS instance ranging from size 10 to 30 using BF-DCQO and QAOA across different numbers of layers: \(p=1\), \(p=4\), \(p=8\), and \(p=12\). The data for QAOA is sourced from~\cite{Shaydulin2024}. Note that the BF-DCQO times reflect idealized QPU runtime estimates, excluding transpilation.}
    \label{fig:qaoa_vs_bfdcqo}
\end{figure}

The quantum adiabatic algorithm (QAA) evolves an easily prepared ground state of an initial Hamiltonian $H_i$ to the unknown ground state of a target Hamiltonian $H_f$, which should correspond to the optimal solution of an optimization problem of interest.
This evolution is described by a time-dependent Hamiltonian of the form
\begin{align}\label{eqn:qaa}
H_\text{ad}(t)=[1-\lambda(t)]H_i+\lambda(t)H_f\: ,
\end{align}
where $\lambda(t)\in[0, 1]$ guides the adiabatic path.
The corresponding time evolution can be discretized and run on a digital quantum computer.
In QAA, the system remains in the instantaneous ground state in the limit $\partial_t\lambda(t)\to 0$.
In the digitized counterdiabatic quantum optimization (DCQO) protocol counterdiabatic terms are added to the time-dependent Hamiltonian of Eq.~\eqref{eqn:qaa} ~\cite{hegade2021shortcuts, Chandarana2022, hegade2022digitized}, which becomes 
\begin{align}
    H_\text{cd}=H_\text{ad}(t)+\partial_t\lambda(t)A_\lambda \:.
\end{align}
Here, $A_\lambda$ is the adiabatic gauge potential~\cite{Kolodrubetz2017}, which may be obtained or approximated in different ways~\cite{Claeys2019, Hatomura2021}.
The additional counterdiabatic term aims to suppress non-adiabatic transitions that move the state away from the instantaneous ground state. 
Hence, approaches based on counterdiabatic terms can improve solution quality~\cite{Chandarana2022, cadavid2024biasfield}. 

An extension of DCQO is bias-field DCQO (BF-DCQO), which is an iterative algorithm~\cite{cadavid2024biasfield, romero2024bias, iskay}. 
In BF-DCQO, the information from the final evolved state is fed back as a bias for a new initial state, such that the initial Hamiltonian for another DCQO iteration is adapted accordingly. 
In addition to that, we consider a classical post-processing scheme where we perform a local search \cite{dupont_Quantum-enhanced_greedy, dupont2024quantumoptimizationmaximumcut, wurtz2024solvingnonnativecombinatorialoptimization} on the $n_{keep}$ lowest energy bitstrings obtained from BF-DCQO at each iteration. 
This local search can be seen as a zero-temperature simulated annealing with $n_{s}$ sweeps.     

\begin{figure}[htb!]
    \centering
    \includegraphics[width=0.7\linewidth]{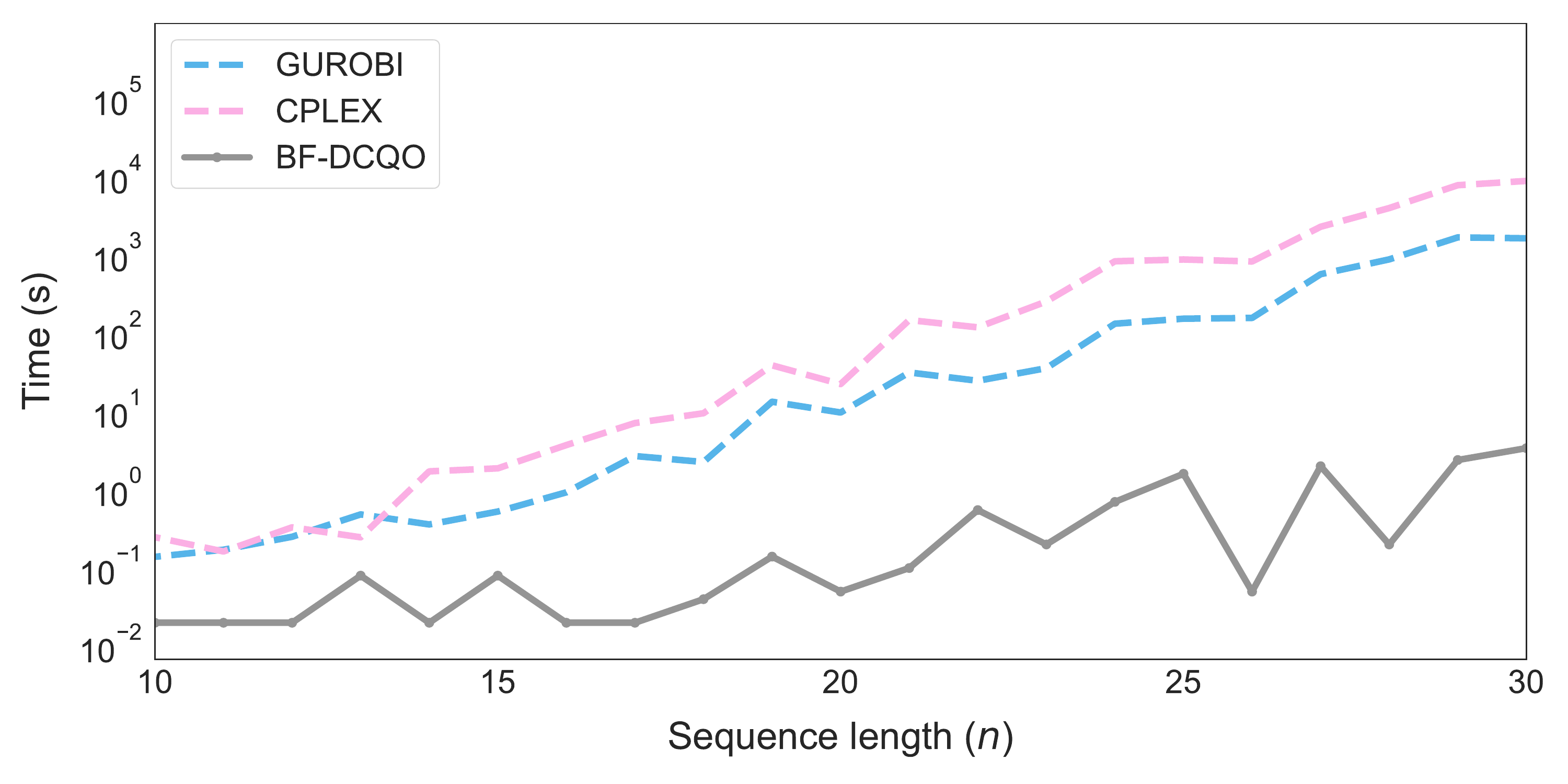}
    \customcaption{Comparison of the time-to-solution for BF-DCQO, \gurobi, and CPLEX. The data for the classical optimizers was sourced from~\cite{Shaydulin2024}. Note that the BF-DCQO times reflect idealized QPU runtime estimates, excluding transpilation.}
    \label{fig:csolvers_vs_bfdcqo}
\end{figure}

We conduct classical statevector simulations of quantum circuits from BF-DCQO for sequence lengths up to \(n=30\) using Qiskit 1.1.1, NumPy 2.0.2, and SciPy 1.14.1. We consider a maximum of $10$ iterations, and $n_{keep}$ was set to 50$\%$ of the measured states at each iteration. Furthermore, $n_{s}=5$ sweeps were performed on each bitstring.
In the context of these simulation results, we predict QPU execution times by making optimistic assumptions, i.e., that one shot on quantum hardware takes $2\cdot 10^{-4}$ seconds, and the classical post-processing requires $\sim 10^{-7}$ seconds per sweep (for 30 qubits), based on the measured resources on an Apple M2 Pro chip.
In Figure~\ref{fig:qaoa_vs_bfdcqo}, we present the scaling of the time to measure the optimal solution for the classically simulated BF-DCQO combined with greedy classical post-processing for different sequence lengths. 
Additionally, we show the time-to-solution of QAOA with different numbers of layers $p$ based on~\cite{Shaydulin2024}. 
The time-to-solution is defined as \(\tau \log(1-0.99)/\log(1-p_{gs})\) with $\tau$ the execution time per shot, $p_{gs}$ is the success probability of finding the optimal solution, and the ratio \(\log(1-0.99)/\log(1-p_{gs})\) accounts for the number of shots required to measure the optimal solution at least once with $99\%$ probability. We emphasize that the reported runtimes for BF-DCQO were calculated based on early termination once the ground state was reached, rather than running all 10 iterations. In most cases, the ground state was obtained within 3 iterations; however, for cases beyond 28 qubits, up to 7 iterations were required. If we choose not to terminate after reaching the ground state, the success probability continues to be amplified by the bias field in successive iterations, contributing to a further reduction in the overall time-to-solution. Additionally, termination criteria can be defined by setting a threshold on the energy difference between successive iterations.
Note that the BF-DCQO circuit requires as many entangling gates as QAOA ($p=2$). In the studied range of system sizes, BF-DCQO scales better than QAOA ($p=1$, $p=4$) and is comparable to QAOA ($p=8$, $p=12$). Additionally, the scaling is also better with respect to exact classical solvers like \gurobi and CPLEX~\cite{Shaydulin2024}, as shown in Figure~\ref{fig:csolvers_vs_bfdcqo}. 

\begin{table}[htb!]
    \centering
    \begin{tabular}{c c c c c c c}\toprule
        Run &  $p_{gs}$ & Total Runtime (s) & Pre-Proc. (s)  &  QPU Runtime (s) & Post-Proc. (s)\\  
        \midrule
        1 & 0.10 & 32.23 & 0.213 & 32  & 0.01 \\\midrule
        2 & 0.23 & 32.25 & 0.241 & 32  & 0.01 \\\midrule
        3 & 0.37 & 32.23 & 0.216 & 32  & 0.01 \\\midrule
        4 & 0.45 & 32.24 & 0.226 & 32  & 0.01 \\\midrule
        5 & 0.10 & 32.24 & 0.225 & 32  & 0.01 \\\midrule
        6 & 0.25 & 96.68 & 0.665 & 96 & 0.01 \\\midrule
        7 & 0.49 & 64.44 & 0.427 & 64 & 0.01 \\\midrule
        8 & 0.13 & 64.42 & 0.408 & 64  & 0.01 \\\midrule
        9 & 0.08 & 128.87& 0.854 & 128 & 0.01 \\\midrule
        10& 0.12 & 96.64 & 0.630 & 96 & 0.01 \\\bottomrule
    \end{tabular}
    \customcaption{
    Summary of results for ten experimental trials of BF-DCQO on the size $n=20$ LABS problem executed on \textit{ibm$\_$marrakesh}. The overall runtime consists of pre-processing time, which includes circuit construction and transpilation, QPU runtime, which includes the payload circuit preparation and execution time, and post-processing time, which includes the local search post-processing. Further $p_{gs}$ shows the probability of sampling the best solution--here the optimal one.}
    \label{tab:labs_bench}
\end{table}

To test the performance experimentally, we selected the LABS problem with a sequence length of \(n=20\), which exceeds the largest LABS problem implemented on a quantum computer to date (\(\text{i.e.}\;  n=18\)~\cite{Shaydulin2024}), and present the results in the submission format introduced in Table~\ref{tbl:metrics} in Table~\ref{tbl:labs:metrics}.

\begin{table}[!ht]
\small{
\begin{center}
\begin{tabular*}{\linewidth}{l|l}

\toprule
\textbf{Problem Identifier}         & LABS ($n=20$) \\
\makecell[l]{\textbf{Submitter}\\~} & \makecell[l]{Narendra N.~Hegade (Kipu Quantum),\\ 
                                                   Alejandro Gomez Cadavid (Kipu Quantum, UPV/EHU).}\\
\textbf{Date}                       & January 20, 2025 \\

\midrule
\textbf{Reference} & \cite{GomezHegade25LabsBfDcqoBenchmark} \\

\midrule
\textbf{Best Objective Value} & 26 (optimal)\\ 
\textbf{Optimiality Bound}    & N/A \\ 

\midrule
\textbf{Modeling Approach}        & HUBO \\
\textbf{\#Decision Variables}     & 20 \\ 
\textbf{\#Binary Variables}       & 20 \\ 
\textbf{\#Integer Variables}      & 0 \\ 
\textbf{\#Continuous Variables}   & 0 \\ 
\textbf{\#Non-Zero Coefficients}  & 90+525 (quadratic + quartic terms) \\ 
\textbf{Coefficients Type}        & integer \\ 
\textbf{Coefficients Range}       & \{2, 4\} \\ 

\midrule 
\makecell[l]
{\textbf{Workflow}\\~\\~}   & \makecell[l]{Each iteration of the algorithm calls: \\
                                           1) Sampling with BF-DCQO \\ 
                                           2) Local search sweeps} \\
\textbf{Algorithm Type}     & Stochastic \\ 
\textbf{\#Runs}             & 10 \\
\textbf{\#Feasible Runs}    & 10 (unconstrained problem) \\
\textbf{\#Successful Runs}  & 10 \\
\textbf{Success Threshold}      & 0 (requiring optimal solutions) \\

\midrule
\textbf{Hardware Specifications}      & CPU: Apple M2 Pro, QPU: \emph{ibm\_marrakesh}.
\\\midrule
\textbf{Total Runtime}    & 61.22s \\ 
\textbf{CPU Runtime}      & 0.42s \\
\textbf{GPU Runtime}      & N/A \\ 
\textbf{QPU Runtime}      & 60.80s \\
\textbf{Other HW Runtime} & N/A\\

\bottomrule
\end{tabular*}
\end{center}
\customcaption{\label{tbl:labs:metrics} Report of benchmark submission for the LABS instance with $n = 20$, also available in \RN{}. All runtimes are reported as the average over the 10 runs.}
} 
\end{table}

Since the Hamiltonian corresponding to LABS is dense with both 2-local and 4-local terms, its realization is challenging due to the large number of entangling gates required. Therefore, we restrict the counterdiabatic terms to include only 2-local terms. We used the 156-qubit \textit{ibm$\_$marrakesh} device with \texttt{optimization level 2} transpilation, dynamical decoupling based on an XpXm sequence, and Qiskit 1.3.1. 
To facilitate the experimental implementation, we adopted the following strategy for initializing the bias fields. 
First, we created a random bitstring and applied a local greedy search with \(n_{\text{s}} = 5\). 
Then, we use the resulting bitstring to compute the initial bias fields. 
We set the initial number of shots to \(n_{\text{shots}} = 500\) and \(n_{\text{s}} = 2\). 
The \(n_{\text{keep}}\) parameter is chosen as \(50\%\) and \(n_{\text{shots}}\) is increased by \(500\) after each iteration. 
In Table \ref{tab:labs_bench}, we present the results for ten different bias-field initializations. For the selected set of parameters, all the initializations resulted in a non-zero success probability. The \(n=20\) LABS problem has an eight-fold degenerate ground state, which can be divided into two distinct independent groups based on spin-flip and reflection symmetry. In our results, we measure optimal sequences from both groups.
The average overall runtime to measure the optimal solution was \(61.2\) seconds. However, excluding the payload quantum circuit preparation, the average runtime becomes \(9.1\) seconds, which is still better than the two exact classical methods shown in Figure~\ref{fig:csolvers_vs_bfdcqo}. 

In summary, this section presents a benchmarking study of BF-DCQO against CPLEX, GUROBI, and QAOA for the LABS problem with sequence lengths of up to 30. Our empirical results demonstrate that BF-DCQO’s runtime scaling to reach the optimal solution is comparable to that of QAOA with \(p=12\) layers, while requiring six times fewer entangling gates. Prior empirical studies have shown that QAOA with \(p=12\) outperforms the best-known classical approach, Memetic Tabu Search, by simulating sequence lengths of up to \(n = 40\). Nevertheless, further research is needed to systematically assess whether BF-DCQO can ultimately achieve a definitive quantum speedup over state-of-the-art classical methods.

\subsection{Minimum Birkhoff Decomposition}
\renewcommand{\customcaptiontext}{(Birkhoff) }
\label{sec:min_birk_exp}

\begin{figure}[ht!]
\centering
\centering
\includegraphics[width=0.49\textwidth]{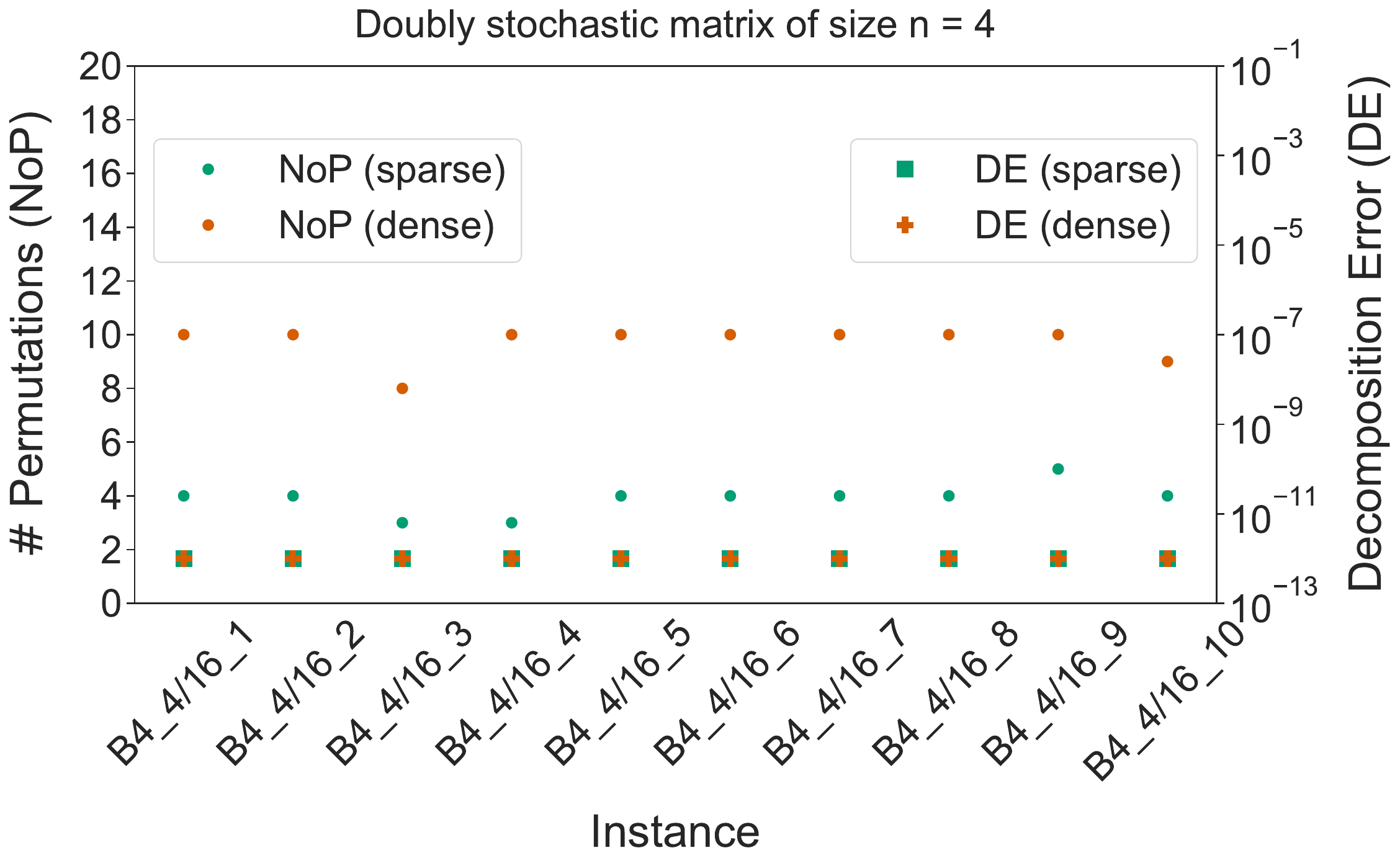}
\includegraphics[width=0.49\textwidth]{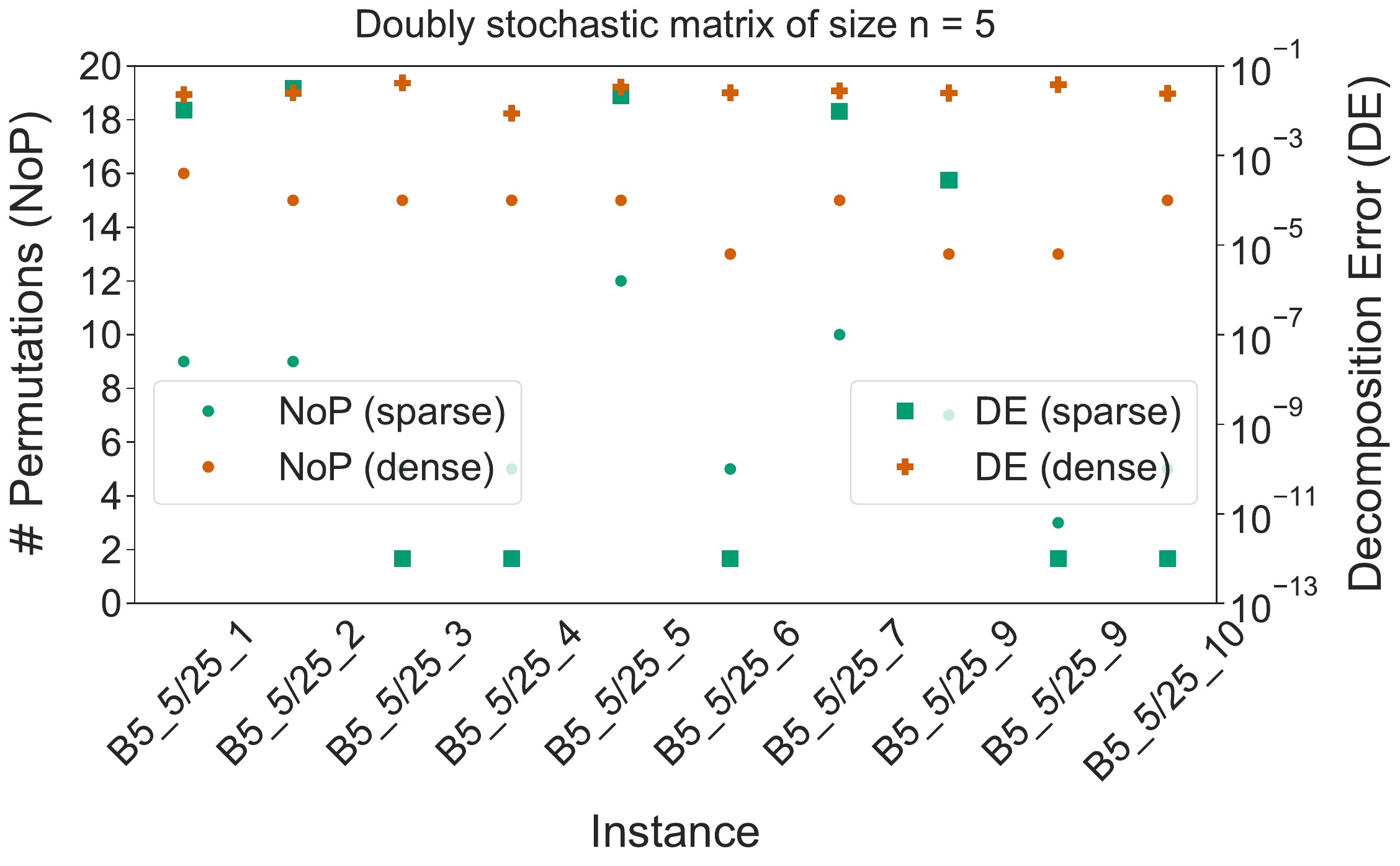}
\includegraphics[width=0.49\textwidth]{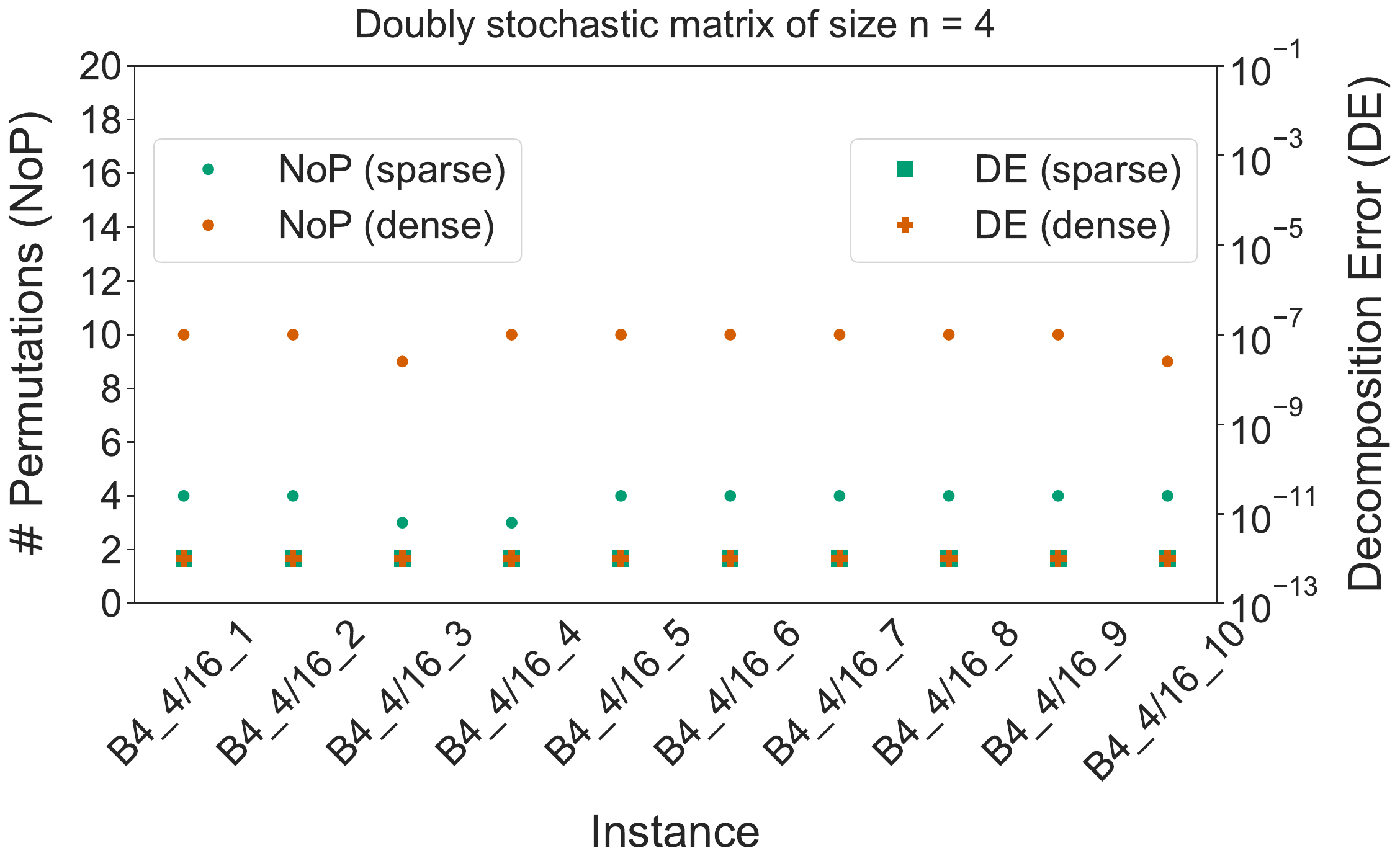}
\includegraphics[width=0.49\textwidth]{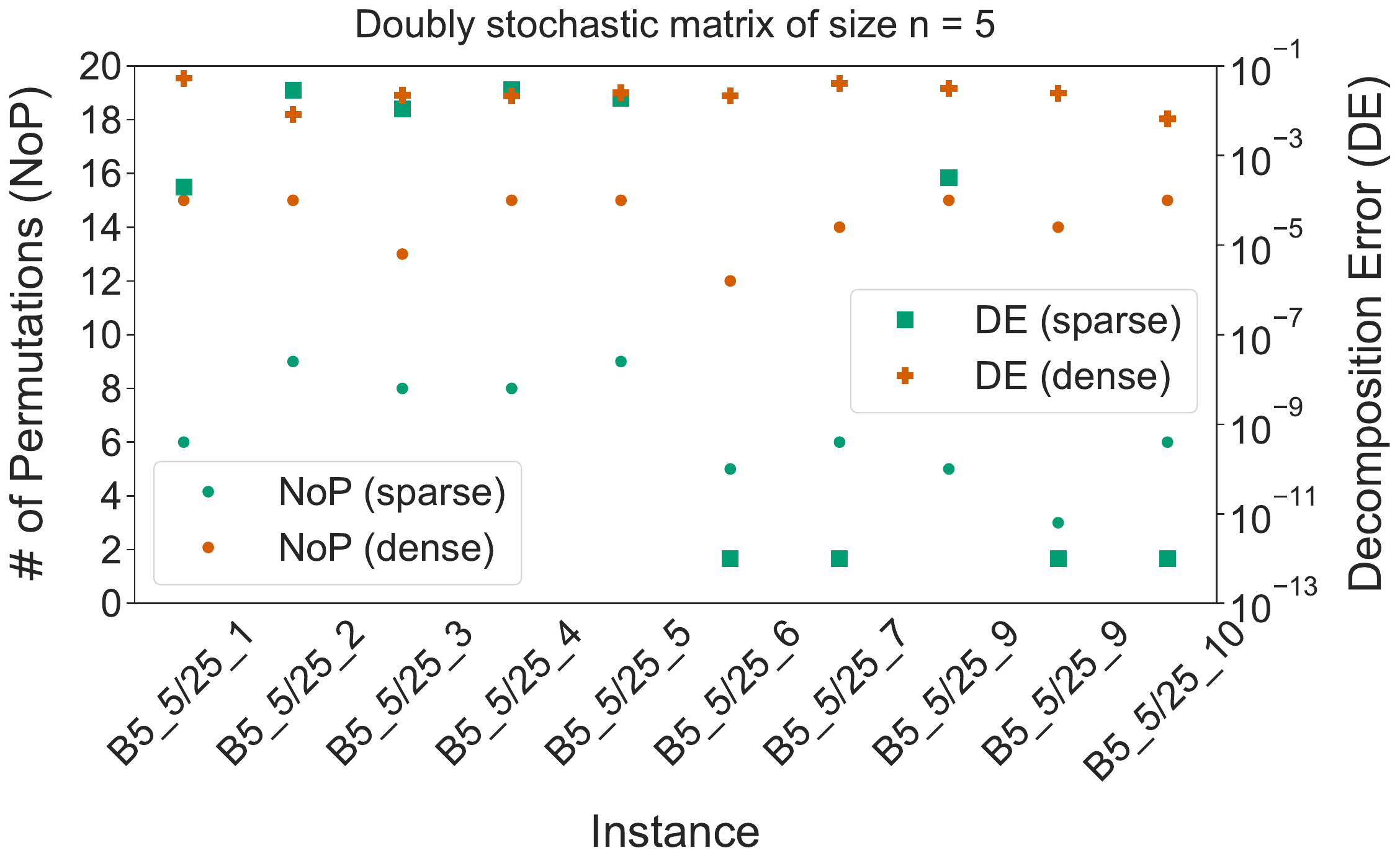}
\customcaption{ Quantum  \emph{simulation} (top) \emph{hardware} (bottom) results for $n=4$ (left) and $n=5$ (right). The \emph{error markers} with $10^{-12}$ error indicate an exact decomposition (i.e., zero error). The x-axis indicates the problem instance BX\_sparse/dense\_Z, where X is the size of the doubly stochastic matrix (i.e., $n$), sparse/dense is the number of permutations used to generate the doubly stochastic matrix ($n$ for sparse and $n^2$ for dense), and Z is the instance number.}
\label{fig:birkplot}
\end{figure}

The following section presents a quantum variational algorithm for the Minimum Birkhoff Decomposition problem, which looks for a decomposition of an $n\times n$ doubly stochastic matrix with $k$ permutation matrices, where $k$ is as small as possible. The solutions found with state-of-the-art classical solvers are given in Section~\ref{sec:Birkhoff_classicalbaseline}. 

The variational quantum algorithm considered here aims to solve the following problem 
\begin{align}
    \underset{\theta \in [0,2\pi]^m}{\text{minimize}} \quad \underset{(P_1,\dots,P_k) \sim \mathcal P(\theta)}{\mathbf E}  \left[ \left\| D - \sum_{i=1}^k c_i  P_i \right\|_2 \right]\:
\label{eq:birk_var_obj}
\end{align}
where $(P_1, \dots,P_k)$ is \emph{a collection} of $k$ different permutation matrices drawn from a parameterized distribution $\mathcal P(\theta)$, and $(c_1,\dots,c_k)$ are weights such that $\sum_{i=1}^k c_i = 1$. 
The distribution $\mathcal P(\theta)$ is prepared with a parameterized quantum circuit where $\theta := (\theta_1,\dots,\theta_m) \in [0,2\pi]^m$ controls $m$ circuit gates.  The variational algorithm runs for a sequence of iterations where $\theta$ is updated in each iteration to minimize the cost function $\mathbf E[\| D - \sum_{i=1}^k c_i  P_i \|_2]$. The cost function is computed in two steps. First, we sample \emph{a collection} of $k$ different permutation matrices with a quantum circuit (\textit{main permutation sampling}) and then compute the weights classically (\textit{black-box optimization}).

To sample $k$ permutation matrices, we use a quantum circuit with four layers. Each layer consists of RY($\theta_j$) gates ($j \in \{1,\dots,m\}$) on every qubit and a CZ pairwise entanglement where qubit $i$ is entangled with qubit $i+1$ for all even values of $i$, and then a second layer where qubit $i$ is entangled with qubit $i+1$ for all odd values of $i$.
We can encode $k$ different permutation matrices as a bit string as follows. First, we write a permutation matrix as a sequence of integers using the Lehmer code \cite{lehmer1960teaching}, and then map the sequence to a \emph{unique} integer in $\{1,\dots,n!\}$. That is, an integer 
$\{1,\dots,n!\}$ corresponds to a permutation matrix. The next step is to map $k$ permutation matrices (i.e., integers) to a unique integer using the combinatorial number system \cite{siddique2016proof}. 
The resulting encoding requires $\lceil \log_2 \binom{n!}{k} \rceil$ qubits, where  $\binom{n!}{k}$ is the number of all possible solutions---i.e., all possible combinations of $k$ permutations. 
For example, with $n=4$ and $k = (n-1)^2 + 1$, there are a total of $\binom{4!}{10} = 1,961,256$ different permutation combinations and the encoding requires $21$ qubits. Similarly, with $k=(n-1)^2 + 1$ and $n=5$, there are $\binom{5!}{17}\approx 1.9 \cdot 10^{20}$ different permutation combinations and we require $68$ qubits. The encoding process described can be reversed to map a bit string back to a unique collection of $k$ permutation matrices. However, since $\log_2 \binom{n!}{k}$ may not be an integer, some bit strings may not correspond to permutation combinations. Thus, if the quantum circuit returns a non-valid bit string, we replace this with a valid bit string selected uniformly at random. 

In the experiments presented below, we employed Optuna 4.1 \cite{akiba2019optuna} as a black-box solver for Eq.~\eqref{eq:birk_var_obj} to optimize the parameters $\theta$. 
After sampling the permutation matrices, the coefficients $c_1,\dots,c_k$ are computed classically in two steps. In the first step (\textit{black-box step 1}), we use CPLEX 22.1.1 \cite{Cplex} to minimize $\| s D - \sum_{i=1}^k u_i P_i\|_2$ with respect to $u_i \in \{0, 1,2,\dots, s\}$ subject to $0 \le u_i \le s$ and $\sum_{i=1}^k u_i = s$ where $s$ is a given integer in the dataset that makes $sD$ and interger-valued matrix. 
In the second step (\textit{black-box step 2}), we use again CPLEX 22.1.1 to find the coefficients $c_i \in [0,1]$ that provide the smallest decomposition subject to $\| D - \sum_{i=1}^k c_i P_i\|_2 \le \| D - \frac{1}{s} \sum_{i=1}^k u_i P_i\|_2$ and $\sum_{i=1}^k c_i = 1$. 
Further details about the data encoding and the algorithmic flow of this variational quantum algorithm can be found in Appendix \ref{app:birkhoff}.

We benchmark the variational algorithm using Qiskit 1.3.1 \cite{qiskit} with a simulator (\emph{Qiskit Aer MPS}) and actual quantum hardware (\emph{ibm$\_$cusco}). All classical execution steps were run on a MacBook Pro (M1 Max CPU and 32 GB RAM).
Each run of the algorithm consists of 10 iterations, where each iteration involves drawing 1024 shots from the quantum computer and hyperparameter optimization of the respective coefficients with a black-box solver, as explained above. 
We run the algorithm once for all instances with size $n = 4$ and $n=5$, and present a performance summary in Figures~\ref{fig:birkplot}. Recall from Section~\ref{subsub:birk_instances} that there are 10 instances per matrix size and density (sparse/dense) and that sparse/dense matrices are generated by sampling, respectively, $n$ and $n^2$ permutation matrices and weights uniformly at random. 
Furthermore, Table  \ref{table:birkhoffsubmissiontable} presents an example of a submission for a particular problem instance (B5\_5\_8) where we run the algorithm 10 times. 

Figure~\ref{fig:birkplot} shows the simulation and hardware results for $n=4$ and $n=5$ with sparse and dense matrices; the dots indicate the number of permutations in the decomposition (left y-axis), and the squares/crosses represent the decomposition error (right y-axis). 
The x-axis indicates the problem instance BX\_Y\_Z, where X is the size of the doubly stochastic matrix (i.e., $n$), Y is the number of permutations used to generate the doubly stochastic matrix ($n$ for sparse and $n^2$ for dense), and Z is the instance number. 
Observe from both figures that when $n=4$ and the target matrix is sparse, the variational algorithm can consistently obtain sparse decompositions of length 3 and 4. For both simulation and hardware results, the solutions coincide with the solution obtained with CPLEX except for one instance (B4\_4\_9; simulation), where the decomposition length is 5 instead of 4 (CPLEX). 
Also, the decomposition is exact for all instances (i.e., zero error). 
With $n=4$ and a dense target matrix, the performance remains similar to the sparse case. Finally, with $n=5$, the performance degrades significantly compared to the $n=4$ case, with only a few instances reaching sparse decompositions with decomposition errors below $10^{-10}$.
Despite solving only a few instances, the result is remarkable for a variational algorithm since there are $\binom{5!}{5}\sim 190$ million combinations of 5 permutation matrices of size $5$.
Moreover, some of the decompositions obtained with the variational algorithm are better than the decompositions obtained with approximate classical algorithms such as Birkhoff+ and Blended FW. For instance, with instance B5\_5\_6, Birkhoff+ obtains an exact decomposition of length 7, while the variational algorithm obtains an exact decomposition of length 5.

\begin{table}[!ht]
\small{
\begin{center}
\begin{tabular*}{\linewidth}{l|l}

\toprule
\textbf{Problem Identifier} & {B5$\_$5$\_$8} ($n = 5$; density: sparse; instance number: 8)\\
\textbf{Submitter} & Mitsuharu Takeori (IBM Research -- Tokyo) \\ &  V\'ictor Valls (IBM Research Europe -- Dublin) \\
\textbf{Date} & January 30, 2025 \\

\midrule
\textbf{Reference} & Appendix~\ref{app:birkhoff} \\

\midrule
\textbf{Best Objective Value} & 5\\ 
\textbf{Optimality Bound}     & N/A \\

\midrule
\textbf{Modeling Approach}       & Quantum black-box optimization with CPLEX evaluated black-box\\
\textbf{\#Decision Variables}     & 68 (main) + 17 (black-box step 1) + 34 (black-box step 2) \\ 
\textbf{\#Binary Variables}       & 68 (main) + 17 (black-box step 2)\\ 
\textbf{\#Integer Variables}      & 17 (black-box step 1) \\ 
\textbf{\#Continuous Variables}   & 17 (black-box step 2)  \\
\textbf{\#Non-Zero Coefficients}  & 35 (black-box step 1) + 18 (black-box step 2)  \\ 
\textbf{Coefficients Type}        & integer \\
\textbf{Coefficients Range}       & $[0, 10000]$ (black-box step 1) \\
      
\midrule 
\textbf{Workflow}          & Each iteration of the variational algorithm calls: \\
                           & 1) Main permutation sampling \\
                           & 2) Black-box optimization step 1 \\
                           & 3) Black-box optimization step 2 \\
\textbf{Algorithm Type}    & Stochastic \\ 
\textbf{\#Runs}            & 10 \\
\textbf{\#Feasible Runs}   & 10 \\
\textbf{\#Successful Runs} & 4 (1st, 2nd, 5th, and 10th run) \\
\textbf{Success Threshold} & $10^{-10}$ \\

\midrule
\textbf{Hardware Specifications}      & CPU: Apple M1 Max (Macbook Pro) , QPU: \emph{ibm\_cusco}.
\\
\midrule
\textbf{Total Runtime}    & 623.8s \\
\textbf{CPU Runtime}      & 603.7s \\
\textbf{GPU Runtime}      & N/A \\
\textbf{QPU Runtime}      & 20.1s \\
\textbf{Other HW Runtime} & N/A \\

\bottomrule
\end{tabular*}
\end{center}
}
\customcaption{Report of benchmark submission for instance {B5\_5\_8} ($n = 5$; density: sparse; instance number: 8) from the Minimum Birkhoff Decomposition problem collection provided in \RN{}. All runtimes are reported as the average over the 10 different runs.
}
\label{table:birkhoffsubmissiontable}
\end{table}

\subsection{Maximum Independent Set}
\renewcommand{\customcaptiontext}{(Independent Set) }

Here, we benchmark standard QAOA~\cite{Farhi2014} on the independent set problem instances defined by graphs $G=(V, E)$ with $|V| = 17$ and $|V| = 52$ variables--see Section~\ref{sec:stable_set_instances}. 
Independent set is a constrained optimization problem which we can, thus, model as a QUBO by adding a penalty term
\begin{align}
\max_x \sum_{i\in V} x_i -\lambda \sum_{(i, j)\in E} x_i x_j.
\end{align}
Here, $\lambda$ is a positive Lagrange multiplier to introduce the constraints.
We convert this QUBO to a Hamiltonian with the variable change $x_i=(1-z_i)/2$ and promote the spin variables $z_i$ to Pauli-$Z$ spin operators $Z_i$.
The resulting Hamiltonian has two contributions: the objective and the constraints
\begin{align}
H_\text{obj}=-\frac{1}{2}\sum_{i\in V} Z_i,\qquad \lambda H_\text{const}=-\frac{\lambda}{4} \sum_{(i,j)\in E}Z_iZ_j-Z_i-Z_j.
\end{align}

We neglect the irrelevant constant energy offsets, i.e., terms that are neither linear nor quadratic in $Z_i$.
These terms are $|V|/2$, and $-\lambda|E|/4$ and stem from the objective and constraints, respectively.
Therefore, the total cost function Hamiltonian is $H=H_\text{obj}+\lambda H_\text{const}$.
To produce bitstrings from the quantum hardware, we sample from a QAOA trial wavefunction $|{\psi(\beta, \gamma)}\rangle=e^{-i\beta H_M}e^{-i\gamma H'}|+\rangle$.
Here, $H_M=\sum_{i\in V} X_i$ is the standard QAOA mixer. $H'$ is the cost operator $H_\text{obj}+\lambda H_\text{const}'$ used to design the trial wavefunction, and we apply $p=1$ repetitions.
Crucially, $H_\text{const}'$ may implement only a subset of the constraints in $H_\text{const}$ to limit the depth of the quantum circuit.
We must now optimize the parameters $\beta$, $\gamma$, and $\lambda$.
In a typical QAOA, this is done in a closed loop with the quantum hardware, which is time and resource-consuming.
Best practices on how to carefully choose transpilation and error suppression strategies for a QAOA benchmark execution can be found on Github~\cite{BestPractices2024}.
In our case, we choose to optimize $\beta$, $\gamma$, and $\lambda$ on classical hardware by evaluating $\langle H\rangle$, which is efficiently possible for a depth-one ansatz~\cite{Egger2021warmstart}.
Optimizing $\lambda$ is a hard task~\cite{alessandroni2024alleviatingquantumbigmproblem}.
Here, we chose to strike a balance between the impact of the actual objective part of the Hamiltonian and the constraint part of the Hamiltonian.
This can be formulated, e.g., as the optimization problem
\begin{align}
\max_\lambda~\min\left(\langle H_\text{obj}\rangle^\star, \lambda\langle H_\text{const}\rangle^\star\right).
\end{align}
More specifically, we are aiming to find the largest $\lambda$, which still gives us the correct solution for the objective function underlying our problem.
Here, the expectation values $\langle\cdot\rangle^\star$ are taken over $|{\psi(\beta^\star, \gamma^\star)}\rangle$ with classically evaluated optimal parameters $\beta^\star$ and $\gamma^\star$ that maximize $\langle H\rangle$ at a fixed $\lambda$.
Notably, $\beta^\star$ and $\gamma^\star$ are found from the inner optimization
\begin{align}
\max_{\beta, \gamma}~\langle \psi(\beta, \gamma)|H_\text{obj} +\lambda H_\text{const}|\psi(\beta, \gamma)\rangle = \max_{\beta, \gamma}~E(\beta,\gamma; \lambda).
\end{align}
It should be noted that the operator in the expectation value is now the full cost Hamiltonian $H$ and not the potentially simplified $H'$.
The parameters $\beta$ and $\gamma$ are optimized with COBYLA, provided by ScipPy. 
Crucially, these QAOA parameters depend on the Lagrange multiplier $\lambda$.
Therefore, the Lagrange multiplier optimization is expressed as
\begin{align}
\max_\lambda~\min\left(\langle H_\text{obj}\rangle^\star, \langle H\rangle^\star - \langle H_\text{obj}\rangle^\star\right)=\max_\lambda F(\lambda).
\end{align}
We optimize $\lambda$ by performing a linear scan over the interval [0, 1] and [0, 0.2] for the 17 and 52 qubit instances, respectively, followed by a refinement using COBYLA from SciPy~\cite{2020SciPy-NMeth}.
The resulting optimized $\lambda$ values are 0.528 and 0.119 for the 17 and 52 qubit instances, respectively.

\begin{figure}[htb!]
    \centering
    \includegraphics[width=0.7\linewidth]{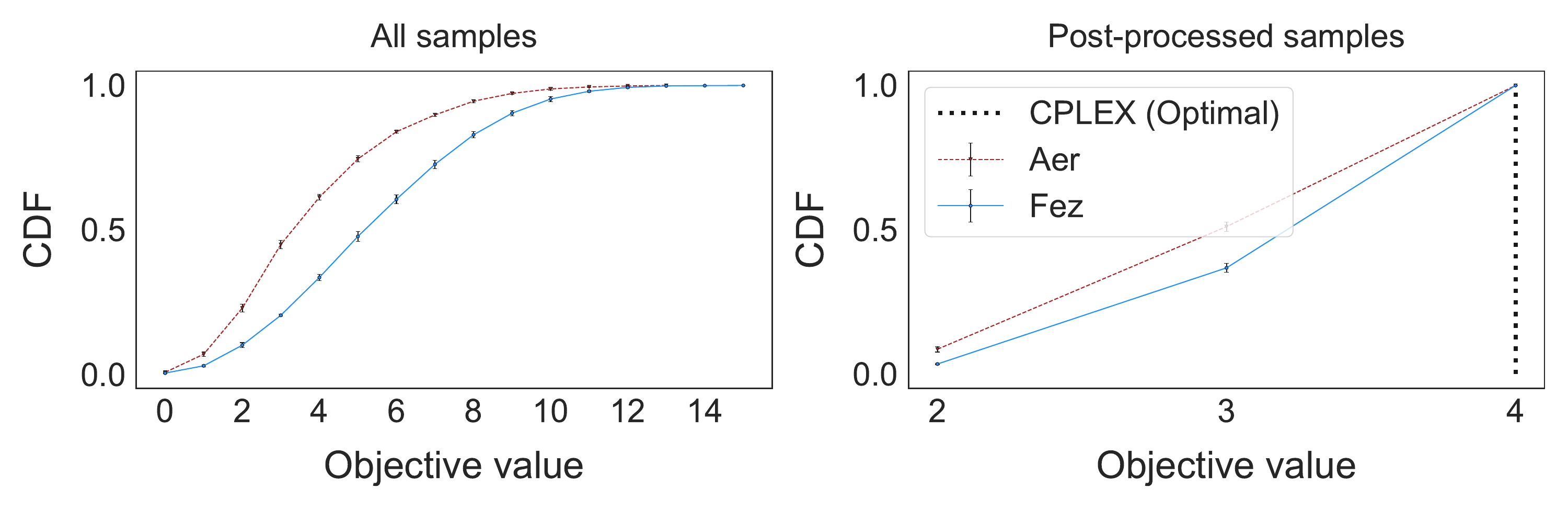}
    
    \includegraphics[width=0.7\linewidth]{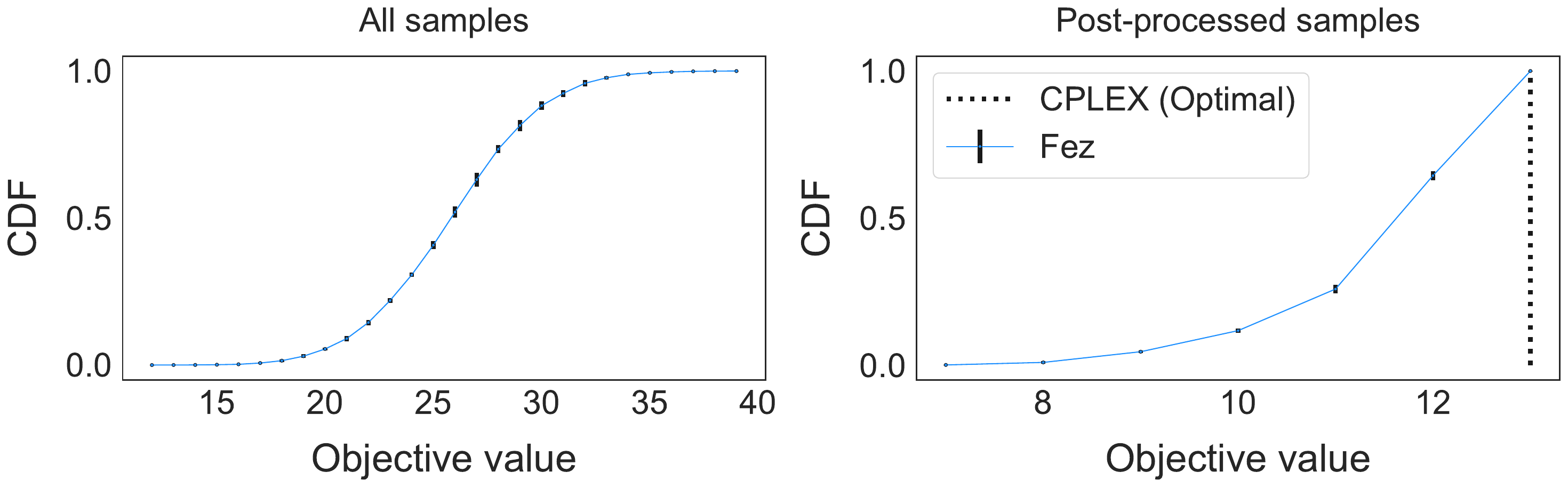}
    \customcaption{Cumulative distribution of the objective value corresponding to the samples for the mamila kangaroo (top) and the aves sparrow (bottom) problem instances before classical post-processing (left) and after post-processing of the quantum samples (right).
    The mamila kangaroo problem is small enough to simulate the QAOA on a classical processor (dashed-dotted line).
    The samples from the \emph{ibm\_fez} quantum processor are shown as the solid blue line.
    }
    \label{fig:stable_bench}
\end{figure}

\begin{table}[htb!]
    \centering
    \begin{tabular}{c r r r r r r}\toprule
       $n$ & Total Runtime & Pre-Proc. (s) &  QPU Runtime (s) & Post-Proc. (s) & Best Obj.Val. \\
          &  &  $(t_\text{opt.};t_\text{circ})$ & 
        &  & 
        \\\midrule
        17 & 77 & 36 (22; 14) & 41 & 0 & 4 (optimal)\\
        52 & 227 & 187 (175;12) & 38 & 2 & 13 (optimal)\\\bottomrule
    \end{tabular}

    \customcaption{ Summary of benchmarking results for mamila-kangaroo and aves-sparrow stable set instances. The process was executed once. The Overall Runtime, reported in seconds, is broken down into (i) the pre-processing time which includes the time $t_\text{opt}$ to obtain $\beta^\star$, $\gamma^\star$, $\lambda^\star$ and $t_\text{circ}$ to create the quantum circuits, (ii) the QPU runtime which includes the payload quantum circuit preparation and execution (queuing time is not included), and (iii) the time taken to post-process the samples.}
    \label{tab:mis_bench}
\end{table}
We would require $\lambda>1$ to make the lowest energy state of $H$ correspond to the optimal solution of the MIS problem.
However, we can only afford large values of lambda if we run deep QAOAs with $p>1$ since they have a better resolution than QAOAs with $p=1$.
Since hardware noise restricts us to shallow circuits, we use $p=1$ and work with small values of $\lambda$.
Once the optimal parameters are found, we create a quantum circuit corresponding to $|\psi(\beta^\star(\lambda^\star), \gamma^\star(\lambda^\star);\lambda^\star)\rangle$ and sample 1024 candidate solutions on \emph{ibm\_fez}, a 156 qubit Heron R2 superconducting qubit processor~\cite{ibm_quantum_platform}.
These circuits are executed in \textit{session mode} as this allows us to report the QPU time as the payload quantum circuit preparation and execution time--as discussed in Section~\ref{sec:resources}. 
Finally, the samples are post-processed by the greedy classical method described in Appendix~\ref{app:mis_psotprocessing}.
This benchmark was executed using Numpy 1.23.5, Qiskit 1.2.2, and SciPy 1.11.3, all running on Python 3.11.5.
We benchmark this approach on stable set instances referred to as \textit{mamila kangoroo} and \textit{aves sparrow} with 17 and 52 variables, which are given in \RN{}.
All constraints of the 17-variable instance are implemented on hardware, i.e., $H'_\text{const}=H_\text{const}$.
This requires a total of 308 CNOT gates.
For the 52-variable instance, we only implement 16.7\% of the constraints by allowing three layers of SWAP 
gates~\cite{weidenfeller2022}.
We, therefore, only implement the constraints that can be built up on a line of qubits on which we apply three layers of SWAP gates as described in Ref.~\cite{weidenfeller2022}.
The resulting circuit has a total of 207 CNOT gates.
This keeps the fidelity of the circuit, approximated by $(1-E_\text{CZ})^{n_\text{CZ}}$, above 50\%.
Here, $E_\text{CZ}$ and $n_\text{CZ}$ are the median error per two-qubit CZ gate and the number of two-qubit CZ gates, respectively.
With the 52-variable instance, we do not sample any feasible solutions. However, after the greedy classical post-processing, we obtained the optimal sample.
The runtime results from the process described above are reported in Table~\ref{tab:mis_bench}.
As expected, most of the runtime is dedicated to optimizing the parameters $\beta$, $\gamma$, and~$\lambda$.
We observe that 60\% and 27\% of the post-processed samples for the 17 and 52 decision variable problems, respectively, are optimal, see Figure~\ref{fig:stable_bench}.
Finally, Table~\ref{table:independentsetsubmissiontable} presents an example of a submission report for the $52-$variable instance.

\begin{table}[htb!]
\small{
\begin{center}
\begin{tabular*}{\linewidth}{l|l}

\toprule
\textbf{Problem Identifier} & 	aves-sparrow-social.gph\\
\textbf{Submitter} & Daniel J. Egger (IBM Research Europe -- Zurich) \\
\textbf{Date} & January 15th, 2025 \\

\midrule
\textbf{Reference} &  \cite{Egger25independentSetQAOAIndSetBenchmark} \\

\midrule
\textbf{Best Objective Value} & 13 \\ 
\textbf{Optimality Bound}     & N/A \\

\midrule
\textbf{Modeling Approach}        & QUBO \\
\textbf{\#Decision Variables}     & 52 \\ 
\textbf{\#Binary Variables}       & 52 \\ 
\textbf{\#Integer Variables}      & 0\\ 
\textbf{\#Continuous Variables}   & 0 \\ 
\textbf{\#Non-Zero Coefficients}  & 506 (the number of edges in the graph + number of nodes) \\
\textbf{Coefficients Type}        & continuous \\
\textbf{Coefficients Range}       & $[0, 2]$ \\

\midrule 
\textbf{Workflow}          & 1) The parameters $\beta$ and $\gamma$ of the depth-one QAOA and the \\
                           & $\quad$ Lagrange multipliers are optimized classically. \\
                           & 2) Samples are drawn from the QPU. \\
                           & 3) Samples from the QPU are classically post-processed. \\
\textbf{Algorithm Type }   & Stochastic \\
\textbf{\#Runs}            & 5 \\
\textbf{\#Feasible Runs}   & 5 \\
\textbf{\#Successful Runs} & 5 \\
\textbf{Success Threshold} & 0 \\

\midrule
\textbf{Hardware Specifications}      & CPU: Intel Core i9-10885H (Levono P1), QPU: \emph{ibm\_fez}.
\\\midrule
\textbf{Total Runtime}    &  252s \\
\textbf{CPU Runtime}      & 192s \\
\textbf{GPU Runtime}      &  N/A \\ 
\textbf{QPU Runtime}      & 60s \\
\textbf{Other HW Runtime} & N/A \\

\bottomrule
\end{tabular*}
\end{center}
}
\customcaption{Benchmark submission, for instance aves-sparrow-social from the independent set benchmarking problem collection provided in the \RN{} repository.
} 
\label{table:independentsetsubmissiontable}
\end{table}

\renewcommand{\customcaptiontext}{}
\section{Conclusion and Discussion}
\label{sec:discussion_conclusion}

It is widely believed that quantum computers can achieve an advantage over classical computers for \emph{some} optimization problem instances in the future. 
While complexity theory suggests that, for \NP-hard problems, the speed-up for finding provably optimal solutions is in general at most quadratic \cite{bennett1997strengths, de2019quantum, durr1996quantum, Gilliam_2021_grover}, better quantum approximation algorithms with an exponential speed-up \cite{jordan2024dqi} or quantum optimization heuristics that outperform classical algorithms on certain instances may still exist.
However, practical demonstrations of quantum advantage in optimization are still missing.

To identify problems where quantum advantage might be possible and to track progress toward achieving it, systematic analysis and benchmarking are key.
To this end, we introduce \RN{}, an open-source quantum optimization benchmarking library.
We present ten classes of combinatorial optimization problems with instances of varying size and complexity for benchmarking classical and quantum optimization algorithms. 
Further, we define a reporting standard that ensures reproducibility and fair comparison while remaining flexible for evaluating different algorithms and hardware platforms.

Most problems are challenging for state-of-the-art classical solvers even at moderate size, and many of them are related to use cases of practical relevance. 
The problem classes exhibit diverse properties, such as different densities and types of constraints, to enable the study of different mathematical formulations and algorithmic approaches. This is essential for identifying problem structures suitable for quantum advantage in optimization.

We provide classical baseline results for all problem classes and quantum baseline results for selected ones to kick off this effort. However, performing exhaustive benchmarks is a massive endeavor and can only be achieved by the broader scientific community. 
Thus, we invite all researchers interested in advancing (quantum) optimization to investigate the provided problem instances and submit solutions using algorithms and hardware of their choice.
We also want to emphasize the importance of improving classical methodologies, amongst other reasons, because quantum advantage can only be claimed if the best-known classical methods are outperformed. 
To accelerate the overall research, we also encourage researchers to report results for approaches that did not work well.
To conclude, \RN{} is intended to be a living repository: novel solutions will be included on a rolling basis, and novel problem instances or adaptations of existing ones may be added as the state-of-the-art evolves.

\vspace{3mm}

\subsubsection*{Acknowledgments}
We want to thank Karen Aardal at TU-Delft for providing results for the Market Split instances using lattice point enumeration and Nils-Christian Kempke at ZIB for providing results by GPU-accelerated Schroeppel Shamir enumeration.
We would like to thank S\'ebastien Designolle at the Zuse Institute Berlin for useful discussions on the Blended FW algorithm used in the classical benchmarking of the minimum  Birkhoff decomposition problem (Section~\ref{sec:Birkhoff_classicalbaseline}). 
We also thank Julien Gacon for valuable discussions and a helpful review, and Takashi Imamichi for assistance with software issues, sharing his insights, and providing feedback.
Additionally, we are grateful to Andrew Wack and Paul Nation for insightful conversations about the benchmarking of quantum algorithms and their feedback on this work.

Part of the work for this article has been conducted in the Research Campus MODAL funded by the German Federal Ministry of Education and Research (BMBF) (fund numbers 05M14ZAM, 05M20ZBM, 05M2025). 
GN is supported by ONR award \# N000142312585. 
YC would like to acknowledge the support from the Quantum Engineering Program project A-8000339-01-00, from the Ministry of Education Singapore under project A-8000014-00-00, and by the National Research Foundation Singapore under its Industry Alignment Fund – Pre-positioning (IAF-PP) Funding Initiative and the Monetary Authority of Singapore under A-0003504-17-00. 
DVB would like to acknowledge the support of the Research Foundation Flanders (FWO) [23AXE35N].



\pagebreak

\subsubsection*{Author Contributions}
This review was written as part of the Quantum Optimization Working Group initiated in July 2023 by IBM Quantum and its partners.

\noindent
TKo: Conceptualization, Software, Investigation, Resources, Data Curation, Writing--Original Draft, Writing--Review and Editing, Supervision;
DEBN: Conceptualization, Writing--Original Draft, Writing--Review and Editing;
YC: Conceptualization, Investigation, Modeling, Resources, Writing--Original Draft;
GC: Conceptualization,  Writing--Review and Editing;
DJE: Investigation, Data Curation, Writing--Original Draft;
RHe: Investigation, Resources, Writing--Original Draft;
NNH: Investigation, Resources, Writing--Original Draft;
AGC: Investigation, Resources, Writing--Original Draft;
RHu: Software, Data Curation;
TIm: Investigation, Data Curation;
TIt: Writing--Original Draft;
TKl: Resources, Writing--Original Draft;
PMX: Writing--Original Draft, Investigation, Software;
NM: Investigation, Data Curation, Writing--Original Draft;
JAMB: Writing - Review and Editing ;
KN: Software, Investigation;
GN: Conceptualization, Methodology, Software, Investigation, Resources, Data Curation, Writing--Original Draft, Writing--Review and Editing;
COM: Conceptualization, Writing-Review and Editing;
JP: Investigation, Resources;
MP: Writing--Original Draft;
AR: Writing--Original Draft, Investigation;
MS: Software, Investigation, Resources, Data Curation;
NS: Software, Investigation;
MT: Investigation;
VV: Methodology, Software, Investigation, Resources, Data Curation, Writing--Original Draft;
DVB: Software, Investigation, Data Curation;
SW: Conceptualization, Writing--Original Draft, Writing--Review and Editing, Supervision;
CZ: Conceptualization, Writing--Original Draft, Writing--Review and Editing, Supervision

\appendix

\clearpage
\section{Building Capabilities via Instance Simplification}
\label{app:instance_simplification}

The problem instances we propose are challenging to even implement on current quantum computers due to noise, which limits the circuit depth. To track the progress of quantum algorithms and hardware today we therefore need a spectrum of simplified problem instances to observe a quantum signal. These instances will often be easy to solve classically. We propose to generate simplified QUBOs such that the time-evolution of the corresponding Hamiltonian becomes less difficult. More specifically, we consider a line of $n$ qubits which achieves all-to-all connectivity with a SWAP network with $n-2$ layers of SWAP gates~\cite{weidenfeller2022}. Crucially, each entry $(i, j)$ of the QUBO matrix can be implemented after $d(i,j)\in\{0, ..., n-2\}$ layers of SWAP gates. The distances $d_{i,j}$ allow us to define an $n\times n$ mask where entry $m_{i, j}(k)$ is one if $d(i, j)\leq k$ and zero otherwise. The number of allowed layers $k$ thus controls the difficulty of implementing $\exp(-i H_C)$ on hardware. We simplify a QUBO matrix with this mask through the mapping $Q_{i,j}\to Q_{i,j}'=m_{i,j}(k)Q_{i,j}$. Crucially, this only makes sense if the resulting QUBO $Q'$ remains connected. In the benchmarking repository, we provide a code to simplify a QUBO matrix following this procedure. Crucially, in this process, the code generates a named field which contains the number of SWAP layers that constrain the QUBO for traceability.

\section{Market Split Classical Results}

\begin{figure}[htb!]
\centering
\includegraphics[width=0.9\textwidth]{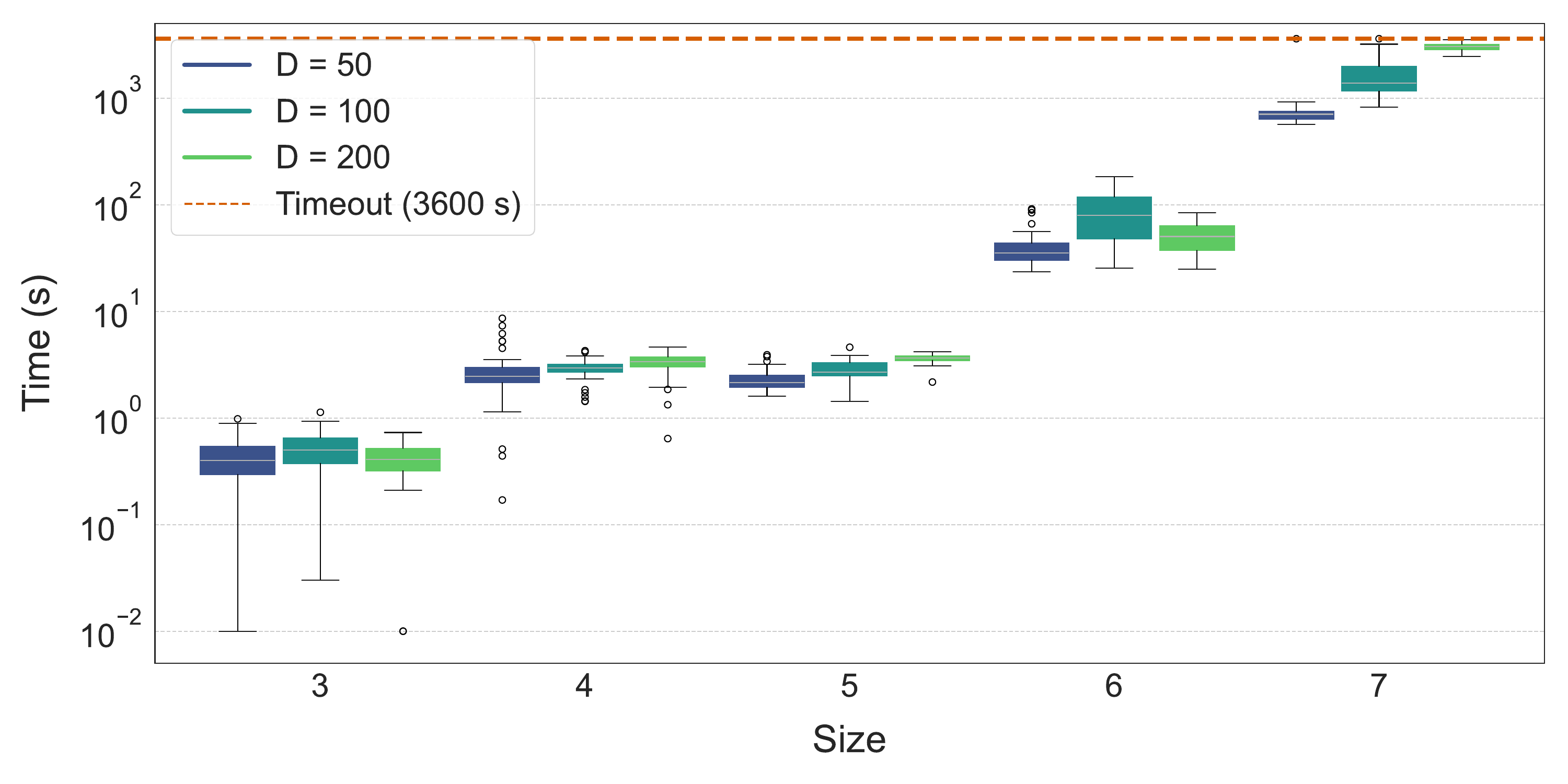}
\includegraphics[width=0.9\textwidth]{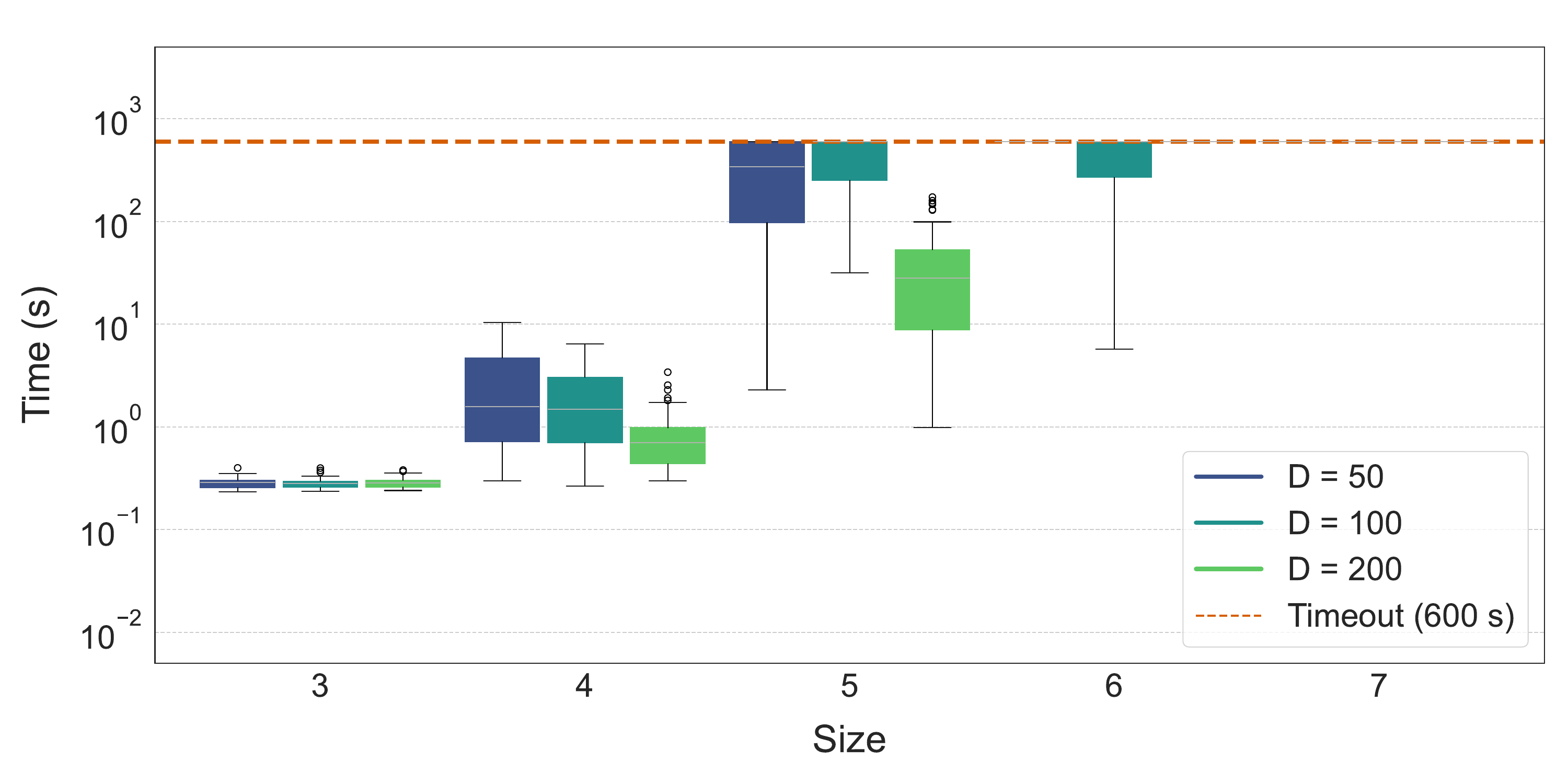}
\customcaption{ 
 Distribution of solution times using (top) \gurobi 11.0.0 with a time limit of 3600s and (bottom) \abstwo with a time limit of 600s for instances with different sizes $m$ and coefficients range $D$. All trials for $m \geq 8$ timed out for \gurobi and for $m \geq 7$ timed out for \abstwo.}
\label{fig:marketshare:baseline:plot:gurobi}
\end{figure}

Fig~\ref{fig:marketshare:baseline:plot:gurobi} shows the results for \gurobi on a 32-core AMD EPYC-7542 processor using 8 cores and \abstwo~\cite{NakanoEtal2023} on an A100 SXM4 80GB, respectively, for different sizes of instances with different parameters $m$.
\abstwo is very fast for smaller instances, but fails to find solutions within one hour for instances larger than size $m=6$. 

\section{Market Split Instance Generation}
\label{app:marketsplit}
To benchmark solvers, it is essential to generate feasible instances of this problem. A naive approach is to randomly sample the entries of $A$ from a uniform distribution and compute $b$ accordingly. However, this approach often results in infeasible instances, which are less useful for benchmarking.
For instances that were too large to be tested, i.e., $m\geq 7$, we constructed instances as follows:

\begin{enumerate}
\item Generate a Random Solution Vector:
Create a random solution vector $x$ where approximately half $(\pm 2)$ of the entries are set to $1$, and the rest are set to $0$.

\item Generate Matrix $A$:
Sample the entries of $A$ uniformly from a specified range $D$.

\item Compute the Right-Hand Side $b$:
Compute each entry of $b$ as:
\[b_i = \left\lfloor \frac{\sum_{j \in [n]} a_{ij}}{2} \right\rfloor\]

\item Adjust for Feasibility:
Ensure the feasibility of the generated instance using the following steps for each row:
\begin{itemize}
    \item Step 1: Repeat until no more improvement is possible:

Identify a pair of entries in the row where one corresponds to $0$ in $x$ and the other to $1$.
Perform a ``switch'' that reduces the absolute slack of the row as much as possible.

\item Step 2: If the row is still not feasible:

Calculate the slack $s$ for the row.
Set $c = \frac{3 \cdot \text{numones}}{2}$, where num\_ones is the number of $1$'s in the row.
Find $c$ pairs of $0$ and $1$, and adjust the corresponding entries by adding $\frac{s}{c}$ to the $1$'s and subtracting it from the $0$'s.

\item Step 3: Repeat this adjustment until the row becomes feasible.

\end{itemize}

In most cases, only a few switches and adjustments are sufficient to achieve feasibility.
\end{enumerate}

\section{Minimum Birkhoff Decomposition \texorpdfstring{\\}{}
Quantum Baseline Benchmark -- Additional Details}
\label{app:birkhoff}
\renewcommand{\customcaptiontext}{ (Birkhoff)}

This section presents additional details that facilitate the reproducibility of the quantum baseline benchmarking results for the minimum Birkhoff decomposition problem presented in Section~\ref{sec:min_birk_exp}.

\subsection*{Model Encoding}
The following example illustrates the encoding strategy used in the variational algorithm, which was employed to generate the results presented in Section~\ref{sec:min_birk_exp}. 

\vspace{3mm}
\textbf{Permutation to bistring.} Consider the following $4 \times 4$ permutation matrix
\begin{align}
\begin{pmatrix}
0 & 1 & 0 & 0 \\
0 & 0 & 1 & 0 \\
1 & 0 & 0 & 0 \\
0 & 0 & 0 & 1 
\end{pmatrix}.
\end{align}
Recall there are $4! = 24$ permutation matrices. We can write the permutation matrix as an array $$p = [ 2, 3, 1, 4 ].$$ The Lehmer code \cite{lehmer1960teaching} maps $p$ to another array $l$ where the $i$-th entry of $l$ is equal to $l_i = \sum_{j=1}^{i-1} 1_{p_j < p_i}$. That is, $l_i$ is equal to the number of values $p_1,\dots,p_{i-1}$  smaller than $p_i$. For the permutation array $p = [ 2, 3, 1, 4 ]$, the Lehmer code mapping is
 $$l = [0,1,0,3].$$ Next, we can map the Lehmer code to an integer in $[0,4!-1]$. In our example, the mapping of $[0,1,0,3]$ to an integer is as follows: $$0! \cdot 0 + 1! \cdot 1 + 2! \cdot 0 + 3! \cdot 3 = 19,$$ where 19 corresponds to the bit string $10011$.

 \vspace{3mm}
\textbf{Bitstring to Permutation}. Consider the bitstring 10011. We can map the bit string to the integer 19, and 19 to the array $[0,1,0,3]$ by reversing the process used above. Similarly, we can map $[0,1,0,3]$  to $p = [ 2, 3, 1, 4 ]$ and so recover the permutation matrix. Finally, note that while we can always map a permutation matrix to a bit string, the inverse is not always possible. An example is the bit string 11111 (31) since there are 24 permutation matrices, and the encoding above maps a permutation matrix to an integer in $\{0,1,2,\dots,4!-1\}$. If such a case occurs, we return a random bit string in the feasible range $\{0,1,2,\dots,n!-1\}$.

 \vspace{3mm}
\textbf{Mapping $k$ unique integers to an integer}. Suppose we have $k= 4$ different integers $[17, 19, 18, 16]$, each representing a unique permutation matrix. Next, we order the integers in ascending order: 
$$a = [16, 17, 18, 19].$$
Map the integers to a unique integer using the combinatorial number system, i.e.,  $b = \sum_{i=1}^k \begin{pmatrix}
    a_i \\ i
\end{pmatrix}$. In our example, 
\begin{align}
\begin{pmatrix}
    16 \\ 1
\end{pmatrix} +
\begin{pmatrix}
    17 \\ 2
\end{pmatrix}
+
\begin{pmatrix}
    18 \\ 3
\end{pmatrix}
+ 
\begin{pmatrix}
    19 \\ 4
\end{pmatrix}
 = 4844.
\end{align}

 \vspace{3mm}
\textbf{Mapping an integer to $k$ unique integers}. Suppose we want to map 4844 to $k = 4$ unique integers. We start by finding the largest integer $x$ such that $\begin{pmatrix}
    x \\ 4
\end{pmatrix}\le 4844$. In this case, $x$ is equal to 19. Repeat the process for $4844 - \begin{pmatrix}
    19 \\ 4
\end{pmatrix} = 968$ and $k-1$, i.e., find $x$ such that $\begin{pmatrix}
    x \\ 3
\end{pmatrix} \le 968$, which is 18. Repeating the process until the remainder is zero gives the integers $[19,18,17,16]$.

\subsection*{Ansatz and black-box optimization}

The variational part of the algorithm samples bit strings using the following variational circuit (Qiskit 1.3.1):
\lstdefinestyle{vicpython}{
language=Python,
basicstyle=\ttfamily\footnotesize
}

\begin{lstlisting}[style=vicpython]
ansatz = n_local(
    num_qubits = n,
    rotation_blocks = "ry", 
    entanglement_blocks = 'cz',
    entanglement = 'pairwise', 
    reps = 4
    )
\end{lstlisting}


    
Furthermore, the optimization of the parameters is carried out with Optuna 4.1 \cite{akiba2019optuna} with the CMAES sampler. There are a total of 10 optimization rounds/iterations, where each round measures 1024 shots with the quantum hardware or simulator, respectively. 
For each shot, the black-box computes the objective value as described in Section~\ref{sec:min_birk_exp}. 
Recall that the black-box optimization takes as an input a collection of $k$ permutation matrices, the target matrix $D$, and an integer $s$ such that $sD$ is integer-valued. The following is a Python (3.10) implementation of the black-box optimization using CPLEX (22.1.1): 

\begin{lstlisting}[style=vicpython]
     ########## Black-box optimization input  ##########
    A # $n^2 \times k$ matrix where columns are permutation matrices
    b # Target permutation matrix as column vector
    scale # integer that makes A * scale an integer matrix
    k # maximum decomposition length
    
    ########## Black-box optimization: STEP 1 ##########
    
    model = Model()
    w = model.integer_var_list(A.shape[1])

    # add constraints    
    model.add_constraint(
    	model.sum(w[index] for index in range(0,A.shape[1])) == scale)
    for i in range(0,A.shape[1]):
        model.add_constraint(w[i] >= 0)
    for i in range(0,A.shape[1]):
        model.add_constraint(w[i] <= scale)

    # add objective
    model.minimize(model.sum_squares(
    	(model.sum(A[i,j]*w[j] for j in range(0,A.shape[1])) - b[i]) 
		for i in range(0,A.shape[0])))
    
    # solve
    solution = model.solve()
    sol = solution.get_value_list(w)
   
    # compute error  for step 2
    dense_val = np.linalg.norm(A @ sol - b,2)
    
    ########## Black-box optimization: STEP 2 ##########
    model = Model()
    w = model.continuous_var_list(k)
    y = model.binary_var_list(k)
    p = np.asarray((P.flatten()))[0]

    # add constraints
    model.add_constraint(model.sum(w[index] 
    	for index in range(0,k)) == 1)
    for i in range(0,k):
        model.add_constraint(w[i] >= 0)
    for i in range(0,k):
        model.add_constraint(w[i] <= 1)
    for i in range(0,k):
        model.add_constraint(w[i] <= y[i])
    
    # add objective
    model.add_constraint((model.sum_squares(
    	(model.sum(w[j]*p[i + j*n*n] for j in range(0,k)) - b[i]) 
		for i in range(0,n*n))) <= dense_val)
    model.minimize(model.sum(y[i] for i in range(0,k)))
    
    # solve
    solution = model.solve()
    
    \end{lstlisting}

\subsection*{Solution for the submission in Table \ref{table:birkhoffsubmissiontable}}
The doubly stochastic matrix in instance \text{B5\_5\_8} is:
\begin{align}
D = 
\begin{bmatrix}
 0.0143  & 0.6351  & 0.0488  & 0.151 &   0.1508 \\
 0.7859 &  0.1998 &  0     & 0.0143 &  0 \\
 0  &   0   &  0.151 &  0.1508 & 0.6982 \\
 0.151 &  0.1508 & 0.6494 & 0.0488 & 0 \\
 0.0488 & 0.0143 & 0.1508 & 0.6351 & 0.151
\end{bmatrix}.
\end{align}
The decomposition obtained in the four successful runs is:

\begin{align}
 D & = 0.0143
    \begin{bmatrix}
    1 & 0 & 0 & 0 & 0 \\
    0 & 0 & 0 & 1 & 0 \\
    0 & 0 & 0 & 0 & 1 \\
    0 & 0 & 1 & 0 & 0 \\
    0 & 1 & 0 & 0 & 0
    \end{bmatrix}
    + 0.6351 
    \begin{bmatrix}
    0 & 1 & 0 & 0 & 0 \\
    1 & 0 & 0 & 0 & 0 \\
    0 & 0 & 0 & 0 & 1 \\
    0 & 0 & 1 & 0 & 0 \\
    0 & 0 & 0 & 1 & 0
    \end{bmatrix}
    + 0.1508
    \begin{bmatrix}
     0 & 0 & 0 & 0 & 1 \\
     1 & 0 & 0 & 0 & 0 \\
     0 & 0 & 0 & 1 & 0 \\
     0 & 1 & 0 & 0 & 0 \\
     0 & 0 & 1 & 0 & 0
    \end{bmatrix}
    \\
    & \ \ + 0.151
    \begin{bmatrix}
     0 & 0 & 0 & 1 & 0 \\
     0 & 1 & 0 & 0 & 0 \\
     0 & 0 & 1 & 0 & 0 \\
     1 & 0 & 0 & 0 & 0 \\
     0 & 0 & 0 & 0 & 1 \\
    \end{bmatrix}
    + 0.0488
    \begin{bmatrix}
     0  & 0 &  1 & 0 & 0 \\
     0  & 1 &  0 & 0 & 0 \\
     0  & 0 &  0 & 0 & 1 \\
     0  & 0 &  0 & 1 & 0 \\
     1  & 0 &  0 & 0 & 0
    \end{bmatrix}.
\end{align}

The decomposition length obtained with the variational quantum algorithm coincides with the decomposition length found with the integer program formulation solved with CPLEX (see formulation in Section~\ref{sec:Birkhoff_classicalbaseline}).

\pagebreak
\section{Maximum Independent Set Classical Post-processing\label{app:mis_psotprocessing}}

Here, to keep the paper self-contained, we provide the classical algorithm with which the samples drawn from the quantum computer are post-processed in the maximum independent set benchmark.
Algorithm~\ref{alg:stable_post} first makes a given sample $x$ feasible by removing nodes from the independent set until all the constraints are satisfied.
Second, Algorithm~\ref{alg:stable_post} adds nodes back into the independent set as long as they do not introduce any violations of the constraints.

\begin{algorithm}
\caption{Greedy post-processing of candidate solutions $x$ to make $x$ feasible, i.e., an independent set, and enlarge the set while remaining feasible.}\label{alg:stable_post}
\begin{algorithmic}
\Require candidate solution $x\in\{0,1\}^n$
\State $v \gets \text{compute\_violations}(x)$ \Comment map vertex to the number of constraints it violates
\State $c\gets V$ \Comment $c$ are vertices we try to add to the stable set \\

\While{sum($v$.values()) $>$ 0} \Comment Make $x$ feasible
\State $v_\text{max} \gets$ vertex that violates the largets number of constraints
\State $x[v_\text{max}]\gets 0$
\State $c$.remove($v_\text{max}$)
\State $v \gets \text{compute\_violations}(x)$
\EndWhile \\

\While{$\text{len}(c)>0$}\Comment Add vertices to the stable set and remain feasible
    \State $u \gets c.\text{pop}()$
    \State $x'\gets x$
    \State $x'[u]\gets 1$
    \If {$\text{is\_feasible}(x')$}
        \State $x\gets x'$
    \EndIf
\EndWhile \\
\Return $x$
\end{algorithmic}
\end{algorithm}

\makeatletter
\let\orig@lbibitem\@lbibitem
\def\@lbibitem[#1]#2{%
  \def\tempkey{#2}\def\target{gitlab_repo}%
  \ifx\tempkey\target
    \orig@lbibitem[QOBLIB]{#2}%
  \else
    \orig@lbibitem[{#1}]{#2}%
  \fi}
\makeatother

\pagebreak   
\bibliographystyle{alpha}
\bibliography{sample}

\end{document}